\documentclass[11pt]{article}
\usepackage{epsfig} 
\usepackage{amssymb}
\usepackage{graphics}
\newcommand{\sect}[1]{ \section{#1} \setcounter{equation}{0} } 

\newcommand{\half}{\mbox{\small{$\frac{1}{2}$}}}

\newcommand{\Nf}{N_{\!f}}
\newcommand{\NF}{N_{\!F}}
\newcommand{\NA}{N_{\!A}}
\newcommand{\Cc}{\mathbb{C}}
\newcommand{\Pp}{\mathbb{P}}

\newcommand{\Dslash}{D \! \! \! \! /}
\newcommand{\MSbar}{\overline{\mbox{MS}}}

\newcommand{\MSbarss}{\overline{\mbox{\scriptsize{MS}}}}
\newcommand{\mMOM}{\mbox{mMOM}}

\newcommand{\MaxSs}{\overline{\mbox{\footnotesize{MaxS}}}}
\newcommand{\MaxSss}{\overline{\mbox{\scriptsize{MaxS}}}}

\newcommand{\MOMts}{\widetilde{\mbox{\footnotesize{MOM}}}}
\newcommand{\MOMtss}{\widetilde{\mbox{\scriptsize{MOM}}}}

\newcommand{\QEDss}{\mbox{\scriptsize{QED}}}

\newcommand{\sQEDss}{\mbox{\scriptsize{sQED}}}

\begin{document}

\title{Four loop renormalization in six dimensions using {\sc Forcer}}

\author{J.A. Gracey, \\ Theoretical Physics Division, \\ 
Department of Mathematical Sciences, \\ University of Liverpool, \\ P.O. Box 
147, \\ Liverpool, \\ L69 3BX, \\ United Kingdom.} 

\date{}

\maketitle 

\vspace{5cm} 
\noindent 
{\bf Abstract.} We employ the {\sc Forcer} algorithm to renormalize a variety
of six dimensional field theories to four loops. In order to achieve this we 
construct the {\sc Forcer} master integrals in six dimensions from their four 
dimensional counterparts by using the Tarasov method. The $\epsilon$ expansion 
of the six dimensional masters are determined up to weight $9$ where
$d$~$=$~$6$~$-$~$2\epsilon$. By applying the {\sc Forcer} routine the four loop
$\MSbar$ renormalization of $\phi^3$ theory is reproduced before gauge theories
are considered. The renormalization of these theories is also determined in the
$\MOMts$ scheme. For instance the absence of $\zeta_4$ and $\zeta_6$ is 
confirmed to five loops in the $\MOMts$ renormalization of $\phi^3$ theory. We 
also evaluate the three loop $\beta$-function of the gauge coupling in six 
dimensional QCD.

\vspace{-16.6cm}
\hspace{13.4cm}
{\bf LTH 1369}

\newpage 

\sect{Introduction.}

There have been significant developments in recent years to progress our
knowledge of the renormalization group functions of four dimensional quantum
field theories and in particular those of various sectors of the Standard
Model. The main development rests in the advances in the algorithms used to 
calculate Feynman integrals and the implementation of these algorithms in 
highly efficient computer algebra and symbolic manipulation languages. The
classic representation of these advances is the establishment of the five loop
$\beta$-function of Quantum Chromodynamics (QCD) which is the field theory that
relates to the strong interactions. Several groups produced the result
contemporaneously, \cite{1,2,3,4,5}, using one of two separate methods to
achieve this impressive level of precision. One is the Laporta algorithm,
\cite{6}, which is a systematic way of solving integration by parts relations
to write all the Feynman integrals of the Green's function of interest in terms
of a relatively small basis set of core or master integrals. The evaluation
of the latter by non-integration by parts methods allows for the Green's
function to be fully determined. While such a systematic approach will always
reach its goal, it is not always the case that this will be achieved in a
reasonable time frame. One other main technique that led to five loop QCD
renormalization group functions, \cite{2,7}, was the development of an 
efficient four loop integration package termed {\sc Forcer}, \cite{8,9}. It is
written in the symbolic manipulation language {\sc Form}, \cite{10,11}. The
{\sc Forcer} package is tailored specifically to evaluate four loop massless 
$2$-point functions in $d$-dimensional spacetime. For realistic computations 
the $\epsilon$ expansion in $d$~$=$~$4$~$-$~$2\epsilon$ dimensions can be 
extracted within the {\sc Forcer} framework up to weight $12$. The routine 
represents a generational advance on its precursor which was {\sc Mincer}, 
\cite{12}. That was an algorithm for evaluating massless three loop $2$-point 
Feynman integrals. Although it was developed in the 1980s it was very much 
ahead of its time given our current knowledge of the mathematics underlying 
Feynman integral computations. However the increase in experimental precision 
in recent years has meant that higher loop order algorithms for automatically 
evaluating Feynman integrals are now necessary. In fitting this requirement 
{\sc Forcer} implements a new integration rule for massless fields that extends
the core rule central to {\sc Mincer}. Moreover this diamond rule, \cite{13}, 
represents an efficient improvement for the more algebraically demanding four
loop Feynman integrals.

While {\sc Forcer} is now the default package of massless four loop $2$-point 
functions and was central to \cite{2,7}, there are theories and problems in 
other spacetime dimensions that it could be useful for. For instance in six 
dimensions there has been interest in gauge theories with and without 
supersymmetry. In \cite{14} the one loop $\beta$-functions of the two coupling 
non-abelian gauge theory was studied to ascertain whether the model was unitary
and perturbatively conformal. Aside from \cite{14} there is interest in the 
ultraviolet completion of four dimensional theories. As an example of this we 
recall that four dimensional scalar $\phi^4$ theory has scalar $\phi^3$ theory 
as its ultraviolet completion in six dimensions. In other words $\phi^4$ theory
with an $O(N)$ symmetry lies in the same universality class in $d$-dimensions 
at the Wilson-Fisher fixed point as $O(N)$ $\phi^3$ theory in six dimensions, 
\cite{15,16,17}. This concept extends beyond scalar theories to include 
fermionic ones such as those with scalar-Yukawa interactions, \cite{18,19,20}, 
scalar and fermionic Quantum Electrodynamics (QED) and QCD, \cite{21,22}. In 
the latter two cases the ultraviolet completion of six dimensional QED and QCD 
begins in two dimensions with the abelian and non-abelian Thirring model 
respectively. The completion of non-abelian gauge theories to six dimensions 
was verified at two loops in \cite{22}. To establish the connection between the
theories in the same universality class requires the explicit values of the 
renormalization group functions of each theory in its critical dimension to as 
high a loop order as is computationally possible. Once these are available the 
respective critical exponents are computed in an $\epsilon$ expansion around 
the critical dimension. The expressions can then be compared with the same 
exponents computed explicitly in $d$-dimensions with respect to a universal 
expansion parameter that is common to all the theories in the universality 
class. For example, for $O(N)$ theories this expansion parameter is usually 
$1/N$ where $N$ is regarded as large. The agreement of the $\epsilon$ expansion
of these $d$-dimensional large $N$ exponents for each critical dimension of the
theories in the same universality class establishes the ultraviolet completion.
There will be other ultraviolet completions aside from the few we mentioned but
in order to concretely establish them requires high loop order computations in 
dimensions beyond four for which {\sc Forcer} would be an indispensable 
integration tool. As it stands {\sc Forcer} evaluates integrals in a similar 
ethos to the Laporta algorithm in that a rule, based on an integration by parts
construction, efficiently reduces all contributing Feynman integrals to a 
basis. The advantage of {\sc Forcer} is that a database of integral relations 
does not have to be solved en route as is the case for packages that implement 
the algorithm of \cite{6}. Rules such as the diamond one are already encoded to
circumvent that necessity. Effectively the ultimate point of the {\sc Forcer} 
routine is the expression of the Green's functions as $d$-dimensional integrals
with the option of subsequently expanding in powers of $\epsilon$ up to weight 
$12$ if needed. Therefore to adapt the {\sc Forcer} routine to six dimensional 
problems requires the $\epsilon$ expansion of the basis integrals relative to 
six dimensions. That is one of the main purposes of this article. We will 
provide the six dimensional {\sc Forcer} master integrals expanded to weight 
$9$ in $d$~$=$~$6$~$-$~$2\epsilon$ dimensions. To achieve this requires a 
straightforward application of the Tarasov method, \cite{23,24}, that relates 
Feynman integrals in $d$-dimensions to a set of integrals in $(d+2)$-dimensions
with the same or reduced topology as that of the original $d$-dimensional 
integral.

Having achieved this extension the use of the {\sc Forcer} package becomes
straightforward with a {\sc Form} module slotted into the automatic integration
of the Green's functions of interest at the point where the four dimensional 
masters would be called. To verify the correctness of the masters we determine,
the extension is used to reproduce the known four loop modified minimal 
subtraction ($\MSbar$) scheme renormalization group functions in scalar 
$\phi^3$ theory with and without $O(N)$ symmetry although we note that higher 
loop $\MSbar$ information is already available, \cite{25,26}. Having 
established this check we apply the algorithm to various abelian and 
non-abelian gauge theories to verify their ultraviolet completeness at a newer 
level. As a by-product we will also study the renormalization group functions 
in several other schemes. While one of these is the canonical $\MSbar$ one we 
will also renormalize $\phi^3$ theory with and without $O(N)$ symmetry in the 
$\MOMts$ scheme as well as introduce a new scheme not unrelated to it. In four 
dimensional studies, \cite{27,28,29,30,31,32,33,34}, the $\MOMts$ scheme has 
the property that at least to five loops the core renormalization group 
functions of QCD do not involve the numbers $\zeta_4$ and $\zeta_6$ where 
$\zeta_n$ is the Riemann zeta function. It is worth recalling that the minimal 
momentum ($\mMOM$) scheme of \cite{35} also shares the same property. The 
definition of that scheme is based specifically on the ghost-gluon vertex. 
However unlike the $\MOMts$ schemes the absence of $\zeta_4$ and $\zeta_6$ is 
only manifest in $\mMOM$ in the Landau gauge. Indeed a no-$\pi$ theorem has 
been constructed that indicates under certain conditions there should be no 
even zetas to all orders, \cite{30,31}. The {\sc Forcer} algorithm in six 
dimensions is essential to verify or otherwise whether this property remains 
purely four dimensional or is true in six dimensions as well since the finite 
part of the various Green's functions are required. Indeed we can exploit the 
known five loop $\MSbar$ renormalization group functions of \cite{25,26} to 
deduce the five loop $\MOMts$ expressions for $\phi^3$ theory. Aside from 
establishing the {\sc Forcer} masters in six dimensions the study of these 
scheme properties and ultraviolet completion is the second main aim of this 
investigation.

The article is organized as follows. Section $2$ summarizes the algorithm used
to construct the six dimensional {\sc Forcer} master integrals up to weight 
$9$. In order to verify that known results are reproduced we focus on such a
check in Section $3$ by considering $\phi^3$ theory. En route we also construct
the $\MOMts$ renormalization group functions for the $O(N)$ cubic theory. In
Section $4$ the four loop renormalization group functions of QED and scalar
QED are determined in both the $\MSbar$ and $\MOMts$ schemes to further
investigate the presence or otherwise of even zetas in renormalization group
functions. This analysis is continued in Section $5$ where six dimensional QCD
is studied with the three loop gauge $\beta$-function being computed. We
provide an overview of our efforts in Section $6$. Finally, there are two
Appendices. The $\epsilon$ expansion of the {\sc Forcer} masters are provided
in Appendix A while Appendix B records the two $\beta$-functions of $O(N)$
$\phi^3$ theory in the $\MOMts$ scheme at five loops. 

\sect{{\sc Forcer} masters in six dimensions.}

As the first phase for studying six dimensional field theories at four loops is 
to employ the {\sc Forcer} package \cite{8,9} we need to discuss the specifics
of what this entails. To do so we recall that it was primarily developed to 
evaluate four loop massless $2$-point functions to high order in the $\epsilon$
expansion in $d$~$=$~$4$~$-$~$2\epsilon$ dimensions. Indeed it extended the
earlier three loop {\sc Mincer} algorithm, \cite{12}, that was the main working
tool for many decades to compute three loop massless $2$-point functions in the
same $\epsilon$ expansion. In many ways {\sc Mincer} was ahead of its time in
that it relied heavily on integration by parts and the use of what is termed
the carpet rule \cite{36} that allowed for an efficient reduction of a class of
topologies. The last step of such an integration algorithm was the substitution
of a small class of core integrals whose expansion near four dimensions were 
deduced without the use of integration by parts. It is only since the 
development of the Laporta algorithm \cite{6} that one can appreciate the 
prescience of the {\sc Mincer} construction. The summary of how {\sc Mincer} 
operates represents the essence of the Laporta approach where integration by 
parts is the engine room of the method which reduces the Feynman graphs 
contributing to a Green's function to a basis set of integrals now known as 
master integrals. Again these have to be determined by techniques other than 
integration by parts. While the ethos of both approaches is the same there are 
several key differences. The Laporta algorithm is applicable not only to 
$2$-point but also to higher $n$-point functions as well as the situation where
the Green's functions involve masses. One of the main limitations to applying 
the Laporta algorithm is technological rather than procedural. By this we mean 
the speed of an actual reduction reduces with the increase of external legs of 
the Green's function, the presence of a larger number of variables, such as 
masses and external momenta, and increase in loop order. What has been 
beneficial in the years after \cite{6} is the improvement of the reduction 
algorithm through pure mathematics results. The other main difference is that 
the {\sc Mincer} algorithm includes a routine that is more efficient at 
reducing the topologies and subtopologies of a Feynman graph. Such a routine is
necessarily connected with the fact that {\sc Mincer} is restricted to 
$2$-point functions, \cite{12,36}. These aspects reflect the tension between 
having a reasonably general algorithm applicable in virtually all desired 
set-ups and one which is customized to a specific set of Green's functions. Put
another way one has a choice of automatic Feynman integration evaluation tools 
which are extremely efficient in their respective domains.

With the need for higher precision in renormalizing quantum field theories and
determining Green's functions that contribute to observables for experiments
there was a clear need to extend {\sc Mincer} to the next order. This was
achieved with the {\sc Forcer} package, \cite{8,9}. The core approach is the 
same as {\sc Mincer} exploiting integration routines tailored to the Feynman 
integrals that contribute to four loop massless $2$-point functions. In 
particular for a subclass of topologies a new routine was required where the 
carpet rule of {\sc Mincer} was not applicable. This rule, termed the diamond 
rule, was provided in \cite{13} and moreover was encoded in an efficient way 
within the final symbolic manipulation routine written in {\sc Form}, 
\cite{10,11}. In assemblying the package several new aspects were included that
were not in {\sc Mincer}. In the intervening years between the two packages 
being developed the understanding as to what the independent master integrals 
were was resolved. In particular there are two three loop master integrals and 
fourteen at four loops in the {\sc Forcer} routine. In each case several of the
masters are elementary integrals such as those where there are nested bubbles. 
However there were others whose $\epsilon$ expansion near four dimensions had 
only been available to a few orders in $\epsilon$. Encoded within {\sc Forcer} 
are the $\epsilon$ expansions of the master integrals which were compiled from 
\cite{37,38} except for lower loop masters where one or two bubble insertions 
have been mapped to a closely related topology, \cite{8,9}. Therefore the 
{\sc Forcer} algorithm has a range of applicability greater than {\sc Mincer}. 
One particular feature of the package is the option to express the value of a 
$2$-point function in terms of its masters as a function of $d$ where the 
masters are not expanded in powers of $\epsilon$ near four dimensions. It is 
this specific feature that we aim to exploit here. In other words {\sc Forcer} 
has the potential to be adapted to the renormalization of six dimensional field
theories to high loop order if that can be achieved by the evaluation of 
massless $2$-point functions. The missing ingredient is the $\epsilon$ 
expansion of the $d$-dimensional four loop master integrals in 
$d$~$=$~$6$~$-$~$2\epsilon$ dimensions. 

In fact the $\epsilon$ expansion of the masters near six dimensions can be
extracted from the masters near four dimensions that are already available in
{\sc Forcer}. The key to determining them is the Tarasov construction
\cite{23,24}. Briefly this is a method that relates an $L$ loop Feynman 
integral in $d$-dimensions to a sum of Feynman integrals in $(d+2)$ dimensions 
which has the the same topology as the original lower dimensional one but with
the powers of $L$ propagators increased by unity. The distribution of the 
increase in propagator powers is determined by the structure of the second 
Symanzik graph polynomial which reflects and identifies the original topology. 
A useful tool for extracting the specific exponent distribution for each of the
master topologies was the {\sc HyperInt} package of \cite{39} written in 
{\sc Maple}. Subsequently each of the integrals in the uplift to $(d+2)$ 
dimensions can then be reduced using integration by parts to produce a sum of 
Feynman integrals one of which is equivalent to the original topology aside 
from being a $(d+2)$ dimensional integral. The remaining integrals in this sum 
correspond to Feynman graphs with fewer propagators in the sense that they are 
derived from the original topology but with propagators deleted. Therefore if 
we take the reference dimension $d$ to be four, knowledge of its $\epsilon$ 
expansion means that provided the same expansion of the graphs with a lower 
number of propagators is known the only unknown is the six dimensional master 
that we seek. This is the procedure we have followed to determine the 
{\sc Forcer} masters in six dimensions. It has already been used in \cite{17} 
to carry out the four loop renormalization of $\phi^3$ theory. In that 
instance, as {\sc Forcer} was not available then, the $\epsilon$ expansion for 
the four dimensional masters required for that computation was provided in 
\cite{37}. Although the {\sc Forcer} master basis was not known at the time of 
\cite{17}, this was not necessary as the basis of \cite{37} was sufficient to 
determine the $\MSbar$ scheme $\phi^3$ renormalization group functions at four 
loops. Other schemes such as $\MOMts$ that required the finite part of 
Green's functions were not considered at that time. The derivation of the 
relation between the four and six dimensional integrals using the Tarasov 
technique relied upon the Laporta algorithm \cite{6} and its implementation in 
the {\sc Reduze} package, \cite{40}. 

While the Laporta approach could have been repeated to deduce all the 
{\sc Forcer} masters in six dimensions we chose a different strategy here. This
was to use the $d$-dimensional aspect of the {\sc Forcer} algorithm itself to
effect the reduction of the $(d+2)$ dimensional integrals that the Tarasov
method produces at the first step. One benefit of this approach is that the
reduction of integrals in {\sc Forcer} is fast and more efficient than 
constructing an appropriate size database of relations between quite intricate 
topologies using the Laporta algorithm. This is because the increase in powers 
of the propagators by four requires a brute force integration by parts whereas 
{\sc Forcer} applies the custom built diamond rule and is designed solely for
$2$-point functions. Moreover as we wish to determine the masters to a high 
order in $\epsilon$ at some point the four dimensional {\sc Forcer} masters 
would have to be imported if the Laporta reduction had been employed. In 
addition employing {\sc Forcer} itself as a de facto reduction tool offers an 
independent way to check on the previous masters. At four loops the 
construction followed an iterative approach. The Tarasov method was applied to 
the lowest level master defined as the one with the smallest number of 
propagators that could not be evaluated by simple integration such as those 
comprised as bubbles. One has to begin at this point since in the reduction 
step these will appear for master topologies with a larger number of 
propagators. Once the lowest order masters have been determined then one moves 
to the next level. In each case one is effectively finding the terms in the 
$\epsilon$ expansion by solving for the unknown coefficients of $\epsilon$ in 
the required master. In Appendix A we have recorded the explicit $\epsilon$ 
expansions of the $16$ {\sc Forcer} masters using the same labelling as that of
\cite{9} and to the same order in weight as the four dimensional masters that 
appear in Appendix C of \cite{9}. The expressions in Appendix A should be 
sufficient for carrying out a five loop renormalization where the higher powers
of $\epsilon$ in the four loop masters are required to extend the calculation 
of the lower loop graphs contributing to a Green's function. Included in our 
Appendix A list are two three loop masters labelled as {\bf no} and {\bf t105}.
In other words the six dimensional masters are provided up to and including 
weight $9$ where $\zeta_9$ would be a representative. Though we note that the 
{\sc Form}  module in the actual {\sc Forcer} code of \cite{9} that corresponds
to the four dimensional masters includes expressions up to weight $12$ which 
were derived from the higher order $\epsilon$ expansion of the masters provided
in \cite{38}.

\sect{$\phi^3$ theories.}

We are now in a position to renormalize a variety of six dimensional theories 
to four loops in the $\MSbar$ scheme as well as the $\MOMts$ one which has been 
examined in four dimensions \cite{27,28,29,30,31,32,33,34}. Such an exercise 
will act partly as a check on the construction of the {\sc Forcer} masters as 
well as confirm an underlying aspect of the $\MOMts$ scheme that is apparent in
four dimensions. In the former instance an error in a master could produce an 
inconsistency with the application of the renormalization group formalism to 
extract and encode the renormalization constants in the respective 
$\beta$-functions and anomalous dimensions. The $\MOMts$ scheme and issues 
related to it have become of interest in recent years. Its prescription is that
in theories with a cubic interaction the $2$-point functions and $3$-point 
vertex functions are renormalized by absorbing the finite part of the 
respective divergent Green's functions into the wave function and coupling 
renormalization constants. In the case of the $3$-point functions the Feynman 
graphs are evaluated where one of the external momenta is nullified. If there 
are several fields leading to more than one $3$-point interaction then there 
will be a $\MOMts$ scheme attached to each of the vertices. We will focus in 
this section on several six dimensional renormalizable scalar cubic theories. 
The first is the basic single field instance with Lagrangian 
\begin{equation}
L ~=~ \frac{1}{2} \left( \partial_\mu \phi \right)^2 ~+~ \frac{g}{6} \phi^3 ~.
\label{lagphi3}
\end{equation}
As far as we are aware the $\MOMts$ scheme renormalization group functions for
(\ref{lagphi3}) are not yet available although the five loop renormalization 
group functions are known in the $\MSbar$ scheme, \cite{17,25,26,41,42,43}. It 
is worth recording these for completeness since they are needed as the 
foundation for deriving the $\MOMts$ five loop equivalent renormalization group
functions. For completeness we recall, \cite{17,25,26,41,42,43},  
\begin{eqnarray}
\beta_{\MSbarss}^{\phi^3}(a) &=&
\frac{3}{4} a^2 - \frac{125}{144} a^3
+ 5 [ 2592 \zeta_3 + 6617 ] \frac{a^4}{20736}
\nonumber \\
&&
+~ [ - 4225824 \zeta_3 + 349920 \zeta_4 + 1244160 \zeta_5
- 3404365 ] \frac{a^5}{746496}
\nonumber \\
&&
+~ [ 41570496 \zeta_3^2 + 356380884 \zeta_3 - 33912351 \zeta_4
+ 295089480 \zeta_5 + 15746400 \zeta_6
\nonumber \\
&& ~~~~
- 576843120 \zeta_7 + 102052031 ] \frac{a^6}{6718464} ~+~ O(a^7)
\nonumber \\
\gamma_{\phi,\MSbarss}^{\phi^3}(a) &=&
-~ \frac{1}{12} a + \frac{13}{432} a^2
+ [ 2592 \zeta_3 - 5195 ] \frac{a^3}{62208}
\nonumber \\
&&
+~ [ 10080 \zeta_3 + 18144 \zeta_4 - 69120 \zeta_5
+ 53449 ] \frac{a^4}{248832}
\nonumber \\
&&
+~ [ - 3499200 \zeta_3^2 - 18368532 \zeta_3 - 4119579 \zeta_4
+ 8691624 \zeta_5 - 8748000 \zeta_6 
\nonumber \\
&& ~~~~
+ 46294416 \zeta_7 - 16492987 ] \frac{a^5}{20155392} ~+~ O(a^6)
\label{phi3msrge}
\end{eqnarray}
where $a$~$=$~$g^2$ and we note the parameters will always be in the same 
scheme as that indicated on the renormalization group function itself. In
situations where otherwise there might be an ambiguous interpretation the 
parameters will carry the scheme label explicitly. We note the same conventions
for the coupling constant as that of \cite{17} are used here rather than those 
of \cite{27}. The expressions of (\ref{phi3msrge}) can readily be mapped to the
conventions of \cite{26} via $a$~$\to$~$-$~$a$. We recall that the $\MOMts$ 
renormalization prescription is to remove the finite parts of both the $2$- and
$3$-point functions at the subtraction point and absorb them into the 
respective wave function and coupling renormalization constants. For the 
$3$-point function one of the two independent external momenta is set to zero 
when the vertex function is evaluated. In this configuration the $3$-point 
function is equivalent to a $2$-point one whence the {\sc Forcer} algorithm can
be applied. We have first checked that the four loop $\MSbar$ results of 
(\ref{phi3msrge}) are reproduced. This also provides an initial check on the 
six dimensional {\sc Forcer} masters in a field theory calculation. 
Consequently it is straightforward to deduce the $\MOMts$ scheme 
renormalization group functions which are 
\begin{eqnarray}
\beta_{\MOMtss}^{\phi^3}(a) &=&
\frac{3}{4} a^2 - \frac{125}{144} a^3
+ [ - 1296 \zeta_3 + 26741 ] \frac{a^4}{10368}
\nonumber \\
&&
+~ [ - 1370736 \zeta_3 + 2177280 \zeta_5 - 2304049 ] \frac{a^5}{186624}
\nonumber \\
&&
+~ [ 389670912 \zeta_3^2 + 3307195440 \zeta_3 + 89151840 \zeta_5
- 5640570432 \zeta_7 
\nonumber \\
&& ~~~
+ 2190456157 ] \frac{a^6}{26873856} ~+~ O(a^7)
\nonumber \\
\gamma_{\phi,\MOMtss}^{\phi^3}(a) &=&
-~ \frac{1}{12} a + \frac{37}{432} a^2
+ [ - 1296 \zeta_3 - 4435 ] \frac{a^3}{31104}
\nonumber \\
&&
+~ [ 122256 \zeta_3 + 155520 \zeta_5 + 135149 ] \frac{a^4}{559872}
\nonumber \\
&&
+~ [ 6718464 \zeta_3^2 - 51538896 \zeta_3 - 108669600 \zeta_5
- 185177664 \zeta_7 
\nonumber \\
&& ~~~
+ 38661817 ] \frac{a^5}{80621568} ~+~ O(a^6) ~.
\label{phi3momtrge}
\end{eqnarray}
Over several years there has been interest in which elements of the $\zeta_n$
sequence appear in the renormalization group functions in four dimensions,
\cite{27,28,29,30,31,32,33,34}. In the $\MSbar$ scheme $\zeta_n$ is present at 
successive loop orders with $n$~$\geq$~$3$ where the loop order that $\zeta_3$
first occurs depends on the underlying theory. It transpires that in the 
$\MOMts$ prescription in four dimensions only $\zeta_{2n+1}$ for 
$1$~$\leq$~$n$~$\leq$~$3$ appears to five loops. In other words $\zeta_4$ and 
$\zeta_6$ are absent. A similar feature arises in six dimensional $\phi^3$ 
theory as is apparent in (\ref{phi3momtrge}). So the basic four dimensional 
property of the $\MOMts$ scheme is preserved in another spacetime dimension.

The results of (\ref{phi3momtrge}) were derived from a basic property of the
renormalization group equation which relates the renormalization group
functions in one scheme to those in another. For a single coupling theory such
as (\ref{lagphi3}) there is a simple relation between the coupling constant in
one scheme to that in another which is derived from the relation of each
coupling to the bare one which formally gives
\begin{equation}
g_{{\MOMtss}}(\mu) ~=~ \frac{Z_g^{\MSbarss}}{Z_g^{{\MOMtss}}}
g_{{\MSbarss}}(\mu) ~.
\label{ccmapdef}
\end{equation}
The right hand side is interpreted as being a function of $g_{{\MSbarss}}$
although $Z_g^{{\MOMtss}}$ is a function of $g_{{\MOMtss}}$. In other words
\begin{equation}
Z_g^{{\MOMtss}} ~\equiv~ Z_g^{{\MOMtss}} 
\left( a_{\MOMtss}(a_{\MSbarss}) \right) 
\end{equation}
with the explicit $g_{{\MSbarss}}$ dependence being deduced via an iterative
process. Such a change of variables is necessary as otherwise the relation
(\ref{ccmapdef}) would have singularities in $\epsilon$. Once the mapping is
available at four loops then the five loop $\beta$-function can be found from
\begin{equation}
\beta^{\phi^3}_{\MOMtss} ( a_{\MOMtss} ) ~=~
\left[ \beta^{\phi^3}_{\mbox{$\MSbarss$}}( a_{\mbox{$\MSbarss$}} )
\frac{\partial a_{\MOMtss}}{\partial a_{\mbox{$\MSbarss$}}}
\right]_{ \MSbarss \rightarrow {\MOMtss} } ~.
\end{equation}
The restriction indicates that the coupling constant, which would otherwise be 
in the $\MSbar$ scheme has to be mapped to the $\MOMts$ scheme from the inverse
of (\ref{ccmapdef}). A similar process is applied to determine the five loop 
field anomalous dimension using the conversion function defined by
\begin{equation}
C^{\phi^3}_\phi(a_{\MSbarss}) ~=~ 
\left[ \frac{Z_\phi^{{\MOMtss}}}{Z_\phi^{\MSbarss}}
\right]_{ \MOMtss \rightarrow {\MSbarss} }
\end{equation}
where the coupling $a$ in conversion functions will always be an $\MSbar$ 
variable, and the relation
\begin{eqnarray}
\gamma^{\phi^3}_{\phi\,\MOMtss} ( a_{\MOMtss} )
&=& \left[ \gamma^{\phi^3}_{\phi\,\MSbarss} \left(a_{\MSbarss}\right)
+ \beta^{\phi^3}_{\MSbarss}\left(a_{\MSbarss}\right)
\frac{\partial ~}{\partial a_{\MSbarss}} \ln C_\phi^{\phi^3}
\left(a_{\MSbarss}\right)
\right]_{ \MSbarss \rightarrow {\MOMtss} }
\end{eqnarray}
uses the same process as that to determine the $\beta$-function.

To assist with verifying (\ref{phi3momtrge}) we record the relevant explicit 
expressions are
\begin{eqnarray}
a_{\MOMtss} &=&
a_{\MSbarss} - \frac{4}{3} a_{\MSbarss}^2
+ [ - 1728 \zeta_3 + 8005 ] \frac{a_{\MSbarss}^3}{1728}
\nonumber \\
&&
+~ [ 659232 \zeta_3 - 116640 \zeta_4 + 2488320 \zeta_5
- 7758845 ] \frac{a_{\MSbarss}^4}{373248} 
\nonumber \\
&&
+~ [ 11384064 \zeta_3^2 + 69803208 \zeta_3 + 6620292 \zeta_4 
- 120916800 \zeta_5 - 2332800 \zeta_6 
\nonumber \\
&& ~~~
- 123451776 \zeta_7 + 262216343 ]
\frac{a_{\MSbarss}^5}{2239488} ~+~ O(a_{\MSbarss}^6)
\end{eqnarray}
and
\begin{eqnarray}
C_\phi^{\phi^3}(a) &=&
1 + \frac{2}{9} a - \frac{511}{1728} a^2
+ [ - 6240 \zeta_3 - 2592 \zeta_4 + 122099 ] \frac{a^3}{124416}
\nonumber \\
&&
+~ [ - 46656 \zeta_3^2 - 2058228 \zeta_3 - 299862 \zeta_4 + 2251152 \zeta_5
+ 583200 \zeta_6 
\nonumber \\
&& ~~~~
- 12882121 ] \frac{a^4}{3359232} ~+~ O(a^5) ~.
\end{eqnarray}
As it sometimes turns out to be useful, for completeness we note the coupling 
constant conversion function is
\begin{eqnarray}
C_g^{\phi^3}(a) &=&
1 + \frac{2}{3} a + [ 1728 \zeta_3 - 5701 ] \frac{a^2}{3456}
\nonumber \\
&&
+~ [ 87264 \zeta_3 + 116640 \zeta_4 - 2488320 \zeta_5
+ 4853645 ] \frac{a^3}{746496}
\nonumber \\
&&
+~ [ - 155271168 \zeta_3^2 - 1372972032 \zeta_3 - 83529792 \zeta_4
+ 1456911360 \zeta_5 + 37324800 \zeta_6 
\nonumber \\
&& ~~~~
+ 1975228416 \zeta_7 - 2620417103 ] \frac{a^4}{71663616} ~+~ O(a^5)
\end{eqnarray}
to the same order where
\begin{equation}
C^{\phi^3}_g(a_{\MSbarss}) ~=~ \left[
\frac{Z_g^{{\MOMtss}}}{Z_g^{\MSbarss}}
\right]_{ \MSbarss \rightarrow {\MOMtss} } ~.
\end{equation}

We can extend the five loop $\MOMts$ study of the cubic scalar theory to the 
case where there is an $O(N)$ symmetry. In this case there are two coupling 
constants since the renormalizable Lagrangian is, \cite{15,16},
\begin{equation}
L ~=~ \frac{1}{2} \left( \partial_\mu \phi^i \right)^2 ~+~
\frac{1}{2} \left( \partial_\mu \sigma \right)^2 ~+~
\frac{g_1}{2} \sigma \phi^i \phi^i ~+~ \frac{g_2}{6} \sigma^3
\label{lagphi3on}
\end{equation}
where we use the same conventions as \cite{17}. The original single field cubic
theory is clearly recovered in the limit $g_1$~$\to$~$0$ in (\ref{lagphi3on}).
To determine the $O(N)$ $\phi^3$ $\MOMtss$ renormalization group functions we
follow the same procedure as before by constructing the four loop conversion
function but with the formalism extended to accommodate two coupling constants.
For instance the relation between the coupling constants in the respective 
schemes is 
\begin{equation}
g_{i\,{\MOMtss}}(\mu) ~=~ \frac{Z_{g_i}^{\MSbarss}}{Z_{g_i}^{{\MOMtss}}}
g_{i\,{\MSbarss}}(\mu) 
\end{equation}
for $i$~$=$~$1$ and $2$ where there is no summation over $i$. Then simply
differentiating with respect to the renormalization scale $\mu$ leads to
\begin{eqnarray}
\beta^{O(N)}_{i\,\MOMtss} ( \mathbf{g}_{\MOMtss} ) &=&
\left[ \sum_{j=1}^2 \beta^{O(N)}_{j\,\MSbarss}( \mathbf{g}_{\MSbarss} )
\frac{\partial g_i^{\MOMtss}}{\partial g_j^{\MSbarss}}
\right]_{ \MSbarss \rightarrow {\MOMtss} } 
\end{eqnarray}
where the scheme of the coupling constants now appears as subscripts. The 
scheme label is included to avoid amibiguity as there is a mapping of 
variables. To deduce the five loop anomalous dimensions the conversion 
functions for the two fields are given by
\begin{equation}
C^{O(N)}_\phi(\mathbf{g_{\MSbarss}}) ~=~ 
\left[ \frac{Z_\phi^{{\MOMtss}}}{Z_\phi^{\MSbarss}}
\right]_{ \MOMtss \rightarrow {\MSbarss} } ~~~,~~~
C^{O(N)}_\sigma(\mathbf{g_{\MSbarss}}) ~=~ 
\left[ \frac{Z_\sigma^{{\MOMtss}}}{Z_\sigma^{\MSbarss}}
\right]_{ \MOMtss \rightarrow {\MSbarss} }
\end{equation}
where the restriction indicates that the argument of each conversion function
is expressed in terms of the $\MSbar$ variables as the reference scheme. The
explicit expressions for the $\MOMts$ anomalous dimensions are derived from 
\begin{eqnarray}
\gamma^{O(N)}_{\phi\,\MOMtss} ( \mathbf{g}_{\MOMtss} ) &=& \left[ 
\gamma^{O(N)}_{\phi\,\MSbarss} ( \mathbf{g}_{\MSbarss} ) ~+~ 
\sum_{i=1}^2 \beta^{O(N)}_{i\,\MSbarss} ( \mathbf{g}_{\MSbarss} )
\frac{\partial ~}{\partial g_{i\,\MSbarss}} \ln \left( C_\phi^{O(N)}
( \mathbf{g}_{\MSbarss} ) \right)
\right]_{ \MSbarss \rightarrow {\MOMtss} } \nonumber \\
\gamma^{O(N)}_{\sigma\,\MOMtss} ( \mathbf{g}^{\MOMtss} ) &=& \left[ 
\gamma^{O(N)}_{\sigma\,\MSbarss} ( \mathbf{g}_{\MSbarss} ) ~+~ 
\sum_{i=1}^2 \beta^{O(N)}_{i\,\MSbarss} ( \mathbf{g}_{\MSbarss} )
\frac{\partial ~}{\partial g_{i\,\MSbarss}} \ln \left( C_\sigma^{O(N)}
( \mathbf{g}_{\MSbarss} ) \right)
\right]_{ \MSbarss \rightarrow {\MOMtss} } \!\!\!\!\!\!\!.
\end{eqnarray}
The final stage of the process is to recall that the $O(N)$ $\phi^3$ $\MSbar$
five loop renormalization group functions are available from \cite{25}. In 
addition we have carried out the explicit four loop renormalization of
(\ref{lagphi3on}) in the $\MOMts$ scheme which allows us to determine the field
conversion functions and coupling constant mappings to four loops. With these
we have established the $O(N)$ five loop $\MOMts$ renormalization group
functions. The main motivation for doing so is to ascertain whether there are
any terms involving $\zeta_4$ or $\zeta_6$ which are absent in the single
coupling case for (\ref{lagphi3}). We find the same outcome, in keeping with
the analysis of \cite{33}, in that $\zeta_4$ or $\zeta_6$ are absent from each
of the four renormalization group functions for all $N$. The full expressions
together with the conversion functions and coupling constant mappings are 
provided in the data file associated with the arXiv version of this article
\cite{44} but we have provided the two $\beta$-functions in Appendix B. By way 
of example the expressions for the $\gamma^{O(N)}_{\sigma\,\MOMtss}(g_1,g_2)$ 
and 
$\beta^{O(N)}_{1\,\MOMtss}(g_1,g_2)$ for $N$~$=$~$2$ are
\begin{eqnarray}
\gamma^{O(2)}_{\sigma\,\MOMtss}(g_1,g_2) &=&
\left[
- \frac{1}{6} g_1^2
- \frac{1}{12} g_2^2
\right] ~+~ 
\left[
- \frac{11}{216} g_1^2 g_2^2
+ \frac{4}{9} g_1^3 g_2
+ \frac{13}{108} g_1^4
+ \frac{37}{432} g_2^4
\right]
\nonumber \\
&&
+ \left[
- \frac{4435}{31104} g_2^6
- \frac{1247}{1296} g_1^4 g_2^2
- \frac{505}{972} g_1^6
- \frac{89}{216} g_1^5 g_2
- \frac{83}{324} g_1^3 g_2^3
- \frac{5}{12} \zeta_3 g_1^4 g_2^2
\right. \nonumber \\
&& \left. ~~~~
- \frac{1}{6} \zeta_3 g_1^6
- \frac{1}{24} \zeta_3 g_2^6
+ \frac{169}{1944} g_1^2 g_2^4
\right]
\nonumber \\
&&
+ \left[
- \frac{26969}{139968} g_1^2 g_2^6
- \frac{439}{648} \zeta_3 g_1^6 g_2^2
- \frac{23}{162} \zeta_3 g_1^2 g_2^6
- \frac{11}{432} \zeta_3 g_1^3 g_2^5
+ \frac{5}{18} \zeta_5 g_2^8
\right. \nonumber \\
&& \left. ~~~~
+ \frac{20}{9} \zeta_5 g_1^4 g_2^4
+ \frac{20}{9} \zeta_5 g_1^8
+ \frac{40}{9} \zeta_5 g_1^6 g_2^2
+ \frac{127}{324} \zeta_3 g_1^8
+ \frac{139}{54} \zeta_3 g_1^7 g_2
\right. \nonumber \\
&& \left. ~~~~
+ \frac{283}{1296} \zeta_3 g_2^8
+ \frac{1279}{1296} \zeta_3 g_1^4 g_2^4
+ \frac{2486}{729} g_1^7 g_2
+ \frac{24283}{46656} g_1^3 g_2^5
+ \frac{55901}{69984} g_1^6 g_2^2
\right. \nonumber \\
&& \left. ~~~~
+ \frac{68407}{69984} g_1^8
+ \frac{101983}{46656} g_1^5 g_2^3
+ \frac{135149}{559872} g_2^8
+ \frac{170393}{279936} g_1^4 g_2^4
+ 3 \zeta_3 g_1^5 g_2^3
\right]
\nonumber \\
&&
+ \left[
- \frac{18870089}{40310784} g_1^2 g_2^8
- \frac{10106999}{5038848} g_1^8 g_2^2
- \frac{453335}{157464} g_1^{10}
- \frac{119303}{186624} \zeta_3 g_2^{10}
\right. \nonumber \\
&& \left. ~~~~
- \frac{98207}{7776} \zeta_3 g_1^5 g_2^5
- \frac{43223}{11664} \zeta_3 g_1^6 g_2^4
- \frac{20011}{1944} \zeta_3 g_1^9 g_2
- \frac{18665}{2592} \zeta_5 g_1^4 g_2^6
\right. \nonumber \\
&& \left. ~~~~
- \frac{13975}{10368} \zeta_5 g_2^{10}
- \frac{10157}{3888} \zeta_3 g_1^8 g_2^2
- \frac{8815}{216} \zeta_5 g_1^9 g_2
- \frac{5755}{108} \zeta_5 g_1^7 g_2^3
\right. \nonumber \\
&& \left. ~~~~
- \frac{4555}{1296} \zeta_3 g_1^7 g_2^3
- \frac{2585}{1296} \zeta_5 g_1^{10}
- \frac{2401}{32} \zeta_7 g_1^8 g_2^2
- \frac{1323}{32} \zeta_7 g_1^6 g_2^4
- \frac{755}{144} \zeta_5 g_1^3 g_2^7
\right. \nonumber \\
&& \left. ~~~~
- \frac{539}{32} \zeta_7 g_1^4 g_2^6
- \frac{441}{16} \zeta_7 g_1^{10}
- \frac{147}{64} \zeta_7 g_2^{10}
- \frac{95}{27} \zeta_5 g_1^5 g_2^5
- \frac{4}{27} \zeta_3 g_1^3 g_2^7
\right. \nonumber \\
&& \left. ~~~~
- \frac{1}{6} \zeta_3^2 g_1^5 g_2^5
- \frac{1}{12} \zeta_3^2 g_1^2 g_2^8
- \frac{1}{12} \zeta_3^2 g_1^6 g_2^4
+ \frac{1}{3} \zeta_3^2 g_1^3 g_2^7
+ \frac{1}{12} \zeta_3^2 g_2^{10}
+ \frac{5}{6} \zeta_3^2 g_1^9 g_2
\right. \nonumber \\
&& \left. ~~~~
+ \frac{5}{12} \zeta_3^2 g_1^4 g_2^6
+ \frac{5}{36} \zeta_3^2 g_1^8 g_2^2
+ \frac{7}{12} \zeta_3^2 g_1^{10}
+ \frac{11}{3} \zeta_3^2 g_1^7 g_2^3
+ \frac{1505}{1296} \zeta_5 g_1^2 g_2^8
\right. \nonumber \\
&& \left. ~~~~
+ \frac{3665}{1296} \zeta_5 g_1^8 g_2^2
+ \frac{10151}{5184} g_1^3 g_2^7
+ \frac{10255}{1728} \zeta_5 g_1^6 g_2^4
+ \frac{13667}{11664} \zeta_3 g_1^2 g_2^8
\right. \nonumber \\
&& \left. ~~~~
+ \frac{41861}{11664} \zeta_3 g_1^{10}
+ \frac{204311}{93312} \zeta_3 g_1^4 g_2^6
+ \frac{652231}{104976} g_1^9 g_2
+ \frac{5288099}{20155392} g_1^6 g_2^4
\right. \nonumber \\
&& \left. ~~~~
+ \frac{6725627}{3359232} g_1^5 g_2^5
+ \frac{24943961}{1679616} g_1^7 g_2^3
+ \frac{28840009}{20155392} g_1^4 g_2^6
+ \frac{38661817}{80621568} g_2^{10}
\right]
\nonumber \\
&& +~ O(g_i^{12})
\end{eqnarray}
and 
\begin{eqnarray}
\beta^{O(2)}_{1\,\MOMtss}(g_1,g_2) &=&
\left[
- \frac{1}{24} g_1 g_2^2
+ \frac{1}{2} g_1^2 g_2
+ \frac{1}{4} g_1^3
\right]
\nonumber \\
&&
+ \left[
- \frac{337}{432} g_1^3 g_2^2
- \frac{59}{72} g_1^5
- \frac{1}{12} g_1^2 g_2^3
- \frac{1}{18} g_1^4 g_2
+ \frac{37}{864} g_1 g_2^4
\right]
\nonumber \\
&&
+ \left[
- \frac{4435}{62208} g_1 g_2^6
- \frac{1}{6} \zeta_3 g_1^3 g_2^4
- \frac{1}{48} \zeta_3 g_1 g_2^6
+ \frac{1}{2} \zeta_3 g_1^4 g_2^3
+ \frac{7}{4} \zeta_3 g_1^7
+ \frac{7}{24} \zeta_3 g_1^5 g_2^2
\right. \nonumber \\
&& \left. ~~~~
+ \frac{559}{1728} g_1^2 g_2^5
+ \frac{2189}{2592} g_1^4 g_2^3
+ \frac{2803}{1296} g_1^7
+ \frac{5075}{2592} g_1^5 g_2^2
+ \frac{6367}{1296} g_1^6 g_2
\right. \nonumber \\
&& \left. ~~~~
+ \frac{7079}{7776} g_1^3 g_2^4
- 3 \zeta_3 g_1^6 g_2
\right]
\nonumber \\
&&
+ \left[
- \frac{3188293}{139968} g_1^7 g_2^2
- \frac{1584779}{46656} g_1^8 g_2
- \frac{843827}{139968} g_1^5 g_2^4
- \frac{721603}{23328} g_1^6 g_2^3
\right. \nonumber \\
&& \left. ~~~~
- \frac{482599}{23328} g_1^9
- \frac{72569}{15552} g_1^3 g_2^6
- \frac{38011}{31104} g_1^2 g_2^7
- \frac{2579}{1296} \zeta_3 g_1^7 g_2^2
- \frac{2339}{1296} \zeta_3 g_1^3 g_2^6
\right. \nonumber \\
&& \left. ~~~~
- \frac{1931}{108} \zeta_3 g_1^6 g_2^3
- \frac{1603}{288} \zeta_3 g_1^5 g_2^4
- \frac{1253}{54} \zeta_3 g_1^9
- \frac{1189}{864} \zeta_3 g_1^4 g_2^5
- \frac{613}{24} \zeta_3 g_1^8 g_2
\right. \nonumber \\
&& \left. ~~~~
- \frac{35}{9} \zeta_5 g_1^7 g_2^2
- \frac{5}{2} \zeta_5 g_1^4 g_2^5
+ \frac{1}{48} \zeta_3 g_1^2 g_2^7
+ \frac{5}{36} \zeta_5 g_1 g_2^8
+ \frac{70}{3} \zeta_5 g_1^9
\right. \nonumber \\
&& \left. ~~~~
+ \frac{85}{18} \zeta_5 g_1^3 g_2^6
+ \frac{95}{18} \zeta_5 g_1^5 g_2^4
+ \frac{130}{3} \zeta_5 g_1^6 g_2^3
+ \frac{283}{2592} \zeta_3 g_1 g_2^8
+ \frac{325}{6} \zeta_5 g_1^8 g_2
\right. \nonumber \\
&& \left. ~~~~
+ \frac{135149}{1119744} g_1 g_2^8
+ \frac{202163}{186624} g_1^4 g_2^5
\right]
\nonumber \\
&&
+ \left[
\frac{2392700219}{80621568} g_1^3 g_2^8
+ \frac{3141423697}{5038848} g_1^9 g_2^2
+ \frac{10824707071}{40310784} g_1^7 g_2^4
\right. \nonumber \\
&& \left. ~~~~
- \frac{117194495}{13436928} g_1^4 g_2^7
- \frac{2042725}{10368} \zeta_5 g_1^7 g_2^4
- \frac{713515}{864} \zeta_5 g_1^9 g_2^2
\right. \nonumber \\
&& \left. ~~~~
- \frac{604061}{31104} \zeta_3 g_1^4 g_2^7
- \frac{122227}{96} \zeta_7 g_1^{10} g_2
- \frac{119303}{373248} \zeta_3 g_1 g_2^{10}
- \frac{44255}{864} \zeta_5 g_1^{11}
\right. \nonumber \\
&& \left. ~~~~
- \frac{44009}{32} \zeta_7 g_1^8 g_2^3
- \frac{42133}{192} \zeta_7 g_1^7 g_2^4
- \frac{32977}{96} \zeta_7 g_1^6 g_2^5
- \frac{32179}{64} \zeta_7 g_1^9 g_2^2
\right. \nonumber \\
&& \left. ~~~~
- \frac{26125}{2592} \zeta_5 g_1^3 g_2^8
- \frac{13975}{20736} \zeta_5 g_1 g_2^{10}
- \frac{10017}{16} \zeta_7 g_1^{11}
- \frac{9485}{64} \zeta_7 g_1^5 g_2^6
\right. \nonumber \\
&& \left. ~~~~
- \frac{215}{144} \zeta_5 g_1^2 g_2^9
- \frac{182}{3} \zeta_7 g_1^3 g_2^8
- \frac{147}{128} \zeta_7 g_1 g_2^{10}
+ \frac{1}{24} \zeta_3^2 g_1 g_2^{10}
+ \frac{17}{24} \zeta_3^2 g_1^4 g_2^7
\right. \nonumber \\
&& \left. ~~~~
+ \frac{65}{12} \zeta_3^2 g_1^5 g_2^6
+ \frac{287}{8} \zeta_3^2 g_1^{11}
+ \frac{287}{72} \zeta_3^2 g_1^3 g_2^8
+ \frac{293}{12} \zeta_3^2 g_1^6 g_2^5
+ \frac{1085}{12} \zeta_3^2 g_1^{10} g_2
\right. \nonumber \\
&& \left. ~~~~
+ \frac{1477}{96} \zeta_7 g_1^4 g_2^7
+ \frac{1601}{72} \zeta_3^2 g_1^7 g_2^4
+ \frac{1679}{24} \zeta_3^2 g_1^8 g_2^3
+ \frac{3725}{432} \zeta_5 g_1^6 g_2^5
\right. \nonumber \\
&& \left. ~~~~
+ \frac{7391}{10368} \zeta_3 g_1^2 g_2^9
+ \frac{15881}{432} \zeta_3 g_1^3 g_2^8
+ \frac{18805}{864} \zeta_5 g_1^4 g_2^7
+ \frac{29225}{108} \zeta_5 g_1^{10} g_2
\right. \nonumber \\
&& \left. ~~~~
+ \frac{158255}{288} \zeta_5 g_1^8 g_2^3
+ \frac{190625}{5184} \zeta_5 g_1^5 g_2^6
+ \frac{663761}{1296} \zeta_3 g_1^{10} g_2
+ \frac{1065955}{5832} \zeta_3 g_1^7 g_2^4
\right. \nonumber \\
&& \left. ~~~~
+ \frac{2616893}{15552} \zeta_3 g_1^6 g_2^5
+ \frac{3783533}{7776} \zeta_3 g_1^{11}
+ \frac{9270535}{15552} \zeta_3 g_1^8 g_2^3
\right. \nonumber \\
&& \left. ~~~~
+ \frac{16931831}{23328} \zeta_3 g_1^9 g_2^2
+ \frac{17933611}{186624} \zeta_3 g_1^5 g_2^6
+ \frac{31673845}{4478976} g_1^2 g_2^9
\right. \nonumber \\
&& \left. ~~~~
+ \frac{38661817}{161243136} g_1 g_2^{10}
+ \frac{46773469}{373248} g_1^8 g_2^3
+ \frac{82651885}{419904} g_1^{11}
+ \frac{155899681}{419904} g_1^{10} g_2
\right. \nonumber \\
&& \left. ~~~~
+ \frac{351087313}{2239488} g_1^6 g_2^5
+ \frac{427400711}{40310784} g_1^5 g_2^6
+ 72 \zeta_3^2 g_1^9 g_2^2
\right] ~+~ O(g_i^{13}) ~.
\end{eqnarray}
One check on the full $O(N)$ $\MOMts$ results is that the respective 
expressions of (\ref{phi3msrge}) correctly emerged in the $g_1$~$\to$~$0$ 
limit.

Having evaluated the four loop {\sc Forcer} masters to high order in powers of
$\epsilon$ we can study the properties of a scheme that is not unrelated to the
$\MOMts$ one. In the $\MOMts$ prescription the finite part of the Green's
function at the subtraction point is absorbed into the respective
renormalization constants. One natural extension of this prescription is to not
only absorb the finite or $O(1)$ part with respect to $\epsilon$ but also the 
higher order terms in the $\epsilon$ expansion. By introducing such a scheme 
one is in effect absorbing the full structure of the underlying Feynman graphs 
determined as a function of the regularizing parameter. In other words one 
subtracts all properties of the quantum corrections. This may appear to be an 
unusual prescription and there is no clear expectation as to what it means for 
the properties of the resultant renormalization group functions. However it is 
an easy algebraic exercise to pursue in this toy scalar field theory in order 
to explore the consequences. There is the tacit assumption that whatever 
transpires in this example ought to have exact parallels in four dimensional 
theories in much the same way that the even zetas are absent in the $\MOMts$ 
scheme for theories in both four and six dimensions to a specific loop order. 
As this prescription is the polar opposite to the $\MSbar$ one we will term 
this new scheme the maximal subtraction scheme and denote it by $\MaxSs$ where 
the overline retains the nod to the absence of $\zeta_2$ in the $\MSbar$ 
renormalization group functions. Therefore we have repeated the procedure that 
resulted in the $\MOMts$ renormalization group functions of 
(\ref{phi3momtrge}). In other words we have applied the $\MaxSs$ prescription 
to the $2$- and $3$-point functions and deduced the respective coupling and 
wave function renormalization constants before constructing the coupling 
constant map and the wave function conversion function. Retaining the higher 
order terms in $\epsilon$ necessarily involved an additional quantity of 
intermediate algebra. However, we were able to determine that the two $\MaxSs$ 
renormalization group functions are
\begin{eqnarray}
\beta_{\MaxSss}^{\phi^3}(a) &=&
\frac{3}{4} a^2 - \frac{125}{144} a^3
+ [ - 1296 \zeta_3 + 26741 ] \frac{a^4}{10368}
\nonumber \\
&&
+~ [ - 1370736 \zeta_3 + 2177280 \zeta_5 - 2304049 ] \frac{a^5}{186624}
\nonumber \\
&&
+~ [ 389670912 \zeta_3^2 + 3307195440 \zeta_3 + 89151840 \zeta_5
- 5640570432 \zeta_7 
\nonumber \\
&& ~~~
+ 2190456157 ] \frac{a^6}{26873856} ~+~ O(a^7)
\nonumber \\
\gamma_{\phi,\MaxSss}^{\phi^3}(a) &=&
-~ \frac{1}{12} a + \frac{37}{432} a^2
+ [ - 1296 \zeta_3 - 4435 ] \frac{a^3}{31104}
\nonumber \\
&&
+~ [ 122256 \zeta_3 + 155520 \zeta_5 + 135149 ] \frac{a^4}{559872}
\nonumber \\
&&
+~ [ 6718464 \zeta_3^2 - 51538896 \zeta_3 - 108669600 \zeta_5
- 185177664 \zeta_7
\nonumber \\
&& ~~~
+ 38661817 ] \frac{a^5}{80621568} ~+~ O(a^6) ~.
\label{phi3maxsrge}
\end{eqnarray}
What is interesting is that the expressions are formally the same as the
respective $\MOMts$ ones to five loops. This is not unexpected given that even 
zeta contributions were absent in the $\MOMts$ scheme renormalization group
functions but are present in the $O(\epsilon^0)$ part of the $\MOMts$ and
$\MaxSs$ renormalization constants. The higher order $\MaxSs$ $\epsilon$ terms 
will involve the subsequent terms of the $\zeta$-series as well as rationals. 
Such higher order zeta contributions cannot for instance appear in the 
renormalization group functions before a certain loop order. This is preserved 
naturally in the $\MaxSs$ scenario. Where the difference in the $\MOMts$ and 
$\MaxSs$ renormalization group functions would be manifest is in the ultimate 
step that results in (\ref{phi3maxsrge}). That step is to set the regularizing 
parameter, $\epsilon$, to zero. Prior to that the renormalization group 
functions are $\epsilon$ dependent with the coefficients of $\epsilon$ being in 
a direct relationship with the constant and $O(\epsilon)$ terms of the 
respective renormalization constants. In the case of the $\MOMts$ scheme the 
dependence on $\epsilon$ would be linear in contrast to the structure of the 
$\MaxSs$ renormalization group functions. In that instance the coefficients of 
$a$ would be an infinite series in $\epsilon$. However for practical reasons in
our construction we restricted our analysis to weight $9$ as that was the 
weight we determined the six dimensional {\sc Forcer} masters to. What this 
exercise has revealed in addition to the above is that the $\MOMts$ and 
$\MaxSs$ schemes are synonymous in the critical dimension of $\phi^3$ theory. 
Moreover at least to five loops, the $\MOMts$ scheme is equivalent to the 
scheme where the full Feynman integrals themselves are completely subtracted at
the renormalization point which is something we believe has not been examined 
previously.

\sect{Abelian gauge theories.}

Having established the usefulness of the four loop {\sc Forcer} masters in the
cubic scalar theory we devote this section to extending their application to 
gauge theories in six dimensions. In particular our focus here will be on 
abelian gauge theories as these are of interest in \cite{21}. The aim is to 
extend the renormalization of both Quantum Electrodynamics (QED) and scalar QED
(sQED) to four loops. First we recall the QED Lagrangian in six dimensions is, 
\cite{21},
\begin{equation}
\left. L^{(6)} \right|_{\mbox{\footnotesize{QED}}} ~=~
i \bar{\psi}^i \Dslash \psi^i ~-~ 
\frac{1}{4} \left( \partial_\mu F_{\nu\sigma} \right)
\left( \partial^\mu F^{\nu\sigma} \right) ~-~
\frac{1}{2\alpha} \left( \partial_\mu \partial^\nu A_\nu \right)
\left( \partial^\mu \partial^\sigma A_\sigma \right) ~.
\label{lagqedd6}
\end{equation}
where $g$ is the gauge coupling constant, $\alpha$ is the gauge parameter,
$F_{\mu\nu}$~$=$~$\partial_\mu A_\nu$~$-$~$\partial_\mu A_\nu$, $A_\mu$ is the
photon and $\psi^i$ is the electron with $1$~$\leq$~$i$~$\leq$~$\Nf$. The 
linear gauge fixing term is such that the photon propagator takes the form
\begin{equation}
\langle A_\mu(p) A_\nu(-p) \rangle ~=~ -~
\frac{1}{(p^2)^2} \left[ \eta_{\mu\nu} ~-~
(1 - \alpha) \frac{p_\mu p_\nu}{p^2} \right] ~.
\end{equation}
We note that the second order pole in the propagator is not infrared 
pathological in six dimensions as it would be in four dimensions. Moreover it
is not an obstruction to extending the three loop renormalization group
functions to the next order using {\sc Forcer}. Since we have followed the
same algorithm that produced the renormalization group functions of the 
previous section for $\phi^3$ theory we record the equivalent QED four loop 
results are 
\begin{eqnarray}
\beta_{\MSbarss}^{\QEDss}(g,\alpha) &=&
-~ \frac{2 \Nf}{15} g^3 - \frac{19 \Nf}{27} g^5
+ 17 [ - 111 \Nf + 200 ] \frac{\Nf g^7}{12150}
\nonumber \\
&&
+~ [ 170362 \Nf^2 + 17107200 \zeta_3 \Nf - 14025425 \Nf
- 7500000 ] \frac{\Nf g^9}{8201250} ~+~ O(g^{11})
\nonumber \\
\gamma_{A,\MSbarss}^{\QEDss}(g,\alpha) &=&
-~ \frac{4 \Nf}{15} g^2 - \frac{38 \Nf}{27} g^4
+ 17 [ - 111 \Nf + 200 ] \frac{\Nf g^6}{6075}
\nonumber \\
&&
+~ [ 170362 \Nf^2 + 17107200 \zeta_3 \Nf - 14025425 \Nf
- 7500000 ] \frac{\Nf g^8}{4100625} ~+~ O(g^{10})
\nonumber \\
\gamma_{\alpha,\MSbarss}^{\QEDss}(g,\alpha) &=& 
-~ \gamma_{A,\MSbarss}^{\QEDss}(g,\alpha)
\nonumber \\
\gamma_{\psi,\MSbarss}^{\QEDss}(g,\alpha) &=&
[ 3 \alpha + 5 ] \frac{g^2}{6} + 2 [ 32 \Nf - 125 ]\frac{g^4}{135}
\nonumber \\
&&
+~ [ 2864 \Nf^2 - 648000 \zeta_3 \Nf + 730375 \Nf + 1944000 \zeta_3
- 1033000 ] \frac{g^6}{243000}
\nonumber \\
&&
+~ [ - 518400 \zeta_3 \Nf^3 - 25824 \Nf^3 - 43156800 \zeta_3 \Nf^2
+ 17496000 \zeta_4 \Nf^2 + 28663075 \Nf^2 
\nonumber \\
&& ~~~
- 1560999600 \zeta_3 \Nf
- 52488000 \zeta_4 \Nf + 3061800000 \zeta_5 \Nf - 1179131300 \Nf
\nonumber \\
&& ~~~
+ 1552770000 \zeta_3 - 3061800000 \zeta_5
+ 1003751250 ] \frac{g^8}{16402500} ~+~ O(g^{10})
\nonumber \\
\gamma_{m,\MSbarss}^{\QEDss}(g) &=&
-~ \frac{5}{3} g^2 - [ 68 \Nf + 25 ] \frac{g^4}{135} \nonumber \\
&& +~ [ 13456 \Nf^2 + 648000 \Nf \zeta_3 - 818575 \Nf + 1215000 \zeta_3
- 726875 ] \frac{g^6}{121500} \nonumber \\
&& +~ [ 518400 \Nf^3 \zeta_3 - 216384 \Nf^3 + 31492800 \Nf^2 \zeta_3
- 17496000 \Nf^2 \zeta_4 - 6336415 \Nf^2
\nonumber \\
&& ~~~ 
-~ 137011500 \Nf \zeta_3 - 32805000 \Nf \zeta_4 + 656100000 \Nf \zeta_5 
- 318912625 \Nf
\nonumber \\
&& ~~~ 
+~ 2574281250 \zeta_3 - 4538025000 \zeta_5 
+ 845045625 ] \frac{g^8}{8201250} ~+~ O(g^{10})
\label{rgeqedd6ms}
\end{eqnarray}
in the $\MSbar$ scheme. We note that the $\beta$-function is derived from the 
Ward-Takahashi identity 
\begin{equation}
\beta_1(g) ~=~ \frac{g}{2} \gamma_A(g,\alpha)
\label{wti}
\end{equation}
to four loops. The previous three loop results of \cite{22} are recovered 
partially verifying our procedure. Another check is that the electron anomalous
dimension has the same feature as four dimensional QED in that the only 
dependence on the gauge parameter $\alpha$ is in the one loop term. Interesting
properties of this renormalization group function in four dimensional QED were 
initially discussed in \cite{45,46} in the on-shell renormalization scheme. 
There it was noted that the anomalous dimension was gauge independent. 
Subsequently it was shown in \cite{47} that in the $\MSbar$ scheme the gauge 
parameter of a linear covariant gauge fixing appears only in the one loop term 
of the electron anomalous dimension in four dimensional QED. That analysis made
use of a Landau-Khalatnikov-Fradkin (LKF) transformation \cite{48,49} and this
formalism should be equally applicable to establish the same property in scalar
QED. In more recent years the absence of $\alpha$ in the two and higher order 
corrections in four dimensions was examined from the Hopf algebra point of view
in \cite{50,51,52}. It seems clear that the LKF transformation approach as well
as the graphical and algebraic arguments could be extended to the six 
dimensional theory of (\ref{lagqedd6}). The expression for 
$\gamma_{\psi,\MSbarss}^{\QEDss}(g,\alpha)$ in (\ref{rgeqedd6ms}) would support
that expectation. Any proof that $\alpha$ only appears at one loop in the 
$\MSbar$ electron anomalous dimension though should probably be constructed in 
a way that is applicable to all even dimensions beginning from four. This is 
simply because it is known that the same property is present at two loops in 
eight dimensional QED \cite{22}. However such a proof is beyond the scope of 
the current article.

In light of the discussion concerning the location of the set of numbers 
$\zeta_n$ of the previous section we recall that the $\MSbar$ $\beta$-function 
has the same property as its four dimensional counterpart in that $\zeta_3$ is 
absent at three loops but appears for the first time at four loops which is a 
consequence of gauge symmetry. By contrast $\zeta_3$ is present in the three 
loop $\MSbar$ $\beta$-function of $\phi^3$ theory as well as in the $\MOMts$ 
scheme. Therefore we have repeated the renormalization of (\ref{lagqedd6}) in 
the $\MOMts$ scheme to ascertain whether $\zeta_3$ first appears at three or 
four loops. We found the analogous expressions to (\ref{rgeqedd6ms}) are
\begin{eqnarray}
\beta_{\MOMtss}^{\QEDss}(g,\alpha) &=&
-~ \frac{2 \Nf}{15} g^3 - \frac{19 \Nf}{27} g^5
+ [ 2592 \zeta_3 \Nf - 2889 \Nf + 1700 ] \frac{\Nf g^7}{6075}
\nonumber \\
&&
+~ [ 165888 \zeta_3 \Nf^2 - 306876 \Nf^2 + 4976640 \zeta_3 \Nf
- 4665600 \zeta_5 \Nf 
\nonumber \\
&& ~~~
-~ 692335 \Nf - 500000 ] 
\frac{\Nf g^9}{546750} ~+~ O(g^{11})
\nonumber \\
\gamma_{A , \MOMtss}^{\QEDss}(g,\alpha) &=&
-~ \frac{4 \Nf}{15} g^2 - \frac{38 \Nf}{27} g^4
+ 2 [ 2592 \zeta_3 \Nf - 2889 \Nf + 1700 ] \frac{\Nf g^6}{6075}
\nonumber \\
&&
+~ [ 165888 \zeta_3 \Nf^2 - 306876 \Nf^2 + 4976640 \zeta_3 \Nf
- 4665600 \zeta_5 \Nf
\nonumber \\
&& ~~~
-~ 692335 \Nf - 500000 ]
\frac{\Nf g^8}{273375} ~+~ O(g^{10})
\nonumber \\
\gamma_{\alpha , \MOMtss}^{\QEDss}(g,\alpha) &=& 
-~ \gamma_{A , \MOMtss}^{\QEDss}(g,\alpha)
\nonumber \\
\gamma_{\psi , \MOMtss}^{\QEDss}(g,\alpha) &=&
[ 3 \alpha + 5 ] \frac{g^2}{6} + 2 [ 4 \Nf - 25 ] \frac{g^4}{27}
\nonumber \\
&&
+~ [ 976 \Nf^2 - 675 \alpha \Nf + 8555 \Nf + 38880 \zeta_3 - 20660 ]
\frac{g^6}{4860}
\nonumber \\
&&
+~ [ 54675 \alpha^2 \Nf - 97200 \zeta_3 \alpha^2 \Nf 
+ 103680 \zeta_3 \alpha \Nf^2 - 144720 \alpha \Nf^2
- 324000 \zeta_3 \alpha \Nf 
\nonumber \\
&& ~~~
+ 81000 \alpha \Nf + 150528 \Nf^3
- 1731456 \zeta_3 \Nf^2 + 3643432 \Nf^2 - 72595440 \zeta_3 \Nf
\nonumber \\
&& ~~~
+ 139968000 \zeta_5 \Nf - 59802795 \Nf + 82814400 \zeta_3 - 163296000 \zeta_5
\nonumber \\
&& ~~~
+ 53533400 ] \frac{g^8}{874800} ~+~ O(g^{10}) 
\nonumber \\
\gamma_{m,\MOMtss}^{\QEDss}(g,\alpha) &=& -~ \frac{5}{3} g^2
- [ \Nf + 5 ] \frac{g^4}{27} \nonumber \\
&& +~ [ 1215 \alpha \Nf - 1184 \Nf^2 + 5184 \Nf \zeta_3 - 12613 \Nf
+ 48600 \zeta_3 - 29075 ] \frac{g^6}{4860}
\nonumber \\
&& +~ [ 194400 \alpha^2 \Nf \zeta_3 - 103275 \alpha^2 \Nf
- 207360 \alpha \Nf^2 \zeta_3 + 136080 \alpha \Nf^2 
\nonumber \\
&& ~~~ 
+~ 1069200 \alpha \Nf - 4992 \Nf^3 + 2094336 \Nf^2 \zeta_3 - 4752992 \Nf^2 
+ 10886400 \Nf \zeta_3
\nonumber \\
&& ~~~ 
+~ 9331200 \Nf \zeta_5 - 29410965 \Nf + 274590000 \zeta_3 - 484056000 \zeta_5
\nonumber \\
&& ~~~ 
+~ 90138200 ] \frac{g^8}{874800} ~+~ O(g^{10}) ~.
\label{rgeqedd6momt}
\end{eqnarray}
Clearly $\zeta_3$ arises for the first time at three loops more in keeping with
$\phi^3$ theory and similar to the properties of the $\MOMts$ scheme
renormalization group functions in four dimensional gauge theories. One other
feature of (\ref{rgeqedd6momt}) is that unlike 
$\gamma_{\psi,\MSbarss}^{\QEDss}(g,\alpha)$ there is $\alpha$ dependence in 
$\gamma_{\psi,\MOMtss}^{\QEDss}(g,\alpha)$ beyond one loop. However it first
appears at three loops, like $\MOMts$ schemes in four dimensional QED, rather 
than at two loops for the MOM schemes of Celmaster and Gonsalves, \cite{53,54}.
The contrast in this structural difference may be attributed to the difference 
in the underlying renormalization prescription. For instance in the $\MOMts$ 
suite of schemes the subtraction for $3$-point functions is carried out at a
point where one of the external legs has its momentum nullified. By constrast 
for the MOM schemes of \cite{53,54} the prescription is that the vertex 
subtraction is enacted at the fully symmetric point where the squared momenta 
of all three external legs are non-zero and equal. So it would appear that the 
properties of the vertex kinematics has a bearing on the structure of the 
renormalization group functions and in particular this is manifest in the 
electron renormalization in an abelian theory.

It is worth examining whether these observations are peculiar to six 
dimensional QED or a more general feature. Therefore to explore this we have
repeated the above QED analysis but for the version where the fermions are
replaced by a scalar field $\phi^i$. The corresponding Lagrangian is 
\begin{equation}
L^{\sQEDss} ~=~ \overline{D_\mu \phi^i} D^\mu \phi^i ~-~
\frac{1}{4} \partial_\mu F_{\nu\sigma} \partial^\mu F^{\nu\sigma} ~-~
\frac{1}{2\alpha} \left( \partial_\mu \partial^\nu A_\nu \right)
\left( \partial^\mu \partial^\sigma A_\sigma \right) \nonumber \\
\label{lagsqedd6}
\end{equation}
which is the ultraviolet compeletion of four dimensional scalar QED where 
$1$~$\leq$~$i$~$\leq$~$\Nf$. Carrying out the renormalization process we find 
that the $\MSbar$ renormalization group functions are
\begin{eqnarray}
\beta_{\MSbarss}^{\sQEDss}(g) &=&
-~ \frac{\Nf}{60} g^3 - \frac{37 \Nf}{216} g^5
- [ 1017 \Nf + 59600 ] \frac{\Nf g^7}{97200}
\nonumber \\
&&
+~ [ 214131600 \zeta_3 \Nf - 407787250 \Nf - 282521 \Nf^2 + 118098000 \zeta_3
\nonumber \\
&& ~~~~
- 456939750 ] \frac{\Nf g^9}{1049760000} ~+~ O(g^{11})
\nonumber \\
\gamma_{A,\MSbarss}^{\sQEDss}(g,\alpha) &=&
-~ \frac{\Nf}{30} g^2- \frac{37 \Nf}{108} g^4
- [ 1017 \Nf + 59600 ] \frac{\Nf g^6}{48600}
\nonumber \\
&&
+~ [ 214131600 \zeta_3 \Nf - 407787250 \Nf - 282521 \Nf^2 + 118098000 \zeta_3 
\nonumber \\
&& ~~~~
- 456939750 ] \frac{\Nf g^8}{524880000} ~+~ O(g^{10})
\nonumber \\
\gamma_{\alpha,\MSbarss}^{\sQEDss}(g,\alpha) &=& 
-~ \gamma_{A,\MSbarss}^{\sQEDss}(g,\alpha)
\nonumber \\
\gamma_{\phi,\MSbarss}^{\sQEDss}(g,\alpha) &=&
[3 \alpha - 10] \frac{g^2}{6} + [ - 98 \Nf + 1375] \frac{g^4}{1080} 
\nonumber \\
&&
+~ [ 662 \Nf^2 + 648000 \zeta_3 \Nf - 1430875 \Nf + 1458000 \zeta_3 
+ 516500 ] \frac{g^6}{972000} 
\nonumber \\
&&
+~ [ 12960 \zeta_3 \Nf^3 - 2382 \Nf^3 + 8631360 \zeta_3 \Nf^2
- 3499200 \zeta_4 \Nf^2 - 12293840 \Nf^2 
\nonumber \\
&& ~~~~
- 1435890240 \zeta_3 \Nf - 7873200 \zeta_4 \Nf + 2536920000 \zeta_5 \Nf 
- 470052370 \Nf 
\nonumber \\
&& ~~~~
+ 1663578000 \zeta_3 + 2274480000 \zeta_5
- 3242042625 ] \frac{g^8}{104976000} 
\nonumber \\
&& +~ O(g^{10}) ~.
\label{rgesqedd6ms}
\end{eqnarray}
The lower loop expressions are in agreement with \cite{55,56}. As another check
we have computed the critical exponents for the scalar electron and the photon
from (\ref{rgesqedd6ms}) in powers of $1/\Nf$ at the Wilson-Fisher fixed point
of the $\beta$-function in $d$~$=$~$6$~$-$~$2\epsilon$ dimensions. These were
compared with the direct large $\Nf$ expansion of the same quantities computed
in the underlying universal theory in $d$-dimensions in \cite{55}. Expanding 
the results of \cite{55} to $O(\epsilon^3)$ we find exact agreement. From a 
careful comparison it is evident the properties of (\ref{rgeqedd6ms}) that were 
highlighted earlier are the same for (\ref{rgesqedd6ms}). The same situation 
occurs for the four loop $\MOMts$ renormalization group functions which we 
determined as
\begin{eqnarray}
\beta_{\MOMtss}^{\sQEDss}(g,\alpha) &=&
-~ \frac{\Nf}{60} g^3 - \frac{37 \Nf}{216} g^5
+ [ 648 \zeta_3 \Nf - 1701 \Nf - 59600 ]\frac{\Nf g^7}{97200}
\nonumber \\
&&
+~ [ 6480 \zeta_3 \Nf^2 - 18039 \Nf^2 + 2610792 \zeta_3 \Nf 
- 933120 \zeta_5 \Nf 
\nonumber \\
&& ~~~~
- 3049985 \Nf + 787320 \zeta_3 - 3046265 ]\frac{\Nf g^9}{6998400} ~+~ O(g^{11})
\nonumber \\
\gamma_{A , \MOMtss}^{\sQEDss}(g,\alpha) &=&
-~ \frac{\Nf}{30} g^2 - \frac{37 \Nf}{108} g^4 
+ [ 648 \zeta_3 \Nf - 1701 \Nf - 59600 ] \frac{\Nf g^6}{48600}
\nonumber \\
&&
+~ [ 6480 \zeta_3 \Nf^2 - 18039 \Nf^2 + 2610792 \zeta_3 \Nf 
- 933120 \zeta_5 \Nf 
\nonumber \\
&& ~~~~
- 3049985 \Nf + 787320 \zeta_3 - 3046265 ] 
\frac{\Nf g^8}{3499200} ~+~ O(g^{10})
\nonumber \\
\gamma_{\alpha , \MOMtss}^{\sQEDss}(g,\alpha) &=& 
-~ \gamma_{A , \MOMtss}^{\sQEDss}(g,\alpha)
\nonumber \\
\gamma_{\psi , \MOMtss}^{\sQEDss}(g,\alpha) &=&
[ 3 \alpha - 10 ] \frac{g^2}{6} + [ - 4 \Nf + 275 ] \frac{g^4}{216} 
\nonumber \\
&&
+~ [ 270 \alpha \Nf - 49 \Nf^2 - 648 \zeta_3 \Nf - 1684 \Nf + 14580 \zeta_3 
+ 5165 ] \frac{g^6}{9720} 
\nonumber \\
&&
+~ [ 194400 \alpha^2 \zeta_3 \Nf - 97200 \alpha^2 \Nf 
- 25920 \alpha \zeta_3 \Nf^2 + 17280 \alpha \Nf^2 - 1296000 \alpha \zeta_3 \Nf 
\nonumber \\
&& ~~~~
+ 2349000 \alpha \Nf - 1176 \Nf^3 + 222912 \zeta_3 \Nf^2 - 1008229 \Nf^2 
- 160535520 \zeta_3 \Nf 
\nonumber \\
&& ~~~~
+ 218116800 \zeta_5 \Nf - 7940520 \Nf + 110905200 \zeta_3 + 151632000 \zeta_5 
\nonumber \\
&& ~~~~
- 216136175 ] \frac{g^8}{6998400} ~+~ O(g^{10}) ~.
\label{rgesqedd6momt}
\end{eqnarray}
The emergence of the same $\zeta_n$ and $\alpha$ structures for both theories
only reinforces the notion that the kinematics of the vertex subtraction 
point have an influence on the scheme dependent parts of the renormalization
group functions. Although this was already apparent in the MOM schemes of
\cite{53,54} the new insight here is that the $\alpha$ dependence of the MOM
and $\MOMts$ schemes is dependent on whether there is a nullification of an
external vertex leg or not. Finally we note that like $\phi^3$ theory the
abelian gauge theory $\MOMts$ scheme renormalization group functions are devoid
of even zetas to the order we have calculated to in complete agreement with 
four dimensional studies \cite{27,28,29,30,31,32,33,34}. While there are clear 
similarities with the zeta structure of four dimensional theories a formal 
proof of the absence of even zetas in six dimensions may only be possible if, 
for instance, the approach using the KLF formalism for massless correlation 
functions of \cite{57} could be generalized to six dimensions. To the high loop
orders discussed here and other places, \cite{27,28,29,30,31,32,33,34}, this 
may not be as straightforward as it would seem at first. For instance in the 
context of the explicit calculations carried out here one would have to 
establish that there is no primitive graph that arises as a {\sc Forcer} master
at very high loop order whose simple pole residue is an even power of $\pi$. 
However there is some evidence in four dimensions, \cite{58}, that there may be
one or more such primitive graphs at weight $12$ whose residue involves 
$\zeta_{12}$. If such an integral remained in the determination of the 
renormalization constant of a massless correlation function then the no-$\pi$ 
theorem of \cite{30,31,32} may need to be revisited. Similar reasoning should 
equally apply to six dimensions.

\sect{Six dimensional QCD.}

Next we turn to the six dimensional version of QCD that has been studied at one
and two loops in \cite{14,22,59}. We will use the Lagrangian of \cite{22} which
is
\begin{eqnarray}
L^{(6)} &=& -~ \frac{1}{4} \left( D_\mu G_{\nu\sigma}^a \right)
\left( D^\mu G^{a \, \nu\sigma} \right) ~+~
\frac{g_2}{6} f^{abc} G_{\mu\nu}^a \, G^{b \, \mu\sigma} \,
G^{c \,\nu}_{~~\,\sigma} \nonumber \\
&& -~ \frac{1}{2\alpha} \left( \partial_\mu \partial^\nu A^a_\nu \right)
\left( \partial^\mu \partial^\sigma A^a_\sigma \right) ~-~
\bar{c}^a \Box \left( \partial^\mu D_\mu c \right)^a ~+~
i \bar{\psi}^{iI} \Dslash \psi^{iI}
\label{lagqcdd6}
\end{eqnarray}
where the gauge coupling $g_1$ is embedded in the covariant derivative and
field strength $G^a_{\mu\nu}$ and the indices lie in the ranges
$1$~$\leq$~$i$~$\leq$~$\Nf$, $1$~$\leq$~$a$~$\leq$~$\NA$ and 
$1$~$\leq$~$I$~$\leq$~$\NF$ with $\NF$ and $\NA$ corresponding to the dimension
of the fundamental and adjoint representations of the colour group
respectively and $\Nf$ is the number of quarks. The operator associated with
$g_2$ is sometimes referred to as a spectator. In \cite{14,59} the Lagrangian 
took different forms. This is because there are three possible gauge invariant 
dimension six operators but only two are independent in the action which can be
deduced from the Bianchi identity and integration by parts. The renormalization
group functions of one formulation of the Lagrangian can be translated to those
of another via, \cite{14},
\begin{equation} 
\left( D^\mu G_{\mu\sigma}^a \right) 
\left( D_\nu G^{a \, \nu\sigma} \right) ~=~
\frac{1}{2} \left( D_\mu G_{\nu\sigma}^a \right) 
\left( D^\mu G^{a \, \nu\sigma} \right) ~+~
g_1 f^{abc} G_{\mu\nu}^a \, G^{b \, \mu\sigma} \, G^{c \,\nu}_{~~\,\sigma}
\label{opidd6}
\end{equation} 
which is used to connect the respective coupling constants. The gauge fixing 
terms have been chosen to ensure the gluon and ghost dimensions match. The one 
and two loop renormalization of (\ref{lagqcdd6}) was carried out in the 
$\MSbar$ scheme by a direct computation of Feynman integrals. Subsequently the 
one loop renormalization for a more general version of (\ref{lagqcdd6}) was 
performed using the heat kernel expansion in \cite{14}. Included for instance 
in that generalization were scalar fields as well as others that matched the 
field content of the supersymmetric extension. Taking the limit of the one loop
results of \cite{14} to recover the field content of (\ref{lagqcdd6}) and using
the relation between the coupling constants of both formulation that results 
from (\ref{opidd6}) produced agreement between the $\beta$-function at this one
loop order. Given that we now have a six dimensional version of {\sc Forcer} it
is possible to extend results for the renormalization group functions to three 
loops. This required the renormalization of the gluon, ghost and quark 
$2$-point functions and, to extract the gauge coupling $\beta$-function, the 
ghost-gluon vertex. Additionally we renormalized the quark mass operator. The 
number of graphs that were evaluated are recorded in Table $1$. To initiate the
automatic Feynman integration process requires the compilation of the 
electronic representation of the graphs. This was effected with {\sc Qgraf}, 
\cite{60}. At the outset we need to be clear and record the fact that with 
{\sc Forcer} it is not possible to deduce the three loop $\beta$-function of 
the non-gauge coupling $g_2$. To do so requires the renormalization of the 
triple gluon vertex. The structure of the triple gluon vertex Feynman rule will
involve both $g_1$ and $g_2$ in contrast to the ghost-gluon vertex that only 
contains $g_1$. So by determining the renormalization of $g_1$ from the ghost 
vertex that of $g_2$ can only be deduced from the triple gluon vertex 
renormalization. However as the cubic term of (\ref{lagqcdd6}) involves the 
product of $G^a_{\mu\nu}$ nullifying any external leg of the triple gluon 
vertex Feynman rule the terms involving $g_2$ vanish identically. Therefore 
{\sc Forcer} cannot be employed. Moreover rewriting the cubic operator of 
(\ref{lagqcdd6}) to redefine it in terms of the other two operators of 
(\ref{opidd6}) produces the same outcome. Any nullification of an external 
gluon momentum in the resultant triple gluon Feynman rule excludes access to 
the non-gauge coupling constant. The only route to find the $\beta$-function
for $g_2$ is to evaluate the triple gluon vertex at a non-exceptional momentum 
configuration such as the symmetric point one that was employed in \cite{22}. 
At three loops this is clearly beyond the scope of the present article.

{\begin{table}[ht]
\begin{center}
\begin{tabular}{|c||c|c|c|c|}
\hline
Green's function & $L$~$=$~$1$ & $L$~$=$~$2$& $L$~$=$~$3$ & Total \\
\hline
$A^a_\mu \, A^b_\nu$ & $3$ & $18$ & $267$ & $288$ \\
$c^a \bar{c}^b$ & $1$ & $~\,6$ & $~\,78$ & $~\,85$ \\
$\psi \bar{\psi}$ & $1$ & $~\,6$ & $~\,78$ & $~\,85$ \\
$c^a \bar{c}^b A^c_\sigma$ & $2$ & $33$ & $702$ & $737$ \\
\hline
Total & $7$ & $63$ & $\!\!1125$ & $\!\!1195$ \\
\hline
\end{tabular}
\end{center}
\begin{center}
{Table 1. Number of Feynman diagrams computed for various Green's functions at
$L$ loops.}
\end{center}
\end{table}}

Having explained the background to the renormalization of (\ref{lagqcdd6}) we
record the $\MSbar$ renormalization group functions. For the gauge coupling
$\beta$-function we have
\begin{eqnarray}
\beta_1(g_1,g_2) &=&
-~ \left[ 249 C_A + 16 \Nf T_F \right] \frac{g_1^3}{120}
\nonumber \\
&&
+ \left[ 
-~ 50682 C_A^2 g_1^3 
+ 2439 C_A^2 g_1^2 g_2 
+ 3129 C_A^2 g_1 g_2^2 
- 315 C_A^2 g_2^3 
- 1328 C_A \Nf T_F g_1^3 
\right. \nonumber \\
&& \left. ~~~
- 624 C_A \Nf T_F g_1^2 g_2 
+ 96 C_A \Nf T_F g_1 g_2^2 
- 3040 C_F \Nf T_F g_1^3 \right] \frac{g_1^2}{4320}
\nonumber \\
&&
+ \left[ 
-~ 7464290865 C_A^3 g_1^6 
+ 1091499579 C_A^3 g_1^5 g_2 
+ 809468904 C_A^3 g_1^4 g_2^2 
\right. \nonumber \\
&& \left. ~~~
- 141762510 C_A^3 g_1^3 g_2^3 
- 15628455 C_A^3 g_1^2 g_2^4 
+ 1812375 C_A^3 g_1 g_2^5 
\right. \nonumber \\
&& \left. ~~~
+ 94500 C_A^3 g_2^6 
- 495581080 C_A^2 \Nf T_F g_1^6 
- 66132288 C_A^2 \Nf T_F g_1^5 g_2 
\right. \nonumber \\
&& \left. ~~~
+ 63346632 C_A^2 \Nf T_F g_1^4 g_2^2 
- 1733040 C_A^2 \Nf T_F g_1^3 g_2^3 
- 652320 C_A^2 \Nf T_F g_1^2 g_2^4 
\right. \nonumber \\
&& \left. ~~~
- 411047360 C_A C_F \Nf T_F g_1^6 
- 37440000 C_A C_F \Nf T_F g_1^5 g_2 
\right. \nonumber \\
&& \left. ~~~
+ 14592000 C_A C_F \Nf T_F g_1^4 g_2^2 
+ 987520 C_A \Nf^2 T_F^2 g_1^6 
- 4732416 C_A \Nf^2 T_F^2 g_1^5 g_2 
\right. \nonumber \\
&& \left. ~~~
+ 728064 C_A \Nf^2 T_F^2 g_1^4 g_2^2 
+ 17408000 C_F^2 \Nf T_F g_1^6 
\right. \nonumber \\
&& \left. ~~~
- 9661440 C_F \Nf^2 T_F^2 g_1^6 \right] \frac{g_1}{62208000} ~+~ O(g_i^9) ~.
\end{eqnarray}
As all the results in this section will be in the $\MSbar$ scheme we do not
include the scheme label on the renormalization group functions. For the 
renormalization of the fields we have the gauge parameter dependent expressions
\begin{eqnarray}
\gamma_A(g_1,g_2,\alpha) &=&
\left[ 20 \alpha C_A - 199 C_A - 16 \Nf T_F \right] \frac{g_1^2}{60} 
\nonumber \\
&&
+ \left[ 
130 \alpha^2 C_A^2 g_1^3 
+ 1095 \alpha C_A^2 g_1^3 
- 81412 C_A^2 g_1^3 
+ 2178 C_A^2 g_1^2 g_2 
+ 5658 C_A^2 g_1 g_2^2 
\right. \nonumber \\
&& \left. ~~~
- 630 C_A^2 g_2^3 
- 1568 C_A \Nf T_F g_1^3 
- 1248 C_A \Nf T_F g_1^2 g_2 
+ 192 C_A \Nf T_F g_1 g_2^2 
\right. \nonumber \\
&& \left. ~~~
- 6080 C_F \Nf T_F g_1^3 \right] \frac{g_1}{4320} 
\nonumber \\
&&
+ \left[ 
-~ 1215000 \zeta_3 \alpha^3 C_A^3 g_1^6 
+ 1800375 \alpha^3 C_A^3 g_1^6 
+ 891000 \zeta_3 \alpha^2 C_A^3 g_1^6 
- 314125 \alpha^2 C_A^3 g_1^6 
\right. \nonumber \\
&& \left. ~~~
- 4617000 \zeta_3 \alpha C_A^3 g_1^6 
+ 82449225 \alpha C_A^3 g_1^6 
- 1944000 \zeta_3 \alpha C_A^3 g_1^5 g_2 
\right. \nonumber \\
&& \left. ~~~
- 7416000 \alpha C_A^3 g_1^5 g_2 
- 1842000 \alpha C_A^3 g_1^4 g_2^2 
+ 5906400 \alpha C_A^2 \Nf T_F g_1^6 
\right. \nonumber \\
&& \left. ~~~
+ 90477000 \zeta_3 C_A^3 g_1^6 
- 6130578488 C_A^3 g_1^6 
- 66744000 \zeta_3 C_A^3 g_1^5 g_2 
\right. \nonumber \\
&& \left. ~~~
+ 829769679 C_A^3 g_1^5 g_2 
- 2592000 \zeta_3 C_A^3 g_1^4 g_2^2 
+ 718180404 C_A^3 g_1^4 g_2^2 
\right. \nonumber \\
&& \left. ~~~
+ 3888000 \zeta_3 C_A^3 g_1^3 g_2^3 
- 127466010 C_A^3 g_1^3 g_2^3 
- 13886955 C_A^3 g_1^2 g_2^4 
\right. \nonumber \\
&& \left. ~~~
+ 1812375 C_A^3 g_1 g_2^5 
+ 94500 C_A^3 g_2^6 
+ 62208000 \zeta_3 C_A^2 \Nf T_F g_1^6 
\right. \nonumber \\
&& \left. ~~~
- 449504664 C_A^2 \Nf T_F g_1^6 
- 61797888 C_A^2 \Nf T_F g_1^5 g_2 
+ 56482632 C_A^2 \Nf T_F g_1^4 g_2^2 
\right. \nonumber \\
&& \left. ~~~
- 1733040 C_A^2 \Nf T_F g_1^3 g_2^3 
- 652320 C_A^2 \Nf T_F g_1^2 g_2^4 
- 82944000 \zeta_3 C_A C_F \Nf T_F g_1^6 
\right. \nonumber \\
&& \left. ~~~
- 307447360 C_A C_F \Nf T_F g_1^6 
- 37440000 C_A C_F \Nf T_F g_1^5 g_2 
\right. \nonumber \\
&& \left. ~~~
+ 14592000 C_A C_F \Nf T_F g_1^4 g_2^2 
- 734848 C_A \Nf^2 T_F^2 g_1^6 
- 4732416 C_A \Nf^2 T_F^2 g_1^5 g_2 
\right. \nonumber \\
&& \left. ~~~
+ 728064 C_A \Nf^2 T_F^2 g_1^4 g_2^2 
+ 17408000 C_F^2 \Nf T_F g_1^6 
\right. \nonumber \\
&& \left. ~~~
- 9661440 C_F \Nf^2 T_F^2 g_1^6 \right] \frac{1}{31104000} ~+~ O(g_i^8)
\nonumber \\
\gamma_c(g_1,g_2,\alpha) &=&
\left[ \alpha - 5 \right] \frac{C_A g_1^2}{12} 
\nonumber \\
&&
+ \left[ 
-~ 55 \alpha^2 C_A g_1^2 
+ 60 \alpha C_A g_1^2 
- 19952 C_A g_1^2 
+ 2700 C_A g_1 g_2 
+ 600 C_A g_2^2 
\right. \nonumber \\
&& \left. ~~~
- 1088 \Nf T_F g_1^2 \right] \frac{C_A g_1^2}{8640} 
\nonumber \\
&&
+ \left[ 
-~ 2673000 \zeta_3 \alpha^3 C_A^2 g_1^4 
+ 3385125 \alpha^3 C_A^2 g_1^4 
- 891000 \zeta_3 \alpha^2 C_A^2 g_1^4 
\right. \nonumber \\
&& \left. ~~~
+ 867625 \alpha^2 C_A^2 g_1^4 
- 2187000 \zeta_3 \alpha C_A^2 g_1^4 
+ 27779025 \alpha C_A^2 g_1^4 
\right. \nonumber \\
&& \left. ~~~
- 1944000 \zeta_3 \alpha C_A^2 g_1^3 g_2 
- 1764000 \alpha C_A^2 g_1^3 g_2 
- 336000 \alpha C_A^2 g_1^2 g_2^2 
\right. \nonumber \\
&& \left. ~~~
+ 2457600 \alpha \Nf T_F C_A g_1^4 
- 90477000 \zeta_3 C_A^2 g_1^4 
- 1333712377 C_A^2 g_1^4 
\right. \nonumber \\
&& \left. ~~~
+ 66744000 \zeta_3 C_A^2 g_1^3 g_2 
+ 261729900 C_A^2 g_1^3 g_2 
+ 2592000 \zeta_3 C_A^2 g_1^2 g_2^2 
\right. \nonumber \\
&& \left. ~~~
+ 91288500 C_A^2 g_1^2 g_2^2 
- 3888000 \zeta_3 C_A^2 g_1 g_2^3 
- 14296500 C_A^2 g_1 g_2^3 
\right. \nonumber \\
&& \left. ~~~
- 1741500 C_A^2 g_2^4 
- 62208000 \zeta_3 \Nf T_F C_A g_1^4 
- 46076416 C_A \Nf T_F g_1^4 
\right. \nonumber \\
&& \left. ~~~
- 4334400 C_A \Nf T_F g_1^3 g_2 
+ 6864000 C_A \Nf T_F g_1^2 g_2^2 
+ 82944000 \zeta_3 C_F \Nf T_F g_1^4 
\right. \nonumber \\
&& \left. ~~~
- 103600000 C_F \Nf T_F g_1^4 
+ 1722368 \Nf^2 T_F^2 g_1^4 \right] \frac{C_A g_1^2}{62208000} ~+~ O(g_i^8)
\nonumber \\
\gamma_\psi(g_1,g_2,\alpha) &=&
\left[ 3 \alpha + 5 \right] \frac{C_F g_1^2}{6} 
\nonumber \\
&&
+ \left[ 
75 \alpha^2 C_A g_1^2 
+ 1830 \alpha C_A g_1^2 
+ 43617 C_A g_1^2 
- 600 C_A g_2^2 
- 8000 C_F g_1^2 
\right. \nonumber \\
&& \left. ~~~
+ 2048 \Nf T_F g_1^2 \right] \frac{C_F g_1^2}{4320}
\nonumber \\
&&
+ \left[ 
-~ 972000 \zeta_3 \alpha^3 C_A^2 g_1^4 
+ 1296375 \alpha^3 C_A^2 g_1^4 
+ 1215000 \zeta_3 \alpha^2 C_A^2 g_1^4 
\right. \nonumber \\
&& \left. ~~~
- 1231875 \alpha^2 C_A^2 g_1^4 
+ 2430000 \zeta_3 \alpha C_A^2 g_1^4 
+ 29389350 \alpha C_A^2 g_1^4 
\right. \nonumber \\
&& \left. ~~~
- 1944000 \zeta_3 \alpha C_A^2 g_1^3 g_2 
- 1971000 \alpha C_A^2 g_1^3 g_2 
- 747000 \alpha C_A^2 g_1^2 g_2^2 
\right. \nonumber \\
&& \left. ~~~
+ 2420400 \alpha C_A \Nf T_F g_1^4 
+ 74601000 \zeta_3 C_A^2 g_1^4 
+ 816942603 C_A^2 g_1^4 
\right. \nonumber \\
&& \left. ~~~
- 44712000 \zeta_3 C_A^2 g_1^3 g_2 
- 1730025 C_A^2 g_1^3 g_2 
+ 7776000 \zeta_3 C_A^2 g_1^2 g_2^2 
\right. \nonumber \\
&& \left. ~~~
- 55844625 C_A^2 g_1^2 g_2^2 
+ 4357125 C_A^2 g_1 g_2^3 
+ 300375 C_A^2 g_2^4 
- 62208000 \zeta_3 C_A C_F g_1^4 
\right. \nonumber \\
&& \left. ~~~
- 178896000 C_A C_F g_1^4 
+ 3240000 C_A C_F g_1^3 g_2 
+ 4800000 C_A C_F g_1^2 g_2^2 
\right. \nonumber \\
&& \left. ~~~
+ 20736000 \zeta_3 C_A \Nf T_F g_1^4 
+ 41385824 C_A \Nf T_F g_1^4 
+ 7520400 C_A \Nf T_F g_1^3 g_2 
\right. \nonumber \\
&& \left. ~~~
- 2580000 C_A \Nf T_F g_1^2 g_2^2 
+ 62208000 \zeta_3 C_F^2 g_1^4 
- 33056000 C_F^2 g_1^4 
\right. \nonumber \\
&& \left. ~~~
- 20736000 \zeta_3 C_F \Nf T_F g_1^4 
+ 23372000 C_F \Nf T_F g_1^4 
\right. \nonumber \\
&& \left. ~~~
+ 91648 \Nf^2 T_F^2 g_1^4 \right] \frac{C_F g_1^2}{7776000} ~+~ O(g_i^8) ~.
\end{eqnarray}
Finally we note that the quark mass dimension is given by 
\begin{eqnarray}
\gamma_m(g_1,g_2) &=&
-~ \frac{5}{3} C_F g_1^2 
+ \left[ 
-~ 11301 C_A g_1^2 
+ 300 C_A g_2^2 
- 200 C_F g_1^2 
- 544 \Nf T_F g_1^2 \right] \frac{C_F g_1^2}{1080}
\nonumber \\
&&
+ \left[ 
38880000 \zeta_3 C_A^2 g_1^4 
- 424927488 C_A^2 g_1^4 
- 11664000 \zeta_3 C_A^2 g_1^3 g_2 
\right. \nonumber \\
&& \left. ~~~
+ 54029025 C_A^2 g_1^3 g_2 
+ 24506625 C_A^2 g_1^2 g_2^2 
- 2197125 C_A^2 g_1 g_2^3 
\right. \nonumber \\
&& \left. ~~~
- 435375 C_A^2 g_2^4 
- 97200000 \zeta_3 C_A C_F g_1^4 
+ 30950400 C_A C_F g_1^4 
\right. \nonumber \\
&& \left. ~~~
+ 23328000 \zeta_3 C_A C_F g_1^3 g_2 
- 22356000 C_A C_F g_1^3 g_2 
+ 240000 C_A C_F g_1^2 g_2^2 
\right. \nonumber \\
&& \left. ~~~
- 20736000 \zeta_3 C_A \Nf T_F g_1^4 
- 8608304 C_A \Nf T_F g_1^4 
- 4712400 C_A \Nf T_F g_1^3 g_2 
\right. \nonumber \\
&& \left. ~~~
+ 1716000 C_A \Nf T_F g_1^2 g_2^2 
+ 38880000 \zeta_3 C_F^2 g_1^4 
- 23260000 C_F^2 g_1^4 
\right. \nonumber \\
&& \left. ~~~
+ 20736000 \zeta_3 C_F \Nf T_F g_1^4 
- 26194400 C_F \Nf T_F g_1^4 
\right. \nonumber \\
&& \left. ~~~
+ 430592 \Nf^2 T_F^2 g_1^4 \right] \frac{C_F g_1^2}{3888000} ~+~ O(g_i^8)
\end{eqnarray}
which, like the $\beta$-function, is independent of $\alpha$ as it ought to be 
in the $\MSbar$ scheme, \cite{61}. While part of the focus in previous sections
examined the $\MOMts$ scheme in six dimensions, it is not possible to repeat 
that for (\ref{lagqcdd6}). This is primarily because interest in the $\MOMts$ 
scheme concerned the absence of even zetas commencing from four loops. As we do
not have a full set of three loop renormalization group functions yet, let 
alone at four loops, that investigation clearly has to be postponed to a later 
point. However any study of the $\MOMts$ schemes similar to \cite{27,34} may 
only be able to focus on the vertices that determine $\beta_1(g_1,g_2)$ since 
the renormalization of $g_2$ cannot be directly accessed when one of the 
external momenta of a gluon $3$-point vertex is nullified. For similar reasons 
we have not constructed the equivalent of the $\mMOM$ scheme of \cite{35} to 
four loops. One check on our {\sc Forcer} computation is that the two loop 
expressions of \cite{22} have been reproduced and we recall that \cite{22} used
the Laporta approach. Another check is that $\beta_1(g_1,g_2)$ satifies the 
Slavnov-Taylor identity being given by a linear combination of the Landau gauge
gluon and ghost anomalous dimensions. More specifically
\begin{equation}
\beta_a(g_1,g_2) ~=~ \half g_1 \left[ \gamma_A(g_1,g_2,0) 
+ 2 \gamma_c(g_1,g_2,0) \right]
\end{equation}
as the ghost-gluon vertex of (\ref{lagqcdd6}) is finite in the Landau gauge for
the same reasons as its four dimensional counterpart, \cite{62}.

A separate check on the three loop contributions to the gluon, ghost, quark and
quark mass anomalous dimensions lies in comparing with known critical exponents
of the fields computed to several orders in the $1/\Nf$ expansion. These
renormalization group invariants have been computed as a function of $d$ at the
Wilson-Fisher fixed point defined as the non-trivial solution of
$\beta_i(g_1,g_2)$~$=$~$0$ closest to the origin. Evaluating the anomalous
dimensions at the Wilson-Fisher critical point produces critical exponents 
which will be functions of $d$ and $\Nf$ as well as the group Casimirs. They
can be expanded as a double Taylor series in powers of $1/\Nf$, where $\Nf$ is
large, and $\epsilon$ where $d$~$=$~$d_c$~$-$~$2\epsilon$. Here $d_c$ is the
critical dimension of any of the quantum field theories that lie in the same
universality class. In this situation these are the two dimensional 
non-abelian Thirring model, QCD in four dimensions and (\ref{lagqcdd6}) in
six dimensions as well as the tower of theories that lie in eight dimensions
and beyond. Therefore the expansion of the large $\Nf$ exponents when 
$d_c$~$=$~$6$ have to be consistent with the analogous critical anomalous
dimensions evaluated at the same fixed point and expanded in the same double
Taylor series. This is the background to the three loop large $\Nf$ check
extending the previous two loop check of \cite{22}. It is in fact possible to 
carry out such an analysis even though the three loop terms of 
$\beta_2(g_1,g_2)$ have not been found. The evidence for this is deduced from
examining the $g_2$ and $\Nf$ dependence of $\gamma_A(g_1,g_2,\alpha)$, 
$\gamma_c(g_1,g_2,\alpha)$, $\gamma_\psi(g_1,g_2,\alpha)$ and 
$\gamma_m(g_1,g_2)$. It is apparent that the coupling constant associated with
the spectator interaction, which is $g_2$, first appears at two loops. In 
addition at three loops there are no $g_2^2$ terms in
$\gamma_\psi(g_1,g_2,\alpha)$ and $\gamma_m(g_1,g_2)$ meaning that their
associated exponents can be expanded to $O(1/\Nf^2)$ without the three loop
correction to the $g_2$ critical coupling and then compared with the $\epsilon$
expansions of the large $\Nf$ exponents of \cite{63}. For
$\gamma_A(g_1,g_2,\alpha)$ and $\gamma_c(g_1,g_2,\alpha)$ only the $O(1/\Nf)$
$d$-dimensional exponents are known, \cite{64}. More concretely, following the 
large $\Nf$ approach of \cite{21,22} we set
\begin{equation}
g_1 ~=~ \frac{i}{2} \sqrt{\frac{15\epsilon}{T_F\Nf}} x ~~~,~~~
g_2 ~=~ \frac{i}{2} \sqrt{\frac{15\epsilon}{T_F\Nf}} y
\end{equation}
and solve $\beta_i(g_1,g_2)$~$=$~$0$ to find the Wilson-Fisher fixed point is
located at  
\begin{eqnarray}
x &=& 1 ~+~ \left[ -~ \frac{249}{32} C_A + \left[ \frac{475}{48} C_F
+ \frac{5855}{768} C_A \right] \epsilon 
- \left[ \frac{3145}{384} C_F + \frac{104729}{18432} C_A \right] \epsilon^2
\right] \frac{1}{T_F\Nf} \nonumber \\
&& +~ \left[ \frac{186003}{2048} C_A^2 - \left[ \frac{197125}{512} C_A C_F
+ \frac{7530655}{32768} C_A^2 \right] \epsilon
\right. \nonumber \\
&& \left. ~~~~+
\left[ \frac{549125}{1536} C_F^2 + \frac{12040925}{18432} C_A C_F 
+ \frac{1150323311}{3145728} C_A^2 \right] \epsilon^2 \right]
\frac{1}{T_F^2\Nf^2} ~+~ O \left( \frac{\epsilon^3}{T_F^3\Nf^3} \right)
\nonumber \\
y &=& \frac{13}{4} ~+~ \left[ -~ \frac{51327}{2048} C_A
+ \left[ \frac{2325}{64} C_F + \frac{62385}{4096} C_A \right] \epsilon \right]
\frac{1}{T_F\Nf} ~+~ O \left( \frac{\epsilon^2}{T_F^2\Nf^2} \right) 
\end{eqnarray}
in the large $\Nf$ expansion where the order symbols indicate the truncation
order of both the $\epsilon$ and large $\Nf$ expansions. Substituting into the 
various anomalous dimensions we have
\begin{eqnarray}
\gamma_A(g_1^\star,g_2^\star,0) &=& \epsilon ~-~
\left[ \frac{25}{8} \epsilon + \frac{85}{24} \epsilon^2
+ \frac{841}{288} \epsilon^3 \right] \frac{C_A}{T_F\Nf} ~+~
O \left( \frac{\epsilon^3}{T_F^2\Nf^2} \right) \nonumber \\ 
\gamma_c(g_1^\star,g_2^\star,0) &=& 
\left[ \frac{25}{16} \epsilon - \frac{85}{48} \epsilon^2
- \frac{841}{576} \epsilon^3 \right] \frac{C_A}{T_F\Nf} ~+~
O \left( \frac{\epsilon^3}{T_F^2\Nf^2} \right) \nonumber \\ 
\gamma_\psi(g_1^\star,g_2^\star,0) &=& 
\left[ - \frac{25}{8} \epsilon + \frac{20}{3} \epsilon^2
- \frac{179}{288} \epsilon^3 \right] \frac{C_F}{T_F\Nf} \nonumber \\
&& +~ \left[ \frac{6225}{128} C_A \epsilon 
- \left[ \frac{5625}{64} C_F + \frac{102755}{768} C_A \right] \epsilon^2
\right. \nonumber \\
&& \left. ~~~~~
+ \left[ \left[ \frac{199}{32} - \frac{1125}{8} \zeta_3 \right] C_A
+ \left[ \frac{60125}{384} + \frac{1125}{8} \zeta_3 \right] C_F \right] 
\epsilon^3 \right] \frac{C_F}{T_F^2 \Nf^2} 
+ O \left( \frac{\epsilon^3}{T_F^3\Nf^3} \right) \nonumber \\
\gamma_m(g_1^\star,g_2^\star) &=& \left[ \frac{25}{4} \epsilon
- \frac{85}{12} \epsilon^2 - \frac{841}{144} \epsilon^3 \right] 
\frac{C_F}{T_F\Nf} \nonumber \\
&& + \left[ -~ \frac{6225}{64} C_A \epsilon 
+ \left[ \frac{3875}{32} C_F + \frac{161185}{768} C_A \right] \epsilon^2
\right. \nonumber \\
&& \left. ~~~~
+ \left[ \left[ \frac{98741}{1536} + \frac{1125}{4} \zeta_3 \right] C_A
- \left[ \frac{5275}{192} + \frac{1125}{4} \zeta_3 \right] C_F \right] 
\epsilon^3 \right] \frac{C_F}{T_F^2 \Nf^2} + 
O \left( \frac{\epsilon^3}{T_F^3\Nf^3} \right) \nonumber \\
\label{qcdd6expnf}
\end{eqnarray}
for the exponents where $g_i^\star$ are the critical couplings evaluated at $x$
and $y$. These exponents are in full agreement with the $\epsilon$ expansion of
the large $\Nf$ exponents of \cite{63,64} near six dimensions. One caveat to 
this is that for the first three exponents of (\ref{qcdd6expnf}) the check is 
resricted to the Landau gauge since that is a fixed point of the 
renormalization group equations.

With the present renormalization group equations we can examine one property
which has parallels in the four dimensional counterpart of (\ref{lagqcdd6}). It
is known that aside from the Banks-Zaks fixed point of \cite{65,66}, treating 
the gauge parameter as a second coupling constant opens up a richer critical 
point phase plane, \cite{67}, which has been studied more recently in four 
dimensions in \cite{68}. This is the observation that there is a non-zero
critical value of the gauge parameter that leads to an infrared stable fixed
point in the plane of gauge coupling and parameter. Moreover one can redefine
the renormalization group function for $\alpha$ in such a way that this 
critical value is exposed at leading order, \cite{69}. For (\ref{lagqcdd6}) we 
recall the parallel transformation is derived by first redefining the gauge 
field by, \cite{69},
\begin{equation}
\hat{A}^a_\mu ~=~ g_1 A^a_\mu ~.
\end{equation}
With this the relevant sector of the Lagrangian becomes
\begin{equation}
L^{(6)}_{\mbox{\footnotesize{gluonic}}} ~=~
-~ \frac{1}{4g_1^2} \left( \hat{D}_\mu \hat{G}_{\nu\sigma}^a \right)
\left( \hat{D}^\mu \hat{G}^{a \, \nu\sigma} \right) ~-~ 
\frac{1}{2\alpha g_1^2} 
\left( \partial_\mu \partial^\nu \hat{A}^a_\nu \right)
\left( \partial^\mu \partial^\sigma \hat{A}^a_\sigma \right) ~+~ \ldots 
\end{equation}
where the hatted quantities are the same as the unhatted ones but expressed as
a function of $\hat{A}^a_\mu$. This identifies the transverse and longitudinal
operators with separate independent coupling constants. If we compute the 
renormalization group function for the combination $\alpha g_1^2$ given in 
\cite{69} we find
\begin{equation}
\hat{\gamma}_\alpha(g_1,g_2,\alpha) ~=~ \frac{2}{g_1} \beta_1(g_1,g_2) ~+~
\gamma_\alpha(g_1,g_2,\alpha)
\label{gengaugerge}
\end{equation}
which gives 
\begin{eqnarray}
\hat{\gamma}_\alpha(g_1,g_2,\alpha) &=&
-~ [ 2 \alpha + 5 ] \frac{C_A g_1^2}{6}
\nonumber \\
&& +~ \left[
2700 C_A g_1 g_2
+ 600 C_A g_2^2
- 130 \alpha^2 C_A g_1^2
- 1095 \alpha C_A g_1^2
\right. \nonumber \\
&& \left. ~~~~
- 19952 C_A g_1^2
- 1088 \Nf T_F g_1^2 \right] \frac{C_A g_1^2}{4320}
\nonumber \\
&& +~ \left[ 
1215000 \zeta_3 \alpha^3 C_A^2 g_1^4
- 1800375 \alpha^3 C_A^2 g_1^4
- 891000 \zeta_3 \alpha^2 C_A^2 g_1^4
+ 314125 \alpha^2 C_A^2 g_1^4
\right. \nonumber \\
&& \left. ~~~~
+ 4617000 \zeta_3 \alpha C_A^2 g_1^4
- 82449225 \alpha C_A^2 g_1^4
+ 1944000 \zeta_3 \alpha C_A^2 g_1^3 g_2
\right. \nonumber \\
&& \left. ~~~~
+ 7416000 \alpha C_A^2 g_1^3 g_2
+ 1842000 \alpha C_A^2 g_1^2 g_2^2
- 5906400 \Nf T_F \alpha C_A g_1^4
\right. \nonumber \\
&& \left. ~~~~
- 90477000 \zeta_3 C_A^2 g_1^4
- 1333712377 C_A^2 g_1^4
+ 66744000 \zeta_3 C_A^2 g_1^3 g_2
\right. \nonumber \\
&& \left. ~~~~
+ 261729900 C_A^2 g_1^3 g_2
+ 2592000 \zeta_3 C_A^2 g_1^2 g_2^2
+ 91288500 C_A^2 g_1^2 g_2^2
\right. \nonumber \\
&& \left. ~~~~
- 3888000 \zeta_3 C_A^2 g_1 g_2^3
- 14296500 C_A^2 g_1 g_2^3
- 1741500 C_A^2 g_2^4
\right. \nonumber \\
&& \left. ~~~~
- 62208000 \zeta_3 C_A \Nf T_F g_1^4
- 46076416 C_A \Nf T_F g_1^4
- 4334400 C_A \Nf T_F g_1^3 g_2
\right. \nonumber \\
&& \left. ~~~~
+ 6864000 C_A \Nf T_F g_1^2 g_2^2
+ 82944000 \zeta_3 C_F \Nf T_F g_1^4
- 103600000 C_F \Nf T_F g_1^4
\right. \nonumber \\
&& \left. ~~~~
+ 1722368 \Nf^2 T_F^2 g_1^4 \right]
\frac{C_A g_1^2}{31104000} ~+~ O(g_i^8) ~.
\end{eqnarray}
Therefore in the $(g_1,\alpha)$-plane there is a critical gauge parameter value
of $\alpha$~$=$~$-$~$\frac{5}{2}$. In four dimensions the analogous value was 
$(-3)$ and it was suggested that defining the combination $\alpha g_1^2$ in
\cite{67,69} as a second coupling then that coupling in effect was an
accounting parameter for the longitudinal modes of the gluon in the
perturbative expansion of gauge variant Green's functions. In fact such a 
pattern of a rational value for a critical gauge parameter extends to the next
case which is the eight dimensional extension of (\ref{lagqcdd6}) and was
renormalized at one loop in \cite{70}. Examining the (\ref{gengaugerge}) for 
eight dimensions gave $(-\frac{7}{3})$ for the critical gauge parameter,
$\alpha_\star$. In fact it is straightforward to deduce the $d$ dependence of 
$\alpha_\star$ and we note
\begin{equation}
\alpha_\star(d) ~=~ -~ \frac{2(d-1)}{(d-2)} ~.
\end{equation}
This is a montonically increasing function of $d$ for $d$~$>$~$2$ with limits
\begin{equation}
\lim_{d\to2^+} \alpha_\star ~=~ -~ \infty ~~~~,~~~~
\lim_{d\to\infty} \alpha_\star ~=~ -~ 2 ~.
\end{equation}
The singular behaviour in strictly two dimensions may be indicative of the
absence of longitudinal modes in the gluon in that dimension. One puzzle that
arose in four dimensions was the relation of the integer value of
$\alpha_\star(4)$ to the value for the Yennie gauge. The latter is
$\alpha$~$=$~$3$ and $\alpha_\star(4)$ has been referred to as the anti-Yennie 
gauge in, for example, \cite{71}. This apparent connection does not translate 
to the six and higher dimensional generalizations of QCD since the defining
criterion for the Yennie gauge is the vanishing of the ghost anomalous
dimension at leading order. Clearly in six dimensions this occurs at 
$\alpha$~$=$~$5$ while in eight dimensions $\alpha$ would be $7$. The 
generalization is 
\begin{equation}
\alpha_{\mbox{\footnotesize{Yennie}}}(d) ~=~ d ~-~ 1 ~.
\label{dyennie}
\end{equation}
It is straightforward to check this directly by computing the one loop
correction to the ghost $2$-point function that eventually produces 
$\gamma_c(g_i,\alpha)$ in the higher dimensional Lagrangians. The key sector of
those is the ghost term given by 
\begin{equation}
L^{(d)}_{\mbox{\scriptsize{ghost}}} ~=~ -~ 
\left( \Box^{n_d} \bar{c}^a \right) 
\left( \partial^\mu D_\mu c \right)^a 
\label{lagqcddgh}
\end{equation}
where $n_d$~$=$~$\half d$~$-$~$2$ when $d$ is integer. Extracting the gluon 
propagator from the relevant sector of
\begin{equation}
L^{(d)}_{\mbox{\scriptsize{gluon ke}}} ~=~ -~ \frac{1}{4} 
\left( D^{\mu_1} \ldots D^{\mu_{n_d}} G^{a \, \mu\nu} \right)^2 ~-~ 
\frac{(-1)^{n_d}}{2\alpha} \left( \partial^\mu A^a_\mu \right) \Box^{n_d} 
\left( \partial^\nu A^a_\nu \right)
\label{lagqcddke}
\end{equation}
and using the ghost-gluon interation from (\ref{lagqcddgh}) the one loop ghost 
$2$-point function can be computed in $d$-dimensions. Expanding it in powers of
$\epsilon$ near each even critical dimension strictly above two produces 
(\ref{dyennie}). We note that for higher dimensions the Lagrangian can always 
be written in terms of one independent gauge invariant operator with two 
gluons. While an alternative operator to the first term of (\ref{lagqcddke}) is
$G^a_{\mu\nu}  \Box^{n_d} G^{a \, \mu\nu}$ the latter can be rewritten in terms
of it using integration by parts in the Lagrangian plus additional gauge 
invariant operators with three or more gluon legs. 

\sect{Discussion.}

One of our main tasks was to provide the missing information that prevented
the symbolic manipulation {\sc Forcer} algorithm being used to study properties
of quantum field theories in six dimensions. This required mapping the 
$\epsilon$ expansion of the known four dimensional {\sc Forcer} masters to six 
dimensions using the Tarasov method, \cite{23,24}, with the six dimensional 
counterparts available now at weight $9$. By way of checks we reproduced the 
available four loop renormalization group functions of $\phi^3$ theory both 
with and without $O(N)$ symmetry. As a consequence of the masters being 
available to high order in powers of $\epsilon$ we were able to explore the 
$\MOMts$ scheme property of the cubic theory to five loops and verify that the 
absence of even zetas to this order is not restricted to four dimensions. As 
the scalar $\phi^3$ interaction is devoid of derivatives it was relatively easy
to explore a new scheme which was $\MaxSs$ whereby all the terms of the 
$\epsilon$ expansion of the sum of the Feynman integrals are removed from the 
two divergent Green's functions. This scheme had the interesting feature that 
in strictly six dimensions the renormalization group functions were equivalent 
to those of the $\MOMts$ scheme. Such a property should also hold in four 
dimensional theories including gauge theories. Having studied $\phi^3$ theory 
and extended the three loop renormalization of six dimensional QED to four 
loops we were able to repeat the same exercise for scalar QED in six 
dimensions. One outcome of the latter was to verify the ultraviolet completion 
of the four dimensional version of QED to six dimensions. In essence this 
demonstrates the usefulness of compiling the new masters and opens the way to 
extend other theories to a similar level. A first step in that direction has 
also been provided here in that all the renormalization group functions of the 
six dimensional ultraviolet completion of QCD have been determined to three 
loops bar one. That outstanding $\beta$-function is not accessible using a 
$2$-point function approach due to the fact that the operator associated with
the coupling involves the product of three field strengths. If that could be 
overcome at three loops then the three loop QCD renormalization group functions
computed here could be extended to four loops. There are other six dimensional 
theories which have been of interest recently whose ultraviolet completeness 
has been studied at one or two loops. For instance in \cite{55,56} the six 
dimensional version of the $\Cc\Pp(N)$ nonlinear $\sigma$ model has been 
studied to two loops and in the large $\Nf$ expansion. For a certain 
configuration of the coupling constants it contains scalar QED. The $\Cc\Pp(N)$
$\sigma$ model contains an additional scalar field and in principle it can now 
be examined using the {\sc Forcer} construction and the ultraviolet completion 
studied that will necessarily build on the earlier work of \cite{72} and the 
large $N$ results of \cite{73} that allow for a connection of theories in 
$d$-dimensions.

One reason we mention this particular theory is by way of caution. In the 
course of the application of the new masters to renormalizing a variety of
theories they have all carried the same feature with respect the 
renormalization of the vertices. This is that when nullifying the momentum of
one external leg of the $3$-point function no infrared singularities arise.
For $\phi^3$ theory the propagators at the nullified vertex will reduce to a
contribution of $\frac{1}{(p^2)^2}$ where $p$ is the momentum of a propagator. 
In six dimensions, unlike four dimensions, this is infrared safe. So we have 
not needed to introduce any infrared rearrangement. For certain theories in six
dimensions this may be necessary and is usually the case for gauge fields since
they have have a higher derivative kinetic term as is evident in 
(\ref{lagqedd6}) and (\ref{lagqcdd6}). For the cases considered here no 
infrared rearrangement was required. For the ghost-gluon vertex of 
(\ref{lagqcdd6}) the extra derivatives in the interaction were the redeeming 
feature. However for the six dimensional version of the $\Cc\Pp(N)$ nonlinear 
$\sigma$ model infrared rearrangement may be necessary given the absence of
derivatives in the basic interactions involving the gauge field. Some evidence 
of such difficulties have been implicit in this article. For instance, the mass
dimension of the scalar field of (\ref{lagsqedd6}) was not provided despite the
fact the electron mass dimension was computed from (\ref{lagqedd6}). The reason
for this has been given previously in \cite{55,73} which is that the scalar
electron mass operator mixes with other operators. One is the square of the
photon field strength. The remainder include the square of the operator 
defining the linear covariant gauge condition and those which are total
derivative operators. Of the latter the gauge variant ones will not play 
a role in the extraction of the eigen-anomalous dimensions of the two gauge 
invariant mass operators. However to extract the renormalization constants for
the photon mass operator, inserting it at zero momentum in a photon $2$-point
will immediately produce an infrared issue which would require either an
infrared rearrangement ahead of a {\sc Forcer} evaluation or determining the
mixing matrix by routing the external monentum in one external leg and out
through the operator itself. In this latter case the mixing with gauge variant
and total derivative operators would have to be included in the construction. 
While this is beyond the scope of the current article we note it as a reminder
and illustration of the potential pitfalls in naively applying Feynman graph 
integration routines. 

\vspace{1cm}
\noindent
{\bf Acknowledgements.} This work was carried out with the support of the STFC
Consolidated Grant ST/T000988/1. For the purpose of open access, the author 
has applied a Creative Commons Attribution (CC-BY) licence to any Author 
Accepted Manuscript version arising. The data representing the main results 
here are accessible in electronic form from the arXiv ancillary directory 
associated with the article \cite{44}. The author gratefully appreciates 
discussions with I. Jack and H. Osborn.

\appendix

\sect{Six dimensional {\sc Forcer} masters.}

In this appendix we record the $\epsilon$ expansion of the $16$ three and four
loop {\sc Forcer} masters in six dimensions. The name of each topology matches 
exactly with those of \cite{8,9} where the respective Feynman graphs are 
defined graphically. We have carried out the expansion of each master to the 
level of $\zeta_9$ which does not equate to the same order in $\epsilon$ for 
each master. Similar to \cite{9} each is expressed in the $G$-scheme. Using 
the same master label as that of the {\tt Masters.prc} file of the {\sc Forcer}
release we have
\begin{eqnarray}
\mbox{{\bf haha}} &=&
-~ \frac{1}{72 \epsilon^2}
+ \left[
- \frac{17}{108}
- \frac{1}{36} \zeta_3
+ \frac{5}{36} \zeta_5
\right] \frac{1}{\epsilon}
- \frac{407}{432}
- \frac{1}{24} \zeta_4
- \frac{1}{36} \zeta_3^2
+ \frac{25}{72} \zeta_6
+ \frac{25}{216} \zeta_3
+ \frac{85}{216} \zeta_5
\nonumber \\
&&
+ \left[
- \frac{24671}{7776}
- \frac{67}{48} \zeta_7
- \frac{17}{216} \zeta_3^2
- \frac{13}{144} \zeta_5
- \frac{1}{12} \zeta_3 \zeta_4
+ \frac{25}{144} \zeta_4
+ \frac{425}{432} \zeta_6
+ \frac{895}{432} \zeta_3
\right] \epsilon
\nonumber \\
&&
+ \left[
- \frac{58795}{7776} \zeta_5
- \frac{6011}{720} \zeta_8
- \frac{3785}{288} \zeta_7
- \frac{287}{144} \zeta_3^2
- \frac{17}{72} \zeta_3 \zeta_4
- \frac{5}{32} \zeta_6
+ \frac{9}{10} \zeta_{5,3}
+ \frac{17}{2} \zeta_3 \zeta_5
\right. \nonumber \\
&& \left. ~~~~
+ \frac{895}{288} \zeta_4
+ \frac{103955}{7776} \zeta_3
+ \frac{113711}{46656}
\right] \epsilon^2
\nonumber \\
&&
+ \left[
- \frac{4168205}{46656} \zeta_5
- \frac{298475}{15552} \zeta_6
- \frac{178393}{7776} \zeta_3^2
- \frac{41341}{432} \zeta_9
- \frac{26929}{432} \zeta_8
- \frac{6235}{192} \zeta_7
\right. \nonumber \\
&& \left. ~~~~
- \frac{287}{48} \zeta_3 \zeta_4
- \frac{127}{54} \zeta_3^3
+ \frac{21}{4} \zeta_{5,3}
+ \frac{385}{18} \zeta_3 \zeta_6
+ \frac{559}{12} \zeta_3 \zeta_5
+ \frac{103955}{5184} \zeta_4
\right. \nonumber \\
&& \left. ~~~~
+ \frac{2837317}{46656} \zeta_3
+ \frac{12816899}{93312}
+ 6 \zeta_4 \zeta_5
\right] \epsilon^3 ~+~ O(\epsilon^4)
\nonumber \\
\mbox{{\bf no1}} &=&
\left[
- \frac{1}{72} \zeta_3
+ \frac{1}{216}
\right] \frac{1}{\epsilon^2}
+ \left[
- \frac{13}{216} \zeta_3
- \frac{1}{48} \zeta_4
+ \frac{65}{1296}
\right] \frac{1}{\epsilon}
- \frac{13}{144} \zeta_4
- \frac{2}{9} \zeta_3
- \frac{1}{36} \zeta_5
+ \frac{179}{648}
\nonumber \\
&&
+ \left[
- \frac{457}{972} \zeta_3
- \frac{97}{432} \zeta_5
- \frac{5}{12} \zeta_3^2
- \frac{5}{144} \zeta_6
- \frac{1}{3} \zeta_4
+ \frac{4109}{5832}
\right] \epsilon
\nonumber \\
&&
+ \left[
- \frac{113837}{34992}
- \frac{457}{648} \zeta_4
- \frac{355}{864} \zeta_6
- \frac{271}{96} \zeta_5
- \frac{239}{144} \zeta_3^2
- \frac{5}{4} \zeta_3 \zeta_4
+ \frac{1079}{144} \zeta_7
+ \frac{32069}{11664} \zeta_3
\right] \epsilon^2
\nonumber \\
&&
+ \left[
- \frac{461663}{7776}
- \frac{233099}{15552} \zeta_5
- \frac{3745}{576} \zeta_6
- \frac{1025}{36} \zeta_3 \zeta_5
- \frac{239}{48} \zeta_3 \zeta_4
+ \frac{9}{10} \zeta_{5,3}
+ \frac{623}{288} \zeta_3^2
\right. \nonumber \\
&& \left. ~~~~
+ \frac{3583}{216} \zeta_7
+ \frac{5533}{240} \zeta_8
+ \frac{32069}{7776} \zeta_4
+ \frac{1051357}{23328} \zeta_3
\right] \epsilon^3
\nonumber \\
&&
+ \left[
- \frac{652648633}{1259712}
- \frac{3871021}{93312} \zeta_5
- \frac{1128935}{31104} \zeta_6
- \frac{50051}{864} \zeta_7
- \frac{16367}{1920} \zeta_8
- \frac{4975}{72} \zeta_3 \zeta_6
\right. \nonumber \\
&& \left. ~~~~
- \frac{1238}{27} \zeta_3 \zeta_5
- \frac{1187}{24} \zeta_4 \zeta_5
- \frac{785}{108} \zeta_3^3
+ \frac{263}{10} \zeta_{5,3}
+ \frac{623}{96} \zeta_3 \zeta_4
+ \frac{87667}{288} \zeta_9
+ \frac{129635}{5184} \zeta_3^2
\right. \nonumber \\
&& \left. ~~~~
+ \frac{1051357}{15552} \zeta_4
+ \frac{159555301}{419904} \zeta_3
\right] \epsilon^4 ~+~ O(\epsilon^5)
\nonumber \\
\mbox{{\bf no2}} &=&
\left[
- \frac{1}{72} \zeta_3
+ \frac{1}{216}
\right] \frac{1}{\epsilon^2}
+ \left[
- \frac{13}{216} \zeta_3
- \frac{1}{48} \zeta_4
+ \frac{37}{648}
\right] \frac{1}{\epsilon}
- \frac{41}{144} \zeta_3
- \frac{13}{144} \zeta_4
- \frac{1}{16} \zeta_5
+ \frac{505}{1296}
\nonumber \\
&&
+ \left[
- \frac{10433}{7776} \zeta_3
- \frac{83}{288} \zeta_5
- \frac{41}{96} \zeta_4
- \frac{35}{288} \zeta_6
- \frac{29}{144} \zeta_3^2
+ \frac{41141}{23328}
\right] \epsilon
\nonumber \\
&&
+ \left[
- \frac{215839}{46656} \zeta_3
- \frac{10433}{5184} \zeta_4
- \frac{3529}{1728} \zeta_5
- \frac{985}{1728} \zeta_6
- \frac{29}{48} \zeta_3 \zeta_4
+ \frac{275}{864} \zeta_3^2
+ \frac{607}{96} \zeta_7
\right. \nonumber \\
&& \left. ~~~~
+ \frac{569563}{139968}
\right] \epsilon^2
\nonumber \\
&&
+ \left[
- \frac{1736987}{93312}
- \frac{354467}{93312} \zeta_3
- \frac{215839}{31104} \zeta_4
- \frac{44501}{3456} \zeta_5
- \frac{15185}{3456} \zeta_6
- \frac{147}{8} \zeta_3 \zeta_5
+ \frac{57}{20} \zeta_{5,3}
\right. \nonumber \\
&& \left. ~~~~
+ \frac{275}{288} \zeta_3 \zeta_4
+ \frac{4873}{192} \zeta_7
+ \frac{9347}{1728} \zeta_3^2
+ \frac{29021}{1920} \zeta_8
\right] \epsilon^3
\nonumber \\
&&
+ \left[
- \frac{1681647517}{5038848}
- \frac{3011489}{62208} \zeta_5
- \frac{1793885}{62208} \zeta_6
- \frac{354467}{62208} \zeta_4
- \frac{6085}{144} \zeta_3 \zeta_5
- \frac{3235}{72} \zeta_3 \zeta_6
\right. \nonumber \\
&& \left. ~~~~
- \frac{783}{16} \zeta_4 \zeta_5
+ \frac{391}{40} \zeta_{5,3}
+ \frac{1117}{216} \zeta_3^3
+ \frac{9347}{576} \zeta_3 \zeta_4
+ \frac{107111}{432} \zeta_9
+ \frac{252305}{3456} \zeta_7
\right. \nonumber \\
&& \left. ~~~~
+ \frac{260863}{3840} \zeta_8
+ \frac{1032319}{31104} \zeta_3^2
+ \frac{174453625}{1679616} \zeta_3
\right] \epsilon^4 ~+~ O(\epsilon^5)
\nonumber \\
\mbox{{\bf no6}} &=&
\frac{1}{864 \epsilon^3}
+ \left[
- \frac{1}{216} \zeta_3
+ \frac{85}{10368}
\right] \frac{1}{\epsilon^2}
\nonumber \\
&&
+ \left[
- \frac{7}{324} \zeta_3
- \frac{1}{144} \zeta_4
+ \frac{1171}{41472}
\right] \frac{1}{\epsilon}
- \frac{7}{216} \zeta_4
- \frac{1}{108} \zeta_5
+ \frac{43}{7776} \zeta_3
+ \frac{575}{1492992}
\nonumber \\
&&
+ \left[
- \frac{15492679}{17915904}
- \frac{191}{1296} \zeta_5
- \frac{5}{36} \zeta_3^2
- \frac{5}{432} \zeta_6
+ \frac{43}{5184} \zeta_4
+ \frac{45173}{93312} \zeta_3
\right] \epsilon
\nonumber \\
&&
+ \left[
- \frac{635174947}{71663616}
- \frac{815}{2592} \zeta_6
- \frac{265}{432} \zeta_3^2
- \frac{5}{12} \zeta_3 \zeta_4
+ \frac{319}{216} \zeta_7
+ \frac{12191}{7776} \zeta_5
+ \frac{45173}{62208} \zeta_4
\right. \nonumber \\
&& \left. ~~~~
+ \frac{499027}{124416} \zeta_3
\right] \epsilon^2
\nonumber \\
&&
+ \left[
- \frac{172136749511}{2579890176}
- \frac{11483}{2592} \zeta_3^2
- \frac{355}{54} \zeta_3 \zeta_5
- \frac{265}{144} \zeta_3 \zeta_4
+ \frac{3}{5} \zeta_{5,3}
+ \frac{4877}{1440} \zeta_8
+ \frac{15107}{2592} \zeta_7
\right. \nonumber \\
&& \left. ~~~~
+ \frac{15185}{3888} \zeta_6
+ \frac{312319}{23328} \zeta_5
+ \frac{499027}{82944} \zeta_4
+ \frac{124450471}{4478976} \zeta_3
\right] \epsilon^3
\nonumber \\
&&
+ \left[
- \frac{13667311599401}{30958682112}
+ \frac{29876391323}{161243136} \zeta_3
- \frac{89813}{3888} \zeta_3^2
- \frac{18185}{648} \zeta_3 \zeta_5
- \frac{11483}{864} \zeta_3 \zeta_4
\right. \nonumber \\
&& \left. ~~~~
- \frac{517}{36} \zeta_4 \zeta_5
- \frac{425}{27} \zeta_3 \zeta_6
+ \frac{11}{4} \zeta_{5,3}
+ \frac{310}{81} \zeta_3^3
+ \frac{5071}{384} \zeta_8
+ \frac{11917}{216} \zeta_9
+ \frac{395303}{6912} \zeta_5
\right. \nonumber \\
&& \left. ~~~~
+ \frac{1285387}{15552} \zeta_7
+ \frac{6020515}{186624} \zeta_6
+ \frac{124450471}{2985984} \zeta_4
\right] \epsilon^4 ~+~ O(\epsilon^5)
\nonumber \\
\mbox{{\bf lala}} &=&
\frac{7}{103680 \epsilon^3}
+ \frac{61}{77760 \epsilon^2}
+ \frac{32939}{7464960 \epsilon}
+ \frac{1}{1296} \zeta_3
+ \frac{277411}{17915904}
\nonumber \\
&&
+ \left[
\frac{1}{864} \zeta_4
+ \frac{781}{155520} \zeta_3
+ \frac{19619333}{1074954240}
\right] \epsilon
\nonumber \\
&&
+ \left[
- \frac{4976176237}{12899450880}
- \frac{147}{64} \zeta_7
+ \frac{781}{103680} \zeta_4
+ \frac{1999}{1728} \zeta_5
+ \frac{113243}{93312} \zeta_3
\right] \epsilon^2
\nonumber \\
&&
+ \left[
- \frac{958677083731}{154793410560}
- \frac{27097}{6480} \zeta_5
- \frac{12547}{1620} \zeta_3^2
- \frac{6189}{640} \zeta_8
- \frac{245}{64} \zeta_7
+ \frac{27}{40} \zeta_{5,3}
+ \frac{45}{8} \zeta_3 \zeta_5
\right. \nonumber \\
&& \left. ~~~~
+ \frac{29965}{10368} \zeta_6
+ \frac{113243}{62208} \zeta_4
+ \frac{207922571}{11197440} \zeta_3
\right] \epsilon^3
\nonumber \\
&&
+ \left[
- \frac{124644920515261}{1857520926720}
+ \frac{24089428751}{134369280} \zeta_3
- \frac{3018899}{41472} \zeta_5
- \frac{835661}{15552} \zeta_3^2
- \frac{651109}{62208} \zeta_6
\right. \nonumber \\
&& \left. ~~~~
- \frac{12547}{540} \zeta_3 \zeta_4
- \frac{4583}{48} \zeta_9
- \frac{2063}{128} \zeta_8
+ \frac{9}{8} \zeta_{5,3}
+ \frac{27}{8} \zeta_4 \zeta_5
+ \frac{75}{8} \zeta_3 \zeta_5
+ \frac{89}{8} \zeta_3^3
\right. \nonumber \\
&& \left. ~~~~
+ \frac{225}{16} \zeta_3 \zeta_6
+ \frac{66137}{960} \zeta_7
+ \frac{207922571}{7464960} \zeta_4
\right] \epsilon^4 ~+~ O(\epsilon^5)
\nonumber \\
\mbox{{\bf nono}} &=&
\frac{7}{20736 \epsilon^3}
+ \frac{277}{138240 \epsilon^2}
+ \left[
\frac{1}{2880} \zeta_3
+ \frac{88751}{14929920}
\right] \frac{1}{\epsilon}
\nonumber \\
&&
- \frac{30421}{59719680}
+ \frac{1}{1920} \zeta_4
+ \frac{1183}{103680} \zeta_3
\nonumber \\
&&
+ \left[
- \frac{333199777}{2149908480}
- \frac{23}{2880} \zeta_5
+ \frac{1183}{69120} \zeta_4
+ \frac{5915}{82944} \zeta_3
\right] \epsilon
\nonumber \\
&&
+ \left[
- \frac{88173683267}{60197437440}
- \frac{1}{48} \zeta_6
+ \frac{55}{4032} \zeta_3^2
+ \frac{511}{1280} \zeta_5
+ \frac{5915}{55296} \zeta_4
+ \frac{39770197}{104509440} \zeta_3
\right] \epsilon^2
\nonumber \\
&&
+ \left[
- \frac{784677709121141}{75848771174400}
+ \frac{33133204459}{14631321600} \zeta_3
- \frac{8088181}{25401600} \zeta_3^2
- \frac{1321}{5760} \zeta_7
+ \frac{55}{1344} \zeta_3 \zeta_4
\right. \nonumber \\
&& \left. ~~~~
+ \frac{2513}{2592} \zeta_6
+ \frac{228563}{82944} \zeta_5
+ \frac{39770197}{69672960} \zeta_4
\right] \epsilon^3
\nonumber \\
&&
+ \left[
- \frac{76520147938928867}{1179869773824000}
- \frac{7099705903}{3556224000} \zeta_3^2
+ \frac{33133204459}{9754214400} \zeta_4
\right. \nonumber \\
&& \left. ~~~~
+ \frac{268698233469613}{18435465216000} \zeta_3
- \frac{8088181}{8467200} \zeta_3 \zeta_4
- \frac{12819}{12800} \zeta_8
+ \frac{51}{400} \zeta_{5,3}
\right. \nonumber \\
&& \left. ~~~~
+ \frac{2183}{2016} \zeta_3 \zeta_5
+ \frac{46385}{6912} \zeta_6
+ \frac{23604521}{2419200} \zeta_7
+ \frac{151421381}{11612160} \zeta_5
\right] \epsilon^4
\nonumber \\
&&
+ \left[
- \frac{5101030757005946137661}{13379723235164160000}
- \frac{38332734677761}{4480842240000} \zeta_3^2
- \frac{7099705903}{1185408000} \zeta_3 \zeta_4
\right. \nonumber \\
&& \left. ~~~~
+ \frac{69388016609}{1016064000} \zeta_7
+ \frac{873900507443}{14631321600} \zeta_5
+ \frac{268698233469613}{12290310144000} \zeta_4
\right. \nonumber \\
&& \left. ~~~~
+ \frac{80588867392532353}{860321710080000} \zeta_3
- \frac{14767793}{1411200} \zeta_3 \zeta_5
- \frac{75241}{15120} \zeta_9
- \frac{7501}{10080} \zeta_3^3
\right. \nonumber \\
&& \left. ~~~~
+ \frac{95}{36} \zeta_3 \zeta_6
+ \frac{4489}{6720} \zeta_4 \zeta_5
+ \frac{145967}{56000} \zeta_{5,3}
+ \frac{23625397}{746496} \zeta_6
+ \frac{163411417}{6912000} \zeta_8
\right] \epsilon^5 ~+~ O(\epsilon^6)
\nonumber \\
\mbox{{\bf cross}} &=&
\frac{1}{69120 \epsilon^2}
+ \frac{47}{497664 \epsilon}
+ \frac{9241}{29859840}
+ \left[
- \frac{1}{288} \zeta_5
+ \frac{1}{405} \zeta_3
+ \frac{383903}{358318080}
\right] \epsilon
\nonumber \\
&&
+ \left[
- \frac{65}{3456} \zeta_5
- \frac{5}{576} \zeta_6
+ \frac{1}{270} \zeta_4
+ \frac{7}{1440} \zeta_3^2
+ \frac{107}{38880} \zeta_3
+ \frac{49253441}{4299816960}
\right] \epsilon^2
\nonumber \\
&&
+ \left[
- \frac{36259}{233280} \zeta_3
- \frac{325}{6912} \zeta_6
- \frac{127}{2880} \zeta_7
+ \frac{7}{480} \zeta_3 \zeta_4
+ \frac{91}{3456} \zeta_3^2
+ \frac{107}{25920} \zeta_4
+ \frac{10889}{207360} \zeta_5
\right. \nonumber \\
&& \left. ~~~~
+ \frac{160435019}{1146617856}
\right] \epsilon^3
\nonumber \\
&&
+ \left[
\frac{275319668651}{206391214080}
- \frac{10696109}{5598720} \zeta_3
- \frac{36259}{155520} \zeta_4
- \frac{1651}{6912} \zeta_7
- \frac{27}{200} \zeta_{5,3}
- \frac{11}{720} \zeta_3 \zeta_5
\right. \nonumber \\
&& \left. ~~~~
+ \frac{91}{1152} \zeta_3 \zeta_4
+ \frac{1153}{9216} \zeta_6
+ \frac{4129}{19200} \zeta_8
+ \frac{52129}{207360} \zeta_3^2
+ \frac{1273231}{2488320} \zeta_5
\right] \epsilon^4
\nonumber \\
&&
+ \left[
\frac{78476268381583}{7430083706880}
- \frac{68574311}{4478976} \zeta_3
- \frac{10696109}{3732480} \zeta_4
- \frac{871}{2160} \zeta_3^3
- \frac{173}{2160} \zeta_9
\right. \nonumber \\
&& \left. ~~~~
- \frac{143}{1728} \zeta_3 \zeta_5
- \frac{117}{160} \zeta_{5,3}
- \frac{1}{16} \zeta_3 \zeta_6
+ \frac{95}{96} \zeta_4 \zeta_5
+ \frac{52129}{69120} \zeta_3 \zeta_4
+ \frac{53677}{46080} \zeta_8
\right. \nonumber \\
&& \left. ~~~~
+ \frac{1164091}{414720} \zeta_7
+ \frac{1266383}{995328} \zeta_6
+ \frac{2105237}{829440} \zeta_3^2
+ \frac{2121787}{9953280} \zeta_5
\right] \epsilon^5 ~+~ O(\epsilon^6)
\nonumber \\
\mbox{{\bf bebe}} &=&
\frac{1}{41472 \epsilon^3}
+ \frac{173}{1382400 \epsilon^2}
+ \left[
\frac{1}{28800} \zeta_3
+ \frac{245651}{746496000}
\right] \frac{1}{\epsilon}
\nonumber \\
&&
- \frac{46303}{4976640000}
+ \frac{1}{19200} \zeta_4
+ \frac{2989}{5184000} \zeta_3
\nonumber \\
&&
+ \left[
- \frac{18489907021}{2687385600000}
- \frac{23}{28800} \zeta_5
+ \frac{2989}{3456000} \zeta_4
+ \frac{13087}{3840000} \zeta_3
\right] \epsilon
\nonumber \\
&&
+ \left[
- \frac{1087166778487}{17915904000000}
- \frac{1}{480} \zeta_6
+ \frac{29}{28800} \zeta_3^2
+ \frac{13087}{2560000} \zeta_4
+ \frac{25751}{1728000} \zeta_5
\right. \nonumber \\
&& \left. ~~~~
+ \frac{318793381}{18662400000} \zeta_3
\right] \epsilon^2
\nonumber \\
&&
+ \left[
- \frac{3908335481909509}{9674588160000000}
+ \frac{33530592221}{373248000000} \zeta_3
- \frac{67019}{5184000} \zeta_3^2
- \frac{221}{11520} \zeta_7
+ \frac{29}{9600} \zeta_3 \zeta_4
\right. \nonumber \\
&& \left. ~~~~
+ \frac{9283}{259200} \zeta_6
+ \frac{404999}{3840000} \zeta_5
+ \frac{318793381}{12441600000} \zeta_4
\right] \epsilon^3
\nonumber \\
&&
+ \left[
- \frac{155391945049543423}{64497254400000000}
+ \frac{2964690079}{6220800000} \zeta_5
+ \frac{33530592221}{248832000000} \zeta_4
\right. \nonumber \\
&& \left. ~~~~
+ \frac{33844757157949}{67184640000000} \zeta_3
- \frac{242977}{3840000} \zeta_3^2
- \frac{67019}{1728000} \zeta_3 \zeta_4
- \frac{9403}{128000} \zeta_8
+ \frac{27}{4000} \zeta_{5,3}
\right. \nonumber \\
&& \left. ~~~~
+ \frac{1153}{14400} \zeta_3 \zeta_5
+ \frac{48989}{192000} \zeta_6
+ \frac{252157}{691200} \zeta_7
\right] \epsilon^4
\nonumber \\
&&
+ \left[
- \frac{472681590682990336861}{34828517376000000000}
- \frac{2341117451}{18662400000} \zeta_3^2
+ \frac{735800616917}{373248000000} \zeta_5
\right. \nonumber \\
&& \left. ~~~~
+ \frac{33844757157949}{44789760000000} \zeta_4
+ \frac{3881209963480309}{1343692800000000} \zeta_3
- \frac{304861}{864000} \zeta_3 \zeta_5
\right. \nonumber \\
&& \left. ~~~~
- \frac{242977}{1280000} \zeta_3 \zeta_4
- \frac{967}{14400} \zeta_3^3
- \frac{643}{1800} \zeta_9
+ \frac{281}{1440} \zeta_3 \zeta_6
+ \frac{667}{9600} \zeta_4 \zeta_5
+ \frac{7767}{80000} \zeta_{5,3}
\right. \nonumber \\
&& \left. ~~~~
+ \frac{12090239}{4608000} \zeta_7
+ \frac{61097399}{69120000} \zeta_8
+ \frac{1071909607}{933120000} \zeta_6
\right] \epsilon^5 ~+~ O(\epsilon^6)
\nonumber \\
\mbox{{\bf nostar6}} &=&
\frac{1}{20736 \epsilon^3}
+ \frac{11}{248832 \epsilon^2}
+ \left[
\frac{1}{576} \zeta_3
+ \frac{3377}{2985984}
\right] \frac{1}{\epsilon}
+ \frac{1}{384} \zeta_4
+ \frac{95}{20736} \zeta_3
+ \frac{924031}{35831808}
\nonumber \\
&&
+ \left[
- \frac{9827}{248832} \zeta_3
- \frac{13}{576} \zeta_5
+ \frac{95}{13824} \zeta_4
+ \frac{40074923}{143327232}
\right] \epsilon
\nonumber \\
&&
+ \left[
\frac{3824875973}{1719926784}
- \frac{1629893}{2985984} \zeta_3
- \frac{9827}{165888} \zeta_4
- \frac{1625}{6912} \zeta_5
- \frac{35}{576} \zeta_6
+ \frac{19}{576} \zeta_3^2
\right] \epsilon^2
\nonumber \\
&&
+ \left[
\frac{930294184801}{61917364224}
- \frac{152188699}{35831808} \zeta_3
- \frac{1629893}{1990656} \zeta_4
- \frac{66521}{27648} \zeta_5
- \frac{12425}{20736} \zeta_6
- \frac{341}{432} \zeta_7
\right. \nonumber \\
&& \left. ~~~~
+ \frac{19}{192} \zeta_3 \zeta_4
+ \frac{8417}{20736} \zeta_3^2
\right] \epsilon^3
\nonumber \\
&&
+ \left[
- \frac{3901210535}{143327232} \zeta_3
+ \frac{68481144449711}{743008370688}
- \frac{152188699}{23887872} \zeta_4
- \frac{19005437}{995328} \zeta_5
\right. \nonumber \\
&& \left. ~~~~
- \frac{1472155}{248832} \zeta_6
- \frac{257761}{69120} \zeta_8
- \frac{21965}{3456} \zeta_7
+ \frac{21}{40} \zeta_{5,3}
+ \frac{2899}{864} \zeta_3 \zeta_5
+ \frac{8417}{6912} \zeta_3 \zeta_4
\right. \nonumber \\
&& \left. ~~~~
+ \frac{981187}{248832} \zeta_3^2
\right] \epsilon^4
\nonumber \\
&&
+ \left[
- \frac{272730778249}{1719926784} \zeta_3
- \frac{3901210535}{95551488} \zeta_4
+ \frac{4741980017952065}{8916100448256}
- \frac{1512777571}{11943936} \zeta_5
\right. \nonumber \\
&& \left. ~~~~
- \frac{138466045}{2985984} \zeta_6
- \frac{7866851}{276480} \zeta_8
- \frac{753571}{13824} \zeta_7
- \frac{47587}{2592} \zeta_9
- \frac{4307}{2592} \zeta_3^3
+ \frac{621}{160} \zeta_{5,3}
\right. \nonumber \\
&& \left. ~~~~
+ \frac{631}{576} \zeta_4 \zeta_5
+ \frac{7105}{864} \zeta_3 \zeta_6
+ \frac{99673}{3456} \zeta_3 \zeta_5
+ \frac{981187}{82944} \zeta_3 \zeta_4
+ \frac{87305845}{2985984} \zeta_3^2
\right] \epsilon^5 ~+~ O(\epsilon^6)
\nonumber \\
\mbox{{\bf nostar5}} &=&
- \frac{1}{5184 \epsilon^4}
- \frac{53}{62208 \epsilon^3}
- \frac{497}{746496 \epsilon^2}
+ \left[
- \frac{59}{5184} \zeta_3
+ \frac{169241}{8957952}
\right] \frac{1}{\epsilon}
\nonumber \\
&&
- \frac{3847}{62208} \zeta_3
- \frac{59}{3456} \zeta_4
+ \frac{7291429}{35831808}
\nonumber \\
&&
+ \left[
- \frac{234397}{746496} \zeta_3
- \frac{3847}{41472} \zeta_4
- \frac{881}{1728} \zeta_5
+ \frac{651732331}{429981696}
\right] \epsilon
\nonumber \\
&&
+ \left[
\frac{151848323951}{15479341056}
- \frac{17417963}{8957952} \zeta_3
- \frac{234397}{497664} \zeta_4
- \frac{6217}{2304} \zeta_5
- \frac{1615}{1296} \zeta_6
+ \frac{2029}{5184} \zeta_3^2
\right] \epsilon^2
\nonumber \\
&&
+ \left[
\frac{10963043240449}{185752092672}
- \frac{468045319}{35831808} \zeta_3
- \frac{17417963}{5971968} \zeta_4
- \frac{916561}{82944} \zeta_5
- \frac{205015}{31104} \zeta_6
\right. \nonumber \\
&& \left. ~~~~
- \frac{45301}{3456} \zeta_7
+ \frac{2029}{1728} \zeta_3 \zeta_4
+ \frac{124469}{62208} \zeta_3^2
\right] \epsilon^3
\nonumber \\
&&
+ \left[
- \frac{36911324041}{429981696} \zeta_3
+ \frac{757459450377103}{2229025112064}
- \frac{468045319}{23887872} \zeta_4
- \frac{146327317}{2985984} \zeta_5
\right. \nonumber \\
&& \left. ~~~~
- \frac{10018315}{373248} \zeta_6
- \frac{2876173}{41472} \zeta_7
- \frac{2334341}{69120} \zeta_8
- \frac{223}{80} \zeta_{5,3}
+ \frac{13361}{864} \zeta_3 \zeta_5
+ \frac{124469}{20736} \zeta_3 \zeta_4
\right. \nonumber \\
&& \left. ~~~~
+ \frac{6561335}{746496} \zeta_3^2
\right] \epsilon^4
\nonumber \\
&&
+ \left[
- \frac{8331112515413}{15479341056} \zeta_3
- \frac{36911324041}{286654464} \zeta_4
- \frac{3041772121}{11943936} \zeta_5
+ \frac{50829818629660001}{26748301344768}
\right. \nonumber \\
&& \left. ~~~~
- \frac{526954985}{4478976} \zeta_6
- \frac{144123917}{829440} \zeta_8
- \frac{135196807}{497664} \zeta_7
- \frac{9290029}{31104} \zeta_9
- \frac{48359}{7776} \zeta_3^3
\right. \nonumber \\
&& \left. ~~~~
- \frac{5397}{320} \zeta_{5,3}
+ \frac{25403}{576} \zeta_4 \zeta_5
+ \frac{95135}{2592} \zeta_3 \zeta_6
+ \frac{258347}{3456} \zeta_3 \zeta_5
+ \frac{6561335}{248832} \zeta_3 \zeta_4
\right. \nonumber \\
&& \left. ~~~~
+ \frac{424883665}{8957952} \zeta_3^2
\right] \epsilon^5 ~+~ O(\epsilon^6)
\nonumber \\
\mbox{{\bf no}} &=&
- \frac{1}{36 \epsilon^2}
+ \left[
- \frac{11}{72}
+ \frac{1}{18} \zeta_3
\right] \frac{1}{\epsilon}
- \frac{631}{1296}
+ \frac{1}{6} \zeta_3
+ \frac{1}{12} \zeta_4
+ \left[
- \frac{797}{864}
- \frac{4}{9} \zeta_5
+ \frac{1}{4} \zeta_4
+ \frac{14}{81} \zeta_3
\right] \epsilon
\nonumber \\
&&
+ \left[
- \frac{215}{324} \zeta_3
- \frac{5}{4} \zeta_6
- \frac{4}{3} \zeta_5
+ \frac{7}{9} \zeta_3^2
+ \frac{7}{27} \zeta_4
+ \frac{31829}{46656}
\right] \epsilon^2
\nonumber \\
&&
+ \left[
- \frac{29345}{5832} \zeta_3
- \frac{1709}{162} \zeta_5
- \frac{215}{216} \zeta_4
- \frac{67}{6} \zeta_7
- \frac{15}{4} \zeta_6
+ \frac{7}{3} \zeta_3 \zeta_4
+ \frac{7}{3} \zeta_3^2
+ \frac{1674437}{93312}
\right] \epsilon^3
\nonumber \\
&&
+ \left[
- \frac{252601}{11664} \zeta_3
- \frac{29345}{3888} \zeta_4
- \frac{20015}{324} \zeta_5
- \frac{17989}{360} \zeta_8
- \frac{965}{36} \zeta_6
- \frac{67}{2} \zeta_7
+ \frac{36}{5} \zeta_{5,3}
\right. \nonumber \\
&& \left. ~~~~
+ \frac{121}{3} \zeta_3 \zeta_5
+ \frac{2147}{162} \zeta_3^2
+ \frac{201342089}{1679616}
+ 7 \zeta_3 \zeta_4
\right] \epsilon^4
\nonumber \\
&&
+ \left[
- \frac{16004225}{209952} \zeta_3
- \frac{1673735}{5832} \zeta_5
- \frac{252601}{7776} \zeta_4
- \frac{17989}{120} \zeta_8
- \frac{14522}{81} \zeta_9
- \frac{10609}{54} \zeta_7
\right. \nonumber \\
&& \left. ~~~~
- \frac{1375}{9} \zeta_6
- \frac{337}{27} \zeta_3^3
+ \frac{13}{2} \zeta_4 \zeta_5
+ \frac{108}{5} \zeta_{5,3}
+ \frac{1745}{18} \zeta_3 \zeta_6
+ \frac{2147}{54} \zeta_3 \zeta_4
+ \frac{21263}{324} \zeta_3^2
\right. \nonumber \\
&& \left. ~~~~
+ \frac{2095164001}{3359232}
+ 121 \zeta_3 \zeta_5
\right] \epsilon^5 ~+~ O(\epsilon^6)
\nonumber \\
\mbox{{\bf fastar2}} &=&
- \frac{1}{31104 \epsilon^4}
- \frac{7}{46656 \epsilon^3}
- \frac{4061}{11197440 \epsilon^2}
+ \left[
- \frac{25}{15552} \zeta_3
+ \frac{164153}{134369280}
\right] \frac{1}{\epsilon}
\nonumber \\
&&
- \frac{175}{23328} \zeta_3
- \frac{25}{10368} \zeta_4
+ \frac{498311}{19906560}
\nonumber \\
&&
+ \left[
- \frac{264677}{5598720} \zeta_3
- \frac{175}{15552} \zeta_4
- \frac{155}{1728} \zeta_5
+ \frac{1468119763}{6449725440}
\right] \epsilon
\nonumber \\
&&
+ \left[
\frac{382027011479}{232190115840}
- \frac{24144799}{67184640} \zeta_3
- \frac{264677}{3732480} \zeta_4
- \frac{3425}{15552} \zeta_6
- \frac{1085}{2592} \zeta_5
+ \frac{1247}{15552} \zeta_3^2
\right] \epsilon^2
\nonumber \\
&&
+ \left[
\frac{29537619574489}{2786281390080}
- \frac{24144799}{44789760} \zeta_4
- \frac{8528651}{3317760} \zeta_3
- \frac{399797}{207360} \zeta_5
- \frac{23975}{23328} \zeta_6
\right. \nonumber \\
&& \left. ~~~~
- \frac{12503}{5184} \zeta_7
+ \frac{1247}{5184} \zeta_3 \zeta_4
+ \frac{8729}{23328} \zeta_3^2
\right] \epsilon^3
\nonumber \\
&&
+ \left[
- \frac{54343033589}{3224862720} \zeta_3
+ \frac{2131076904260791}{33435376680960}
- \frac{8767813}{829440} \zeta_5
- \frac{8528651}{2211840} \zeta_4
- \frac{5264921}{1119744} \zeta_6
\right. \nonumber \\
&& \left. ~~~~
- \frac{1397189}{207360} \zeta_8
- \frac{87521}{7776} \zeta_7
- \frac{3}{10} \zeta_{5,3}
+ \frac{1087}{288} \zeta_3 \zeta_5
+ \frac{8729}{7776} \zeta_3 \zeta_4
+ \frac{10919251}{5598720} \zeta_3^2
\right] \epsilon^4
\nonumber \\
&&
+ \left[
- \frac{12001629023377}{116095057920} \zeta_3
- \frac{54343033589}{2149908480} \zeta_4
+ \frac{146881530580013753}{401224520171520}
- \frac{343024027}{13436928} \zeta_6
\right. \nonumber \\
&& \left. ~~~~
- \frac{91051111}{1866240} \zeta_7
- \frac{22782833}{368640} \zeta_5
- \frac{9780323}{311040} \zeta_8
- \frac{2619709}{46656} \zeta_9
- \frac{37081}{23328} \zeta_3^3
- \frac{7}{5} \zeta_{5,3}
\right. \nonumber \\
&& \left. ~~~~
+ \frac{1519}{192} \zeta_4 \zeta_5
+ \frac{7609}{432} \zeta_3 \zeta_5
+ \frac{70255}{7776} \zeta_3 \zeta_6
+ \frac{10919251}{1866240} \zeta_3 \zeta_4
+ \frac{819297737}{67184640} \zeta_3^2
\right] \epsilon^5
\nonumber \\
&& ~+~ O(\epsilon^6)
\nonumber \\
\mbox{{\bf t155}} &=&
- \frac{1}{15552 \epsilon^4}
- \frac{23}{93312 \epsilon^3}
- \frac{679}{5598720 \epsilon^2}
+ \left[
- \frac{31}{7776} \zeta_3
+ \frac{123281}{22394880}
\right] \frac{1}{\epsilon}
\nonumber \\
&&
- \frac{713}{46656} \zeta_3
- \frac{31}{5184} \zeta_4
+ \frac{15165583}{268738560}
\nonumber \\
&&
+ \left[
- \frac{213343}{2799360} \zeta_3
- \frac{713}{31104} \zeta_4
- \frac{449}{2592} \zeta_5
+ \frac{88785839}{214990848}
\right] \epsilon
\nonumber \\
&&
+ \left[
\frac{307092224429}{116095057920}
- \frac{5868103}{11197440} \zeta_3
- \frac{213343}{1866240} \zeta_4
- \frac{10327}{15552} \zeta_5
- \frac{1645}{3888} \zeta_6
+ \frac{983}{7776} \zeta_3^2
\right] \epsilon^2
\nonumber \\
&&
+ \left[
\frac{21968267826851}{1393140695040}
- \frac{482659289}{134369280} \zeta_3
- \frac{5868103}{7464960} \zeta_4
- \frac{2356007}{933120} \zeta_5
- \frac{37835}{23328} \zeta_6
\right. \nonumber \\
&& \left. ~~~~
- \frac{5669}{1296} \zeta_7
+ \frac{983}{2592} \zeta_3 \zeta_4
+ \frac{22609}{46656} \zeta_3^2
\right] \epsilon^3
\nonumber \\
&&
+ \left[
- \frac{2458665145}{107495424} \zeta_3
+ \frac{300549734350057}{3343537668096}
- \frac{482659289}{89579520} \zeta_4
- \frac{48004847}{3732480} \zeta_5
\right. \nonumber \\
&& \left. ~~~~
- \frac{3427339}{559872} \zeta_6
- \frac{242437}{20736} \zeta_8
- \frac{130387}{7776} \zeta_7
- \frac{3}{4} \zeta_{5,3}
+ \frac{6817}{1296} \zeta_3 \zeta_5
+ \frac{22609}{15552} \zeta_3 \zeta_4
\right. \nonumber \\
&& \left. ~~~~
+ \frac{6051689}{2799360} \zeta_3^2
\right] \epsilon^4
\nonumber \\
&&
+ \left[
- \frac{7956272660107}{58047528960} \zeta_3
- \frac{3353771761}{44789760} \zeta_5
- \frac{2458665145}{71663616} \zeta_4
+ \frac{3693436841274617}{7430083706880}
\right. \nonumber \\
&& \left. ~~~~
- \frac{69073219}{2239488} \zeta_6
- \frac{55868149}{933120} \zeta_7
- \frac{5576051}{124416} \zeta_8
- \frac{2293555}{23328} \zeta_9
- \frac{23959}{11664} \zeta_3^3
- \frac{23}{8} \zeta_{5,3}
\right. \nonumber \\
&& \left. ~~~~
+ \frac{11677}{864} \zeta_4 \zeta_5
+ \frac{24335}{1944} \zeta_3 \zeta_6
+ \frac{156791}{7776} \zeta_3 \zeta_5
+ \frac{6051689}{933120} \zeta_3 \zeta_4
+ \frac{150129569}{11197440} \zeta_3^2
\right] \epsilon^5
\nonumber \\
&&
~+~ O(\epsilon^6)
\nonumber \\
\mbox{{\bf t124}} &=&
- \frac{1}{51840 \epsilon^3}
- \frac{31}{155520 \epsilon^2}
- \frac{2339}{1866240 \epsilon}
- \frac{3667}{559872}
+ \frac{1}{1296} \zeta_3
\nonumber \\
&&
+ \left[
- \frac{2148781}{67184640}
+ \frac{1}{864} \zeta_4
+ \frac{199}{31104} \zeta_3
\right] \epsilon
\nonumber \\
&&
+ \left[
- \frac{1289797}{8398080}
+ \frac{37}{2592} \zeta_5
+ \frac{199}{20736} \zeta_4
+ \frac{70097}{1866240} \zeta_3
\right] \epsilon^2
\nonumber \\
&&
+ \left[
- \frac{447731989}{604661760}
- \frac{17}{1080} \zeta_3^2
+ \frac{175}{5184} \zeta_6
+ \frac{215}{1944} \zeta_5
+ \frac{70097}{1244160} \zeta_4
+ \frac{4467643}{22394880} \zeta_3
\right] \epsilon^3
\nonumber \\
&&
+ \left[
- \frac{104687174539}{29023764480}
- \frac{3131}{25920} \zeta_3^2
- \frac{17}{360} \zeta_3 \zeta_4
+ \frac{13157}{51840} \zeta_7
+ \frac{16205}{62208} \zeta_6
+ \frac{146803}{233280} \zeta_5
\right. \nonumber \\
&& \left. ~~~~
+ \frac{4467643}{14929920} \zeta_4
+ \frac{92262887}{89579520} \zeta_3
\right] \epsilon^4
\nonumber \\
&&
+ \left[
- \frac{1236480137413}{69657034752}
- \frac{129191}{186624} \zeta_3^2
- \frac{3131}{8640} \zeta_3 \zeta_4
- \frac{1829}{3240} \zeta_3 \zeta_5
+ \frac{3}{400} \zeta_{5,3}
+ \frac{40901}{27648} \zeta_6
\right. \nonumber \\
&& \left. ~~~~
+ \frac{123043}{172800} \zeta_8
+ \frac{1191553}{622080} \zeta_7
+ \frac{18642809}{5598720} \zeta_5
+ \frac{23369059}{4423680} \zeta_3
+ \frac{92262887}{59719680} \zeta_4
\right] \epsilon^5
\nonumber \\
&&
+ \left[
- \frac{368047815281083}{4179422085120}
+ \frac{1047221158789}{38698352640} \zeta_3
- \frac{13947499}{3732480} \zeta_3^2
- \frac{129191}{62208} \zeta_3 \zeta_4
\right. \nonumber \\
&& \left. ~~~~
- \frac{82493}{19440} \zeta_3 \zeta_5
- \frac{3901}{4320} \zeta_4 \zeta_5
- \frac{1727}{1296} \zeta_3 \zeta_6
+ \frac{89}{1600} \zeta_{5,3}
+ \frac{2161}{9720} \zeta_3^3
+ \frac{735227}{155520} \zeta_9
\right. \nonumber \\
&& \left. ~~~~
+ \frac{11097827}{2073600} \zeta_8
+ \frac{23369059}{2949120} \zeta_4
+ \frac{70103593}{8957952} \zeta_6
+ \frac{80333051}{7464960} \zeta_7
+ \frac{1165361683}{67184640} \zeta_5
\right] \epsilon^6
\nonumber \\
&&
~+~ O(\epsilon^7)
\nonumber \\
\mbox{{\bf t145}} &=&
- \frac{5}{124416 \epsilon^3}
- \frac{1873}{7464960 \epsilon^2}
- \frac{80371}{89579520 \epsilon}
- \frac{566551}{358318080}
- \frac{19}{31104} \zeta_3
\nonumber \\
&&
+ \left[
- \frac{7463}{1866240} \zeta_3
- \frac{19}{20736} \zeta_4
+ \frac{30715831}{4299816960}
\right] \epsilon
\nonumber \\
&&
+ \left[
- \frac{600293}{22394880} \zeta_3
- \frac{7463}{1244160} \zeta_4
- \frac{341}{10368} \zeta_5
+ \frac{358030103}{3439853568}
\right] \epsilon^2
\nonumber \\
&&
+ \left[
\frac{1513492357733}{1857520926720}
- \frac{17437313}{89579520} \zeta_3
- \frac{600293}{14929920} \zeta_4
- \frac{130957}{622080} \zeta_5
- \frac{1255}{15552} \zeta_6
\right. \nonumber \\
&& \left. ~~~~
+ \frac{493}{15552} \zeta_3^2
\right] \epsilon^3
\nonumber \\
&&
+ \left[
\frac{118328890166123}{22290251120640}
- \frac{1445143903}{1074954240} \zeta_3
- \frac{17437313}{59719680} \zeta_4
- \frac{8077207}{7464960} \zeta_5
- \frac{16619}{20736} \zeta_7
\right. \nonumber \\
&& \left. ~~~~
- \frac{3011}{5832} \zeta_6
+ \frac{493}{5184} \zeta_3 \zeta_4
+ \frac{189527}{933120} \zeta_3^2
\right] \epsilon^4
\nonumber \\
&&
+ \left[
- \frac{36938283067}{4299816960} \zeta_3
+ \frac{1692801725704513}{53496602689536}
- \frac{1445143903}{716636160} \zeta_4
- \frac{172335787}{29859840} \zeta_5
\right. \nonumber \\
&& \left. ~~~~
- \frac{3182639}{622080} \zeta_7
- \frac{738479}{279936} \zeta_6
- \frac{91817}{41472} \zeta_8
- \frac{3}{32} \zeta_{5,3}
+ \frac{3617}{2592} \zeta_3 \zeta_5
+ \frac{189527}{311040} \zeta_3 \zeta_4
\right. \nonumber \\
&& \left. ~~~~
+ \frac{12287453}{11197440} \zeta_3^2
\right] \epsilon^5
\nonumber \\
&&
+ \left[
- \frac{23912983818461}{464380231680} \zeta_3
- \frac{36938283067}{2866544640} \zeta_4
- \frac{11522465077}{358318080} \zeta_5
\right. \nonumber \\
&& \left. ~~~~
+ \frac{63870314319138931}{356644017930240}
- \frac{188849819}{7464960} \zeta_7
- \frac{8795191}{622080} \zeta_8
- \frac{3902891}{279936} \zeta_6
\right. \nonumber \\
&& \left. ~~~~
- \frac{3258785}{186624} \zeta_9
- \frac{6583}{11664} \zeta_3^3
- \frac{19}{32} \zeta_{5,3}
+ \frac{151}{54} \zeta_4 \zeta_5
+ \frac{25895}{7776} \zeta_3 \zeta_6
+ \frac{1386853}{155520} \zeta_3 \zeta_5
\right. \nonumber \\
&& \left. ~~~~
+ \frac{12287453}{3732480} \zeta_3 \zeta_4
+ \frac{283280153}{44789760} \zeta_3^2
\right] \epsilon^6 ~+~ O(\epsilon^7)
\nonumber \\
\mbox{{\bf t105}} &=&
- \frac{1}{648 \epsilon^3}
- \frac{17}{3888 \epsilon^2}
- \frac{121}{23328 \epsilon}
- \frac{7}{324} \zeta_3
+ \frac{3025}{139968}
+ \left[
- \frac{119}{1944} \zeta_3
- \frac{7}{216} \zeta_4
+ \frac{186539}{839808}
\right] \epsilon
\nonumber \\
&&
+ \left[
- \frac{2935}{11664} \zeta_3
- \frac{119}{1296} \zeta_4
- \frac{7}{12} \zeta_5
+ \frac{6682285}{5038848}
\right] \epsilon^2
\nonumber \\
&&
+ \left[
- \frac{97013}{69984} \zeta_3
- \frac{2935}{7776} \zeta_4
- \frac{455}{324} \zeta_6
- \frac{119}{72} \zeta_5
+ \frac{113}{324} \zeta_3^2
+ \frac{203468495}{30233088}
\right] \epsilon^3
\nonumber \\
&&
+ \left[
\frac{5727067633}{181398528}
- \frac{3107395}{419904} \zeta_3
- \frac{97013}{46656} \zeta_4
- \frac{7735}{1944} \zeta_6
- \frac{6485}{1296} \zeta_5
- \frac{245}{27} \zeta_7
+ \frac{113}{108} \zeta_3 \zeta_4
\right. \nonumber \\
&& \left. ~~~~
+ \frac{1921}{1944} \zeta_3^2
\right] \epsilon^4
\nonumber \\
&&
+ \left[
\frac{153930181235}{1088391168}
- \frac{92052137}{2519424} \zeta_3
- \frac{3107395}{279936} \zeta_4
- \frac{159719}{7776} \zeta_5
- \frac{138575}{11664} \zeta_6
- \frac{99647}{4320} \zeta_8
\right. \nonumber \\
&& \left. ~~~~
- \frac{4165}{162} \zeta_7
- \frac{6}{5} \zeta_{5,3}
+ \frac{53}{6} \zeta_3 \zeta_5
+ \frac{1921}{648} \zeta_3 \zeta_4
+ \frac{40817}{11664} \zeta_3^2
\right] \epsilon^5
\nonumber \\
&&
+ \left[
- \frac{2571366511}{15116544} \zeta_3
+ \frac{4014789658453}{6530347008}
- \frac{92052137}{1679616} \zeta_4
- \frac{4418665}{46656} \zeta_5
- \frac{3351145}{69984} \zeta_6
\right. \nonumber \\
&& \left. ~~~~
- \frac{1693999}{25920} \zeta_8
- \frac{123547}{972} \zeta_9
- \frac{35311}{486} \zeta_7
- \frac{1315}{486} \zeta_3^3
- \frac{17}{5} \zeta_{5,3}
+ \frac{89}{4} \zeta_4 \zeta_5
+ \frac{901}{36} \zeta_3 \zeta_5
\right. \nonumber \\
&& \left. ~~~~
+ \frac{3295}{162} \zeta_3 \zeta_6
+ \frac{40817}{3888} \zeta_3 \zeta_4
+ \frac{1194619}{69984} \zeta_3^2
\right] \epsilon^6 ~+~ O(\epsilon^7) ~.
\end{eqnarray}
Here $\zeta_{5,3}$ denotes the multiple zeta that was discovered in \cite{74}.
In \cite{17} the $\epsilon$ expansion for the set of masters that was required 
to renormalize $\phi^3$ to four loops in six dimensions was provided but in the
same master basis as \cite{37}. We have checked that those masters derived from
{\sc Forcer} in this paper are in agreement up to the order in $\epsilon$ given
in \cite{17}. To assist with this comparison we note the mapping of the 
overlapping masters in each basis is {\bf haha}~$\leftrightarrow$~$M_{61}$, 
{\bf no1}~$\leftrightarrow$~$M_{63}$, {\bf no2}~$\leftrightarrow$~$M_{62}$, 
{\bf no6}~$\leftrightarrow$~$M_{51}$, {\bf cross}~$\leftrightarrow$~$M_{36}$, 
{\bf nono}~$\leftrightarrow$~$M_{45}$ and {\bf bebe}~$\leftrightarrow$~$M_{35}$.

\sect{$O(N)$ $\phi^3$ $\beta$-functions.}

In this appendix we record the two $\beta$-functions in $O(N)$ $\phi^3$ theory 
for all $N$ in the $\MOMts$ scheme. We have
\begin{eqnarray}
\beta^{O(N)}_{1\,\MOMtss}(g_1,g_2) &=&
\left[
- \frac{1}{24} g_1 g_2^2
- \frac{1}{24} N g_1^3
+ \frac{1}{2} g_1^2 g_2
+ \frac{1}{3} g_1^3
\right]
\nonumber \\
&&
+ \left[
- \frac{163}{216} g_1^3 g_2^2
- \frac{67}{108} g_1^5
- \frac{43}{432} N g_1^5
- \frac{13}{36} g_1^4 g_2
- \frac{11}{864} N g_1^3 g_2^2
- \frac{1}{12} g_1^2 g_2^3
\right. \nonumber \\
&& \left. ~~~~
+ \frac{11}{72} N g_1^4 g_2
+ \frac{37}{864} g_1 g_2^4
\right]
\nonumber \\
&&
+ \left[
- \frac{13217}{31104} N g_1^5 g_2^2
- \frac{4747}{31104} N^2 g_1^7
- \frac{4435}{62208} g_1 g_2^6
- \frac{233}{1296} N g_1^4 g_2^3
- \frac{91}{20736} N^2 g_1^5 g_2^2
\right. \nonumber \\
&& \left. ~~~~
- \frac{3}{2} \zeta_3 N g_1^6 g_2
- \frac{1}{2} \zeta_3 g_1^7
- \frac{1}{6} \zeta_3 g_1^3 g_2^4
- \frac{1}{6} \zeta_3 g_1^5 g_2^2
- \frac{1}{48} \zeta_3 g_1 g_2^6
+ \frac{1}{2} \zeta_3 g_1^4 g_2^3
\right. \nonumber \\
&& \left. ~~~~
+ \frac{9}{8} \zeta_3 N g_1^7
+ \frac{11}{48} \zeta_3 N g_1^5 g_2^2
+ \frac{101}{1728} N^2 g_1^6 g_2
+ \frac{122}{81} g_1^6 g_2
+ \frac{169}{7776} N g_1^3 g_2^4
\right. \nonumber \\
&& \left. ~~~~
+ \frac{251}{15552} N g_1^7
+ \frac{257}{162} N g_1^6 g_2
+ \frac{559}{1728} g_1^2 g_2^5
+ \frac{749}{864} g_1^3 g_2^4
+ \frac{3121}{2592} g_1^4 g_2^3
\right. \nonumber \\
&& \left. ~~~~
+ \frac{10657}{3888} g_1^7
+ \frac{10985}{3888} g_1^5 g_2^2
\right]
\nonumber \\
&&
+ \left[
- \frac{4216621}{559872} g_1^5 g_2^4
- \frac{2553617}{559872} N g_1^7 g_2^2
- \frac{1951993}{93312} g_1^6 g_2^3
- \frac{1717723}{139968} g_1^7 g_2^2
\right. \nonumber \\
&& \left. ~~~~
- \frac{1279273}{279936} g_1^3 g_2^6
- \frac{907375}{186624} N g_1^6 g_2^3
- \frac{676499}{46656} N g_1^8 g_2
- \frac{383461}{1119744} N^2 g_1^7 g_2^2
\right. \nonumber \\
&& \left. ~~~~
- \frac{229721}{15552} g_1^8 g_2
- \frac{111733}{69984} N^2 g_1^9
- \frac{43283}{2592} g_1^9
- \frac{38011}{31104} g_1^2 g_2^7
\right. \nonumber \\
&& \left. ~~~~
- \frac{26969}{559872} N g_1^3 g_2^6
- \frac{20323}{1296} \zeta_3 N g_1^9
- \frac{9953}{1296} \zeta_3 g_1^5 g_2^4
- \frac{6761}{93312} N^2 g_1^6 g_2^3
\right. \nonumber \\
&& \left. ~~~~
- \frac{5189}{93312} N^3 g_1^9
- \frac{4045}{432} \zeta_3 N g_1^8 g_2
- \frac{1295}{108} \zeta_3 g_1^6 g_2^3
- \frac{749}{432} \zeta_3 g_1^3 g_2^6
\right. \nonumber \\
&& \left. ~~~~
- \frac{677}{373248} N^3 g_1^7 g_2^2
- \frac{535}{432} \zeta_3 g_1^4 g_2^5
- \frac{341}{72} \zeta_3 g_1^8 g_2
- \frac{209}{72} \zeta_3 N g_1^6 g_2^3
\right. \nonumber \\
&& \left. ~~~~
- \frac{155}{18} \zeta_5 N g_1^7 g_2^2
- \frac{119}{1728} \zeta_3 N g_1^4 g_2^5
- \frac{115}{8} \zeta_3 g_1^7 g_2^2
- \frac{25}{18} \zeta_5 N g_1^5 g_2^4
- \frac{25}{36} \zeta_5 N^2 g_1^9
\right. \nonumber \\
&& \left. ~~~~
- \frac{23}{648} \zeta_3 N g_1^3 g_2^6
- \frac{14}{27} \zeta_3 N^2 g_1^8 g_2
- \frac{5}{2} \zeta_5 g_1^4 g_2^5
- \frac{1}{48} \zeta_3 N^2 g_1^6 g_2^3
- \frac{1}{1728} \zeta_3 N^3 g_1^8 g_2
\right. \nonumber \\
&& \left. ~~~~
+ \frac{1}{48} \zeta_3 g_1^2 g_2^7
+ \frac{1}{5184} \zeta_3 N^3 g_1^7 g_2^2
+ \frac{5}{36} \zeta_5 g_1 g_2^8
+ \frac{5}{576} \zeta_3 N^3 g_1^9
+ \frac{7}{1296} \zeta_3 N^2 g_1^5 g_2^4
\right. \nonumber \\
&& \left. ~~~~
+ \frac{25}{9} \zeta_5 g_1^9
+ \frac{35}{3} \zeta_5 N g_1^9
+ \frac{40}{3} \zeta_5 g_1^7 g_2^2
+ \frac{47}{5184} \zeta_3 N^2 g_1^7 g_2^2
+ \frac{65}{3} \zeta_5 N g_1^8 g_2
\right. \nonumber \\
&& \left. ~~~~
+ \frac{65}{6} \zeta_5 g_1^8 g_2
+ \frac{70}{3} \zeta_5 g_1^6 g_2^3
+ \frac{85}{18} \zeta_5 g_1^3 g_2^6
+ \frac{131}{62208} g_1^4 g_2^5
+ \frac{145}{18} \zeta_5 g_1^5 g_2^4
\right. \nonumber \\
&& \left. ~~~~
+ \frac{283}{2592} \zeta_3 g_1 g_2^8
+ \frac{631}{216} \zeta_3 g_1^9
+ \frac{889}{144} \zeta_3 N g_1^7 g_2^2
+ \frac{967}{93312} N^2 g_1^5 g_2^4
+ \frac{1213}{62208} N^3 g_1^8 g_2
\right. \nonumber \\
&& \left. ~~~~
+ \frac{3349}{2592} \zeta_3 N^2 g_1^9
+ \frac{5423}{5184} \zeta_3 N g_1^5 g_2^4
+ \frac{56263}{23328} N^2 g_1^8 g_2
+ \frac{99455}{69984} N g_1^9
\right. \nonumber \\
&& \left. ~~~~
+ \frac{100885}{186624} N g_1^4 g_2^5
+ \frac{135149}{1119744} g_1 g_2^8
+ \frac{818105}{1119744} N g_1^5 g_2^4
+ 10 \zeta_5 N g_1^6 g_2^3
\right]
\nonumber \\
&&
+ \left[
\frac{4787589055}{20155392} N g_1^9 g_2^2
- \frac{44097997}{4478976} N^2 g_1^8 g_2^3
- \frac{43577579}{13436928} g_1^4 g_2^7
\right. \nonumber \\
&& \left. ~~~~
- \frac{23131849}{8957952} N^3 g_1^{11}
- \frac{19776671}{6718464} N g_1^{11}
- \frac{18870089}{161243136} N g_1^3 g_2^8
\right. \nonumber \\
&& \left. ~~~~
- \frac{17762039}{3359232} N^2 g_1^{10} g_2
- \frac{9924857}{2519424} N g_1^5 g_2^6
- \frac{6134743}{2239488} N g_1^4 g_2^7
\right. \nonumber \\
&& \left. ~~~~
- \frac{1465675}{124416} \zeta_3 N g_1^5 g_2^6
- \frac{818735}{2592} \zeta_5 N g_1^9 g_2^2
- \frac{626087}{31104} \zeta_3 g_1^4 g_2^7
\right. \nonumber \\
&& \left. ~~~~
- \frac{475265}{10368} \zeta_5 N^2 g_1^9 g_2^2
- \frac{431969}{1119744} N^2 g_1^7 g_2^4
- \frac{362675}{20736} \zeta_5 N g_1^7 g_2^4
\right. \nonumber \\
&& \left. ~~~~
- \frac{310787}{3359232} N^3 g_1^9 g_2^2
- \frac{254395}{1728} \zeta_5 g_1^7 g_2^4
- \frac{251831}{6718464} N^4 g_1^{11}
- \frac{151631}{15552} \zeta_3 N^2 g_1^{11}
\right. \nonumber \\
&& \left. ~~~~
- \frac{119303}{373248} \zeta_3 g_1 g_2^{10}
- \frac{46865}{384} \zeta_7 N g_1^9 g_2^2
- \frac{42193}{746496} N^3 g_1^8 g_2^3
- \frac{39625}{3456} \zeta_3 N^2 g_1^8 g_2^3
\right. \nonumber \\
&& \left. ~~~~
- \frac{37885}{1296} \zeta_5 N g_1^{11}
- \frac{34315}{864} \zeta_5 N g_1^6 g_2^5
- \frac{30863}{96} \zeta_7 g_1^9 g_2^2
- \frac{27905}{864} \zeta_3 g_1^{10} g_2
\right. \nonumber \\
&& \left. ~~~~
- \frac{27461}{32} \zeta_7 g_1^8 g_2^3
- \frac{23915}{26873856} N^4 g_1^9 g_2^2
- \frac{23303}{96} \zeta_7 g_1^{10} g_2
- \frac{22055}{432} \zeta_5 N g_1^{10} g_2
\right. \nonumber \\
&& \left. ~~~~
- \frac{19439}{48} \zeta_7 N g_1^{10} g_2
- \frac{17101}{96} \zeta_7 g_1^6 g_2^5
- \frac{14935}{1296} \zeta_5 g_1^9 g_2^2
- \frac{13975}{20736} \zeta_5 g_1 g_2^{10}
\right. \nonumber \\
&& \left. ~~~~
- \frac{9605}{2592} \zeta_5 N^2 g_1^7 g_2^4
- \frac{8395}{576} \zeta_5 N^2 g_1^{11}
- \frac{7259}{32} \zeta_7 N g_1^{11}
- \frac{7231}{32} \zeta_7 g_1^{11}
\right. \nonumber \\
&& \left. ~~~~
- \frac{5593}{384} \zeta_7 N g_1^7 g_2^4
- \frac{5453}{32} \zeta_7 g_1^5 g_2^6
- \frac{5095}{15552} \zeta_3 N^3 g_1^{10} g_2
- \frac{4137}{16} \zeta_7 N g_1^8 g_2^3
\right. \nonumber \\
&& \left. ~~~~
- \frac{3045}{16} \zeta_7 g_1^7 g_2^4
- \frac{1535}{144} \zeta_5 g_1^3 g_2^8
- \frac{1361}{144} \zeta_3^2 N g_1^7 g_2^4
- \frac{1323}{16} \zeta_7 N g_1^6 g_2^5
\right. \nonumber \\
&& \left. ~~~~
- \frac{995}{15552} \zeta_3 N^2 g_1^5 g_2^6
- \frac{575}{20736} \zeta_5 N^2 g_1^5 g_2^6
- \frac{441}{8} \zeta_7 N^2 g_1^{10} g_2
- \frac{395}{576} \zeta_5 N g_1^4 g_2^7
\right. \nonumber \\
&& \left. ~~~~
- \frac{283}{10368} \zeta_3 N^3 g_1^8 g_2^3
- \frac{245}{36} \zeta_3^2 N^2 g_1^{11}
- \frac{215}{144} \zeta_5 g_1^2 g_2^9
- \frac{182}{3} \zeta_7 g_1^3 g_2^8
\right. \nonumber \\
&& \left. ~~~~
- \frac{147}{128} \zeta_7 g_1 g_2^{10}
- \frac{109}{12} \zeta_3^2 g_1^{10} g_2
- \frac{83}{48} \zeta_3^2 N g_1^5 g_2^6
- \frac{55}{64} \zeta_5 N^3 g_1^{10} g_2
\right. \nonumber \\
&& \left. ~~~~
- \frac{25}{31104} \zeta_3 N^4 g_1^{10} g_2
- \frac{3}{2} \zeta_3^2 N^2 g_1^8 g_2^3
- \frac{1}{48} \zeta_3^2 N g_1^3 g_2^8
+ \frac{1}{12} \zeta_3^2 N g_1^4 g_2^7
\right. \nonumber \\
&& \left. ~~~~
+ \frac{1}{24} \zeta_3^2 g_1 g_2^{10}
+ \frac{7}{144} \zeta_3^2 N^3 g_1^{11}
+ \frac{11}{12} \zeta_3^2 N^2 g_1^{10} g_2
+ \frac{13}{24} \zeta_3^2 g_1^4 g_2^7
+ \frac{13}{72} \zeta_3^2 N^2 g_1^7 g_2^4
\right. \nonumber \\
&& \left. ~~~~
+ \frac{41}{4} \zeta_3^2 N g_1^6 g_2^5
+ \frac{47}{12} \zeta_3^2 g_1^6 g_2^5
+ \frac{55}{373248} \zeta_3 N^4 g_1^9 g_2^2
+ \frac{71}{8} \zeta_3^2 g_1^5 g_2^6
+ \frac{145}{36} \zeta_3^2 g_1^3 g_2^8
\right. \nonumber \\
&& \left. ~~~~
+ \frac{151}{41472} \zeta_3 N^4 g_1^{11}
+ \frac{185}{72} \zeta_3^2 N^2 g_1^9 g_2^2
+ \frac{245}{24} \zeta_3^2 g_1^9 g_2^2
+ \frac{295}{256} \zeta_5 N^3 g_1^{11}
+ \frac{343}{18} \zeta_3^2 g_1^{11}
\right. \nonumber \\
&& \left. ~~~~
+ \frac{385}{7776} \zeta_3 N^2 g_1^6 g_2^5
+ \frac{435}{64} \zeta_5 N g_1^5 g_2^6
+ \frac{481}{221184} N^2 g_1^5 g_2^6
+ \frac{485}{12} \zeta_3^2 g_1^7 g_2^4
\right. \nonumber \\
&& \left. ~~~~
+ \frac{515}{5184} \zeta_5 N^3 g_1^9 g_2^2
+ \frac{575}{12} \zeta_3^2 N g_1^{10} g_2
+ \frac{635}{1728} \zeta_5 N^2 g_1^6 g_2^5
+ \frac{683}{186624} \zeta_3 N^3 g_1^7 g_2^4
\right. \nonumber \\
&& \left. ~~~~
+ \frac{1093}{41472} \zeta_3 N^3 g_1^9 g_2^2
+ \frac{1247}{24} \zeta_3^2 g_1^8 g_2^3
+ \frac{1421}{128} \zeta_7 N g_1^5 g_2^6
+ \frac{1477}{96} \zeta_7 g_1^4 g_2^7
\right. \nonumber \\
&& \left. ~~~~
+ \frac{1505}{5184} \zeta_5 N g_1^3 g_2^8
+ \frac{1715}{128} \zeta_7 N^2 g_1^{11}
+ \frac{2009}{128} \zeta_7 N^2 g_1^9 g_2^2
+ \frac{3143}{144} \zeta_3^2 N g_1^{11}
\right. \nonumber \\
&& \left. ~~~~
+ \frac{3671}{10368} \zeta_3 N g_1^4 g_2^7
+ \frac{3709}{144} \zeta_3^2 N g_1^9 g_2^2
+ \frac{4805}{108} \zeta_5 N^2 g_1^{10} g_2
+ \frac{7391}{10368} \zeta_3 g_1^2 g_2^9
\right. \nonumber \\
&& \left. ~~~~
+ \frac{9995}{432} \zeta_5 g_1^4 g_2^7
+ \frac{13667}{46656} \zeta_3 N g_1^3 g_2^8
+ \frac{19549}{7776} \zeta_3 N^2 g_1^{10} g_2
+ \frac{21775}{108} \zeta_5 g_1^{10} g_2
\right. \nonumber \\
&& \left. ~~~~
+ \frac{24413}{2239488} N^4 g_1^{10} g_2
+ \frac{26855}{288} \zeta_5 g_1^8 g_2^3
+ \frac{37405}{432} \zeta_5 g_1^6 g_2^5
+ \frac{37705}{1728} \zeta_5 N^2 g_1^8 g_2^3
\right. \nonumber \\
&& \left. ~~~~
+ \frac{60365}{2592} \zeta_5 g_1^5 g_2^6
+ \frac{72995}{1296} \zeta_5 g_1^{11}
+ \frac{106975}{93312} \zeta_3 N^3 g_1^{11}
+ \frac{159395}{864} \zeta_5 N g_1^8 g_2^3
\right. \nonumber \\
&& \left. ~~~~
+ \frac{197227}{53747712} N^3 g_1^7 g_2^4
+ \frac{262375}{3888} \zeta_3 g_1^{11}
+ \frac{684313}{3456} \zeta_3 N g_1^9 g_2^2
+ \frac{704323}{5832} \zeta_3 g_1^7 g_2^4
\right. \nonumber \\
&& \left. ~~~~
+ \frac{740725}{15552} \zeta_3 N g_1^6 g_2^5
+ \frac{782275}{15552} \zeta_3 N g_1^8 g_2^3
+ \frac{843907}{23328} \zeta_3 g_1^3 g_2^8
\right. \nonumber \\
&& \left. ~~~~
+ \frac{1002293}{373248} \zeta_3 N^2 g_1^7 g_2^4
+ \frac{1132363}{15552} \zeta_3 g_1^6 g_2^5
+ \frac{1723973}{93312} \zeta_3 N^2 g_1^9 g_2^2
\right. \nonumber \\
&& \left. ~~~~
+ \frac{3392747}{1679616} N^3 g_1^{10} g_2
+ \frac{4175995}{15552} \zeta_3 N g_1^{10} g_2
+ \frac{4781087}{186624} \zeta_3 N g_1^7 g_2^4
\right. \nonumber \\
&& \left. ~~~~
+ \frac{5235413}{23328} \zeta_3 N g_1^{11}
+ \frac{5594599}{46656} \zeta_3 g_1^5 g_2^6
+ \frac{5964659}{23328} \zeta_3 g_1^9 g_2^2
\right. \nonumber \\
&& \left. ~~~~
+ \frac{6586807}{13436928} N^2 g_1^6 g_2^5
+ \frac{8422631}{15552} \zeta_3 g_1^8 g_2^3
+ \frac{31673845}{4478976} g_1^2 g_2^9
\right. \nonumber \\
&& \left. ~~~~
+ \frac{38661817}{161243136} g_1 g_2^{10}
+ \frac{86076149}{559872} g_1^{10} g_2
+ \frac{107275895}{6718464} N g_1^8 g_2^3
\right. \nonumber \\
&& \left. ~~~~
+ \frac{107949341}{20155392} N^2 g_1^9 g_2^2
+ \frac{119287091}{2239488} N g_1^6 g_2^5
+ \frac{124107581}{6718464} g_1^5 g_2^6
\right. \nonumber \\
&& \left. ~~~~
+ \frac{130062553}{5038848} N^2 g_1^{11}
+ \frac{229662773}{8957952} N g_1^7 g_2^4
+ \frac{324365779}{6718464} g_1^6 g_2^5
\right. \nonumber \\
&& \left. ~~~~
+ \frac{373459423}{3359232} N g_1^{10} g_2
+ \frac{428987345}{3359232} g_1^9 g_2^2
+ \frac{447498265}{3359232} g_1^8 g_2^3
\right. \nonumber \\
&& \left. ~~~~
+ \frac{551172643}{2519424} g_1^7 g_2^4
+ \frac{602892577}{20155392} g_1^3 g_2^8
+ \frac{608352707}{5038848} g_1^{11}
+ 12 \zeta_3^2 N g_1^8 g_2^3
\right]
\nonumber \\
&& 
+~ O(g_i^{13})
\label{beton1til}
\end{eqnarray}
and
\begin{eqnarray}
\beta^{O(N)}_{2\,\MOMtss}(g_1,g_2) &=&
\left[
- \frac{1}{8} N g_1^2 g_2
+ \frac{1}{2} N g_1^3
+ \frac{3}{8} g_2^3
\right]
\nonumber \\
&&
+ \left[
- \frac{149}{144} N g_1^4 g_2
- \frac{125}{288} g_2^5
- \frac{1}{4} N g_1^5
- \frac{1}{24} N g_1^3 g_2^2
+ \frac{7}{288} N g_1^2 g_2^3
\right]
\nonumber \\
&&
+ \left[
- \frac{5063}{10368} N^2 g_1^6 g_2
- \frac{361}{1296} N g_1^2 g_2^5
- \frac{25}{48} N g_1^7
- \frac{17}{8} \zeta_3 N g_1^6 g_2
- \frac{17}{288} N^2 g_1^5 g_2^2
\right. \nonumber \\
&& \left. ~~~~
- \frac{7}{4} \zeta_3 N g_1^3 g_2^4
- \frac{3}{4} \zeta_3 N^2 g_1^7
- \frac{1}{16} \zeta_3 g_2^7
+ \frac{1}{4} \zeta_3 N g_1^2 g_2^5
+ \frac{1}{4} \zeta_3 N^2 g_1^6 g_2
+ \frac{3}{2} \zeta_3 N g_1^7
\right. \nonumber \\
&& \left. ~~~~
+ \frac{35}{16} \zeta_3 N g_1^4 g_2^3
+ \frac{65}{72} N^2 g_1^7
+ \frac{173}{6912} N^2 g_1^4 g_2^3
+ \frac{889}{288} N g_1^5 g_2^2
+ \frac{941}{864} N g_1^3 g_2^4
\right. \nonumber \\
&& \left. ~~~~
+ \frac{1171}{1152} N g_1^4 g_2^3
+ \frac{22051}{5184} N g_1^6 g_2
+ \frac{26741}{20736} g_2^7
\right]
\nonumber \\
&&
+ \left[
- \frac{3333025}{373248} N g_1^4 g_2^5
- \frac{2304049}{373248} g_2^9
- \frac{726439}{46656} N g_1^8 g_2
- \frac{690361}{62208} N g_1^6 g_2^3
\right. \nonumber \\
&& \left. ~~~~
- \frac{473453}{62208} N g_1^3 g_2^6
- \frac{384881}{10368} N g_1^7 g_2^2
- \frac{274939}{62208} N^2 g_1^8 g_2
- \frac{245249}{373248} N^2 g_1^6 g_2^3
\right. \nonumber \\
&& \left. ~~~~
- \frac{176951}{20736} N g_1^5 g_2^4
- \frac{16649}{1728} N^2 g_1^9
- \frac{11933}{1728} \zeta_3 N g_1^4 g_2^5
- \frac{11291}{62208} N^2 g_1^7 g_2^2
\right. \nonumber \\
&& \left. ~~~~
- \frac{10987}{62208} N^2 g_1^4 g_2^5
- \frac{5429}{20736} N^3 g_1^8 g_2
- \frac{4501}{432} \zeta_3 N g_1^7 g_2^2
- \frac{3173}{864} \zeta_3 g_2^9
\right. \nonumber \\
&& \left. ~~~~
- \frac{3023}{144} \zeta_3 N g_1^8 g_2
- \frac{1439}{216} \zeta_3 N g_1^6 g_2^3
- \frac{1091}{216} \zeta_3 N g_1^9
- \frac{773}{192} \zeta_3 N g_1^3 g_2^6
\right. \nonumber \\
&& \left. ~~~~
- \frac{593}{48} \zeta_3 N^2 g_1^7 g_2^2
- \frac{167}{20736} N^3 g_1^7 g_2^2
- \frac{65}{12} \zeta_5 N^2 g_1^8 g_2
- \frac{25}{6} \zeta_5 N^2 g_1^6 g_2^3
\right. \nonumber \\
&& \left. ~~~~
- \frac{25}{36} \zeta_3 N^3 g_1^9
- \frac{17}{6} \zeta_3 N g_1^5 g_2^4
- \frac{15}{2} \zeta_5 N g_1^9
- \frac{5}{2} \zeta_5 N g_1^2 g_2^7
- \frac{1}{16} \zeta_3 N^2 g_1^5 g_2^4
\right. \nonumber \\
&& \left. ~~~~
- \frac{1}{192} \zeta_3 N^3 g_1^6 g_2^3
+ \frac{1}{192} \zeta_3 N^3 g_1^7 g_2^2
+ \frac{5}{2} \zeta_5 N g_1^4 g_2^5
+ \frac{5}{2} \zeta_5 N g_1^5 g_2^4
+ \frac{5}{3} \zeta_5 N^2 g_1^9
\right. \nonumber \\
&& \left. ~~~~
+ \frac{17}{432} \zeta_3 N^2 g_1^4 g_2^5
+ \frac{35}{6} \zeta_5 g_2^9
+ \frac{49}{192} \zeta_3 N^3 g_1^8 g_2
+ \frac{50}{3} \zeta_5 N g_1^3 g_2^6
+ \frac{115}{4} \zeta_5 N g_1^8 g_2
\right. \nonumber \\
&& \left. ~~~~
+ \frac{187}{432} \zeta_3 N g_1^2 g_2^7
+ \frac{673}{124416} N^3 g_1^6 g_2^3
+ \frac{997}{216} \zeta_3 N^2 g_1^9
+ \frac{2713}{1728} \zeta_3 N^2 g_1^6 g_2^3
\right. \nonumber \\
&& \left. ~~~~
+ \frac{6913}{864} \zeta_3 N^2 g_1^8 g_2
+ \frac{7663}{10368} N^3 g_1^9
+ \frac{34279}{31104} N^2 g_1^5 g_2^4
+ \frac{40835}{5184} N g_1^9
\right. \nonumber \\
&& \left. ~~~~
+ \frac{175907}{93312} N g_1^2 g_2^7
+ 5 \zeta_5 N g_1^6 g_2^3
+ 20 \zeta_5 N^2 g_1^7 g_2^2
+ 30 \zeta_5 N g_1^7 g_2^2
\right]
\nonumber \\
&&
+ \left[
\frac{2190456157}{53747712} g_2^{11}
- \frac{934040489}{53747712} N g_1^2 g_2^9
- \frac{125156905}{26873856} N^3 g_1^{10} g_2
\right. \nonumber \\
&& \left. ~~~~
- \frac{21280913}{1119744} N^2 g_1^5 g_2^6
- \frac{20862691}{746496} N^2 g_1^9 g_2^2
- \frac{2572613}{17915904} N^3 g_1^6 g_2^5
\right. \nonumber \\
&& \left. ~~~~
- \frac{1663367}{559872} N^3 g_1^8 g_2^3
- \frac{1233275}{15552} N g_1^{11}
- \frac{1099369}{31104} \zeta_3 N^2 g_1^8 g_2^3
\right. \nonumber \\
&& \left. ~~~~
- \frac{710845}{2592} \zeta_5 N g_1^8 g_2^3
- \frac{704519}{10368} \zeta_3 N g_1^5 g_2^6
- \frac{577133}{31104} \zeta_3 N g_1^2 g_2^9
\right. \nonumber \\
&& \left. ~~~~
- \frac{511715}{10368} \zeta_5 N^2 g_1^8 g_2^3
- \frac{486025}{20736} \zeta_5 N g_1^6 g_2^5
- \frac{364729}{2239488} N^4 g_1^{10} g_2
\right. \nonumber \\
&& \left. ~~~~
- \frac{333119}{15552} \zeta_3 N^2 g_1^{11}
- \frac{235985}{6912} \zeta_5 N^3 g_1^{10} g_2
- \frac{230725}{7776} \zeta_3 N g_1^{11}
\right. \nonumber \\
&& \left. ~~~~
- \frac{219565}{10368} \zeta_3 N^2 g_1^5 g_2^6
- \frac{115265}{1296} \zeta_5 N^2 g_1^6 g_2^5
- \frac{115045}{1728} \zeta_5 N g_1^3 g_2^8
\right. \nonumber \\
&& \left. ~~~~
- \frac{100565}{41472} \zeta_3 N^3 g_1^8 g_2^3
- \frac{71675}{10368} \zeta_3 N^3 g_1^9 g_2^2
- \frac{50585}{288} \zeta_5 N g_1^9 g_2^2
\right. \nonumber \\
&& \left. ~~~~
- \frac{47375}{432} \zeta_5 N^2 g_1^{10} g_2
- \frac{38829}{128} \zeta_7 N g_1^6 g_2^5
- \frac{35805}{128} \zeta_7 N g_1^8 g_2^3
- \frac{26453}{128} \zeta_7 N g_1^4 g_2^7
\right. \nonumber \\
&& \left. ~~~~
- \frac{25775}{6912} \zeta_5 N^2 g_1^4 g_2^7
- \frac{24521}{32} \zeta_7 N g_1^9 g_2^2
- \frac{18985}{432} \zeta_5 N^2 g_1^7 g_2^4
- \frac{16135}{128} \zeta_7 N^2 g_1^{10} g_2
\right. \nonumber \\
&& \left. ~~~~
- \frac{13433}{128} \zeta_7 g_2^{11}
- \frac{12817}{32} \zeta_7 N g_1^{10} g_2
- \frac{12755}{3072} N^3 g_1^{11}
- \frac{9261}{32} \zeta_7 N^2 g_1^7 g_2^4
\right. \nonumber \\
&& \left. ~~~~
- \frac{5733}{32} \zeta_7 N g_1^3 g_2^8
- \frac{2583}{8} \zeta_7 N^2 g_1^9 g_2^2
- \frac{2377}{186624} N^4 g_1^9 g_2^2
- \frac{2023}{4} \zeta_7 N g_1^7 g_2^4
\right. \nonumber \\
&& \left. ~~~~
- \frac{1975}{864} \zeta_5 N^3 g_1^7 g_2^4
- \frac{1545}{64} \zeta_5 N g_1^4 g_2^7
- \frac{931}{32} \zeta_7 N g_1^5 g_2^6
- \frac{595}{54} \zeta_5 N^2 g_1^{11}
\right. \nonumber \\
&& \left. ~~~~
- \frac{441}{32} \zeta_7 N^3 g_1^{11}
- \frac{395}{864} \zeta_3 N^4 g_1^{11}
- \frac{145}{2592} \zeta_5 N^4 g_1^{10} g_2
- \frac{125}{124416} \zeta_3 N^4 g_1^8 g_2^3
\right. \nonumber \\
&& \left. ~~~~
- \frac{121}{48} \zeta_3^2 N g_1^2 g_2^9
- \frac{91}{4} \zeta_7 N^2 g_1^{11}
- \frac{25}{12} \zeta_3^2 N g_1^7 g_2^4
- \frac{19}{18} \zeta_3^2 N^2 g_1^8 g_2^3
\right. \nonumber \\
&& \left. ~~~~
- \frac{17}{144} \zeta_3^2 N^3 g_1^{10} g_2
- \frac{15}{4} \zeta_3^2 N^2 g_1^5 g_2^6
- \frac{11}{12} \zeta_3^2 N^2 g_1^{11}
- \frac{7}{2} \zeta_3^2 N g_1^{11}
- \frac{5}{3} \zeta_3^2 N^3 g_1^9 g_2^2
\right. \nonumber \\
&& \left. ~~~~
+ \frac{1}{3} \zeta_3^2 N^3 g_1^8 g_2^3
+ \frac{5}{12} \zeta_3^2 N^2 g_1^4 g_2^7
+ \frac{13}{18} \zeta_3^2 N^2 g_1^{10} g_2
+ \frac{29}{4} \zeta_3^2 g_2^{11}
\right. \nonumber \\
&& \left. ~~~~
+ \frac{35}{10368} \zeta_3 N^4 g_1^9 g_2^2
+ \frac{55}{3} \zeta_3^2 N^2 g_1^9 g_2^2
+ \frac{85}{648} \zeta_5 N^3 g_1^6 g_2^5
+ \frac{89}{24} \zeta_3^2 N^3 g_1^{11}
\right. \nonumber \\
&& \left. ~~~~
+ \frac{95}{8} \zeta_3^2 N g_1^3 g_2^8
+ \frac{125}{8} \zeta_3^2 N g_1^5 g_2^6
+ \frac{147}{16} \zeta_7 N^3 g_1^{10} g_2
+ \frac{287}{72} \zeta_3^2 N^2 g_1^6 g_2^5
\right. \nonumber \\
&& \left. ~~~~
+ \frac{415}{1728} \zeta_5 N^4 g_1^{11}
+ \frac{441}{8} \zeta_7 N^2 g_1^6 g_2^5
+ \frac{441}{16} \zeta_7 N g_1^2 g_2^9
+ \frac{515}{8} \zeta_3^2 N g_1^9 g_2^2
\right. \nonumber \\
&& \left. ~~~~
+ \frac{705}{8} \zeta_5 N g_1^{11}
+ \frac{811}{144} \zeta_3^2 N g_1^6 g_2^5
+ \frac{1195}{576} \zeta_5 N^3 g_1^9 g_2^2
+ \frac{1235}{432} \zeta_3 N^3 g_1^{11}
\right. \nonumber \\
&& \left. ~~~~
+ \frac{1883}{10368} \zeta_3 N^3 g_1^7 g_2^4
+ \frac{2051}{32} \zeta_7 N g_1^{11}
+ \frac{2317}{144} \zeta_3^2 N g_1^4 g_2^7
+ \frac{2731}{144} \zeta_3^2 N g_1^{10} g_2
\right. \nonumber \\
&& \left. ~~~~
+ \frac{2855}{62208} \zeta_3 N^3 g_1^6 g_2^5
+ \frac{3277}{2592} \zeta_3 N^2 g_1^4 g_2^7
+ \frac{4627}{128} \zeta_7 N^2 g_1^8 g_2^3
+ \frac{4693}{41472} \zeta_3 N^4 g_1^{10} g_2
\right. \nonumber \\
&& \left. ~~~~
+ \frac{7045}{36} \zeta_5 N^2 g_1^9 g_2^2
+ \frac{9403}{144} \zeta_3^2 N g_1^8 g_2^3
+ \frac{11465}{6912} \zeta_5 g_2^{11}
+ \frac{15529}{31104} N^4 g_1^{11}
\right. \nonumber \\
&& \left. ~~~~
+ \frac{16865}{1728} \zeta_5 N^3 g_1^{11}
+ \frac{40805}{864} \zeta_5 N g_1^5 g_2^6
+ \frac{44449}{8957952} N^4 g_1^8 g_2^3
+ \frac{48415}{5184} \zeta_5 N^3 g_1^8 g_2^3
\right. \nonumber \\
&& \left. ~~~~
+ \frac{70135}{1728} \zeta_5 N^2 g_1^5 g_2^6
+ \frac{81655}{5184} \zeta_5 N g_1^2 g_2^9
+ \frac{93265}{1296} \zeta_5 N g_1^{10} g_2
+ \frac{116125}{648} \zeta_3 N^2 g_1^7 g_2^4
\right. \nonumber \\
&& \left. ~~~~
+ \frac{205339}{165888} N^3 g_1^7 g_2^4
+ \frac{279349}{10368} \zeta_3 N^3 g_1^{10} g_2
+ \frac{292661}{17496} N^2 g_1^6 g_2^5
\right. \nonumber \\
&& \left. ~~~~
+ \frac{498613}{7776} \zeta_3 N^2 g_1^{10} g_2
+ \frac{515255}{3888} \zeta_3 N^2 g_1^9 g_2^2
+ \frac{518057}{13824} \zeta_3 N^2 g_1^6 g_2^5
\right. \nonumber \\
&& \left. ~~~~
+ \frac{825475}{1728} \zeta_5 N g_1^7 g_2^4
+ \frac{1057981}{3888} \zeta_3 N g_1^{10} g_2
+ \frac{1206983}{10368} \zeta_3 N g_1^3 g_2^8
\right. \nonumber \\
&& \left. ~~~~
+ \frac{1215581}{15552} \zeta_3 N g_1^7 g_2^4
+ \frac{2205067}{10368} \zeta_3 N g_1^8 g_2^3
+ \frac{2369663}{13824} \zeta_3 N g_1^4 g_2^7
\right. \nonumber \\
&& \left. ~~~~
+ \frac{2895059}{5184} \zeta_3 N g_1^9 g_2^2
+ \frac{7655545}{124416} \zeta_3 g_2^{11}
+ \frac{11130787}{186624} N g_1^7 g_2^4
\right. \nonumber \\
&& \left. ~~~~
+ \frac{11441425}{4478976} N^3 g_1^9 g_2^2
+ \frac{15400453}{5971968} N^2 g_1^4 g_2^7
+ \frac{16013395}{62208} \zeta_3 N g_1^6 g_2^5
\right. \nonumber \\
&& \left. ~~~~
+ \frac{28704085}{373248} N^2 g_1^{11}
+ \frac{34251571}{497664} N g_1^3 g_2^8
+ \frac{94907947}{2239488} N^2 g_1^7 g_2^4
\right. \nonumber \\
&& \left. ~~~~
+ \frac{128402411}{1492992} N g_1^5 g_2^6
+ \frac{213879685}{3359232} N^2 g_1^8 g_2^3
+ \frac{287841901}{6718464} N g_1^{10} g_2
\right. \nonumber \\
&& \left. ~~~~
+ \frac{331932757}{1119744} N g_1^9 g_2^2
+ \frac{470815561}{6718464} N g_1^4 g_2^7
+ \frac{588615251}{3359232} N^2 g_1^{10} g_2
\right. \nonumber \\
&& \left. ~~~~
+ \frac{759191107}{2239488} N g_1^8 g_2^3
+ \frac{2002332845}{26873856} N g_1^6 g_2^5
+ 15 \zeta_3^2 N^2 g_1^7 g_2^4
\right] \,+\, O(g_i^{13}) \,.
\label{beton2til}
\end{eqnarray}
It is evident from both expressions that no terms involve $\zeta_4$ or 
$\zeta_6$. The situation is the same for the other two renormalization group
functions $\gamma^{O(N)}_{\phi\,\MOMtss}(g_1,g_2)$ and 
$\gamma^{O(N)}_{\sigma\,\MOMtss}(g_1,g_2)$ as is apparent by using a search 
tool on the respective expressions recorded in the arXiv data file associated 
with the article \cite{44}.

\end{document}